\documentclass[fleqn,usenatbib]{mnras}

\usepackage{newtxtext,newtxmath}

\usepackage[T1]{fontenc}
\usepackage{deluxetable}
\DeclareRobustCommand{\VAN}[3]{#2}
\let\VANthebibliography\thebibliography
\def\thebibliography{\DeclareRobustCommand{\VAN}[3]{##3}\VANthebibliography}

\usepackage{graphicx}	
\usepackage{amsmath}	

\newcommand\SPSB[2]{^{#1}_{#2}}

\title[Perseus COM Survey]{Survey of Complex Organic Molecules in Starless and Prestellar Cores in the Perseus Molecular Cloud}

\author[S. Scibelli et al.]{
Samantha Scibelli,$^{1,2}$\thanks{E-mail: sscibell@nrao.edu}\thanks{Jansky Fellow of the National Radio Astronomy Observatory.}
Yancy Shirley,$^{2}$ Andr\'es Meg\'ias,$^{3}$ Izaskun Jim\'enez-Serra$^{3}$
\\
$^{1}$National Radio Astronomy Observatory, 520 Edgemont Road, Charlottesville, VA 22903, USA\\ 
$^{2}$Steward Observatory,  University of Arizona, 933 North Cherry Avenue,
Tucson, AZ 85721, USA\\
$^{3}$Centro de Astrobiolog\'ia (CAB), CSIC-INTA, Carretera de Ajalvir, km 4, E-28805 Torrej\'on de Ardoz, Spain
}
\date{Accepted 2024 August 19. Received 2024 August 2; in original form 2024 May 23}

\pubyear{2024}

\begin{document}
\label{firstpage}
\pagerange{\pageref{firstpage}--\pageref{lastpage}}
\maketitle

\begin{abstract}
Cold ($\sim$10 K) and dense ($\sim$10$^{5}$ cm$^{-3}$) cores of gas and dust within molecular clouds, known as starless and dynamically evolved prestellar cores, are the birthplaces of low-mass ($M$ $\leq$ few M$_\odot$) stars. As detections of interstellar complex organic molecules, or COMs, in starless cores has increased, abundance comparisons suggest that some COMs might be seeded early in the star formation process and inherited to later stages (i.e., protostellar disks and eventually comets). To date observations of COMs in starless cores have been limited, with most detections reported solely in the Taurus Molecular Cloud. It is therefore still a question whether different environments affect abundances. We have surveyed 35 starless and prestellar cores in the Perseus Molecular Cloud with the Arizona Radio Observatory (ARO) 12m telescope detecting both methanol, CH$_3$OH, and acetaldehyde, CH$_3$CHO, in 100\% and 49\% of the sample, respectively. In the sub-sample of 15 cores where CH$_3$CHO was detected at $>3\sigma$ ($\sim$18\,mK) with the ARO 12m, follow-up observations with the Yebes 40m telescope were carried out. Detections of formic acid, $t$-HCOOH, ketene, H$_2$CCO, methyl cyanide, CH$_3$CN, vinyl cyanide, CH$_2$CHCN, methyl formate, HCOOCH$_3$, and dimethyl ether, CH$_3$OCH$_3$, are seen in at least $20\%$ of the cores. We discuss detection statistics, calculate column densities, and compare abundances across various stages of low-mass star formation. Our findings have more than doubled COM detection statistics in cold cores and show COMs are prevalent in the gas before star and planet formation in the Perseus Molecular Cloud.
\end{abstract}

\begin{keywords}
astrochemistry -- stars: formation -- ISM: molecules
\end{keywords}



\section{Introduction}

The detection of organic molecules in the interstellar medium (ISM) is intriguing, since these molecules are key to understanding the origins and evolution of organic chemistry, the basis for life on Earth. The complexity of organic molecules detected in the ISM has grown significantly over the years, from the simple diatomic carbon monoxide (CO; \citealt{1970ApJ...161L..43W}) to large rings such as the polycyclic aromatic hydrocarbon benzonitrile (c-C$_6$H$_5$CN; \citealt{2018Sci...359..202M}). 

An organic molecule observed in the ISM is traditionally considered complex if it has more than five atoms and at least one carbon-atom \citep{2009ARA&A..47..427H}. These interstellar complex organic molecules, or COMs, have been observed for a long time to be abundant in dynamic star-forming environments such as towards the Galactic Center \citep{1970ApJ...162L.203B, 1975ApJ...195L.127G, 1975ApJ...197L..29B, 2000ApJ...540L.107H, 2008ApJ...672..352R, 2018MNRAS.478.2962Z, 2020AsBio..20.1048J, 2022ApJ...929L..11R}, as well as towards the star forming region in the Orion Nebula \citep{1974ApJ...191L..79S, 2014ApJ...784L...7K, 2015ApJ...803...97S}. 

Beyond these quintessential regions, it is also the prevalence of COMs, and their relevance to prebiotic chemistry, in the earliest stages of low-mass ($M$ $\leq$ few M$_\odot$) star and planet formation that has sparked the interest of astrochemists and astrobiologists. COMs have now been detected in virtually each stage of the low-mass star formation process, including low-mass protostars \citep{2011ApJ...740...14O, 2015ApJ...804...81T, 2016ApJ...830L..37I, 2017A&A...606A.121L}, protoplanetary disks \citep{2016ApJ...823L..10W, 2018ApJ...862L...2F, 2018ApJ...859..131L}, the atmospheres of solar system moons \citep{2017AJ....154..206L, 2020ApJ...903L..22T}, and comets \citep{2019ECS.....3.1550B, 2019ECS.....3.1854S, 2019AGUFM.P43C3489R}. 

To best probe the initial chemical conditions at the earliest stage of low-mass star formation, it is imperative to study cold (10\,K) and dense (10$^5$ cm$^{-3}$) starless cores and prestellar cores (e.g., \citealt{2007ARA&A..45..339B}). A starless core is used as a generic term for any dense cold core, whereas prestellar cores are those that have overcome turbulence, thermal and magnetic pressure and will eventually dynamically collapse due to gravity and external cloud pressure to form an infant star. Planetesimals are likely to inherit at least some of their organic chemistry from the starless and prestellar core phase, which can in turn drive additional chemical complexity in later stages of star and planet formation \citep{2019MNRAS.490...50D, 2020A&A...639A..87V, 2021NatAs...5..684B, 2021MNRAS.504.5754S}. Observational constraints on the complex molecular content of starless and prestellar cores are therefore important for piecing together the astrochemical origins of the composition of solar and planetary systems \citep{2021PhR...893....1O}.

In hotter regions, during the `warm-up' of ices on the outer layers of grains at temperatures of 20-40\,K, it has been shown that COMs can form and be released into the gas-phase through thermal desorption and/or diffusive surface radical chemistry facilitated by UV radiation from cosmic rays (see reviews \citealt{2020ARA&A..58..727J, 2022arXiv220613270C}). For cold (10\,K) and quiescent starless and prestellar cores, alternative models with gas-phase reactions of smaller species, namely CH$_3$OH and smaller radicals, have been developed to explain how additional COMs form \citep{2013ApJ...769...34V, 2015MNRAS.449L..16B}. Updated models for COMs support this formation process called reactive desorption, where the precursor molecules form on icy surfaces of interstellar grains, and then get ejected into the gas to form larger molecules \citep{2016A&A...585A..24M, 2017ApJ...842...33V}. 

Attempts to use the \citealt{2016A&A...585A..24M} and \citealt{2017ApJ...842...33V} models to match the high abundances of COMs observed toward younger and less dense starless cores, such as L1521E, over-predict CH$_3$OH and under-predict larger COMs (e.g., see \citealt{2021MNRAS.504.5754S}). Models with updated non-diffusive surface reaction networks \citep{2020ApJS..249...26J, 2022ApJS..259....1G}, as well as the addition of cosmic ray chemistry \citep{2018ApJ...861...20S}, supply additional mechanisms for getting these large COMs off the grains into the gas-phase, but have only been tested against few well-known cold cores (i.e., L1544 and TMC-1; see \cite{2017ApJ...842...33V, 2018ApJ...861...20S, 2020ApJS..249...26J}). To guide these modeling efforts and constrain COM formation routes, additional abundance constraints from a larger sample of starless and prestellar cores are needed to compare to models. 

Much of the search for COMs in starless and prestellar cores in low-mass star forming environments has been restricted to the Taurus Molecular Cloud. It is in Taurus where the famous cold (10\,K) dark cloud TMC-1 resides, which has a rich inventory of complex chemistry \citep{1985ApJ...290..609M, 2004PASJ...56...69K, 2015ApJ...802...74S, 2018ApJ...854..116S, 2018Sci...359..202M, 2021Sci...371.1265M, 2021A&A...652L...9C, 2021A&A...649L...4A}. Additionally, within Taurus is the chemically and dynamically evolved prestellar core L1544, which shows high levels of CO depletion and high central densities $\geq 10^7$\,cm$^{-3}$ \citep{1999ApJ...523L.165C, 2007A&A...470..221C, 2010MNRAS.402.1625K, 2019ApJ...874...89C}, and many COM detections (e.g., species such as acetaldehyde, CH$_3$CHO, methyl formate, CH$_3$OCHO, dimethyl ether, CH$_3$OCH$_3$, and vinyl cyanide, CH$_2$CHCN; \citealt{2014ApJ...795L...2V, 2016ApJ...830L...6J}). More recently, there were numerous COM detections in the chemically young core L1521E \citep{2021MNRAS.504.5754S} as well as nitrogen bearing (N-bearing) COMs observed in the dynamically young core L1498 \citep{2021ApJ...917...44J}, both in Taurus. \cite{2023MNRAS.519.1601M} have suggested a scenario in which N-bearing and oxygen bearing (O-bearing) COMs form at different stages in the evolution of starless cores. However, the number of cores studied by these authors was small. Beyond these few case studies, \cite{2020ApJ...891...73S} conducted a survey towards 31 starless and prestellar cores in the L1495 filament of the Taurus Molecular Cloud, finding a prevalence of both methanol, CH$_3$OH, in 100\% of the cores targeted and CH$_3$CHO in 70\%. It is still unknown whether Taurus is a unique environment and if other molecular clouds show a similar prevalence of COMs in a larger sample of starless and prestellar cores.

For this study we have focused on conducting a survey of COMs towards 35 starless and prestellar cores in the Perseus Molecular Cloud. Perseus is one of the most studied active nearby ($\sim$300 pc) star-forming regions, with recent \textit{Herschel} cataloging finding a total 684 starless and prestellar cores and 132 protostars (Class 0, I, and II) in the region  \citep{2021A&A...645A..55P}. It is important to note that it is in Perseus where the first COM detections near the dense star-forming core B1-b were observed, in close proximity (within $\sim$30\,arcsec) to a protostar and outflow \citep{2010ApJ...716..825O, 2012ApJ...759L..43C}. Recent focus has been on the COM content of protostars in Perseus, with studies such as the CALYPSO survey which looked at 26 solar-type protostars \citep{2020A&A...635A.198B}, and the complementary  Perseus ALMA Chemistry Survey, or PEACHES, which probed warm-phase complex chemistry toward 50 embedded protostars in Perseus finding COM emission in 58\% of the sources \citep{2021ApJ...910...20Y}. Here we investigate whether we see a similar prevalence of chemical complexity during the cold gas-phase chemistry within starless and prestellar cores in the same cloud. 

We begin by detailing in section\,\ref{sec:source} the source selection and catalog matching for each of the starless and prestellar cores targeted in the survey. Then, we describe the molecular line observations and data reduction techniques in section\,\ref{sec:obs}. The detection statistics and column density calculations are presented in section\,\ref{sec:results}. In section\,\ref{sec:discussion} we analyze COM abundances correlations, compare COM abundances to starless and prestellar cores in other molecular clouds as well as to objects at other stages of low-mass star formation, and discuss the implications for COM formation routes. We summarize and conclude in section\,\ref{sec:conclusion}.

\section{Source Selection} \label{sec:source}

Each of the 35 starless and prestellar cores targeted first by the ARO 12m were selected by cross-referencing with Perseus core catalogs compiled from Bolocam 1.1 mm continuum maps \citep{2008ApJ...684.1240E}, ammonia, NH$_3$, observations \citep{2008ApJS..175..509R}, as well as the higher resolution \textit{Herschel} core catalog \citep{2021A&A...645A..55P}. We include cores that have corresponding NH$_3$ data because this molecule does not appear to be affected by dramatic freeze-out like CO does until high densities ($>10^6$\,cm$^{-3}$; \citealt{2022ApJ...929...13C}), making it a useful probe of structure as well as kinetic temperature, $T_\mathrm{k}$ in dense prestellar cores \citep{2002ApJ...569..815T, 2015ApJ...805..185S, 2017ApJ...843...63F, 2019ApJ...877...93C, 2019ApJ...886..119C, 2020ApJ...891...84C}. Additionally, the 31 starless and prestellar cores in the Taurus Molecular Cloud that were targeted in the \cite{2020ApJ...891...73S} survey were also selected based on NH$_3$ mapping results from \cite{2015ApJ...805..185S}. For the 35 cores in Perseus we target, $T_\mathrm{k}$ ranges from 9.5 to $<$20\,K and the NH$_3$ column density, $N_\mathrm{NH_3}$, ranges from 0.6 -- 89 $\times$ 10$^{13}$ cm$^{-2}$ \citep{2008ApJS..175..509R}. 

\begin{table*}
	\caption{Perseus Starless and Prestellar Core Catalog}
 \setlength{\tabcolsep}{8pt}
	\label{physparams}
	\begin{tabular}{ccccccccccc} 
    \tablecolumns{10}
     \tablewidth{0pt}
     \tabcolsep=0.2cm
Region & Core \#$^{1}$ &  Core \#$^{2}$ &  Core \#$^{3}$  &  RA$^4$ & DEC$^4$  & $D^{5}$ & \textit{N}(H$_2$)$^6$  & \textit{n}(H$_2$)$^6$ & $T_\mathrm{k}$$^{7}$ \\  \colhead{}& (\textit{Herschel}) &  (NH$_3$)  &  (Bolocam)  &  (J2000)  & (J2000)  & (pc) & 10$^{22} $cm$^{-2}$  &\ 10$^{5}$ cm$^{-3}$ & K \\
\hline
 L1451 & 54 &6 & Per-Bolo4 & 03:25:17.89 & +30:18:56.6 & 283 & 1.10 & 0.40 & 10.3\\
&  67 & 11 & Per-Bolo7 &  03:25:35.11 & +30:13:12.8 & 283 & 1.38 & 0.52 & 10.4$^{*}$ \\
L1455&  130 & 24 & Per-Bolo19 &  03:27:02.78 &	+30:15:21.7 & 286& 0.95 & 0.35 & 10.4  \\
&  231 & 42 & Per-Bolo27 &  03:28:34.16 &	+30:19:38.8& 286 & 1.02 & 0.38 & 10.7 \\
&  256 & 49 & Per-Bolo32 &  03:28:43.14 &	+30:31:08.6& 286 & 1.07 & 0.40 & 10.5$^{*}$ \\
NGC1333 & 264 & 51 & Per-Bolo34 & 03:28:47.14 & +31:15:11.4& 294 & 2.06 & 0.75 & 10.8 \\
&  317 & 69 & Per-Bolo44 &  03:29:04.93 &	+31:18:44.4& 294 & 2.63 & 0.96 & 13.6 \\
&   321  & 72 & Per-Bolo45 & 03:29:07.17 &	+31:17:22.1 & 294& 3.65 & 1.33 & 12.6 \\
& 326 & 73 & Per-Bolo46 & 03:29:08.97 &  +31:15:17.2 & 294& 7.88 & 2.87 & 12.3 \\
& 339 & 79 & Per-Bolo50 & 03:29:15.81 & +31:20:31.6 & 294 & 2.11 & 0.77 & 15.5 \\
& 344 & 82 & Per-Bolo53 & 03:29:18.39  & +31:25:07.2 & 294& 1.46 & 0.53 & 11.9\\
&   355  & 86 & Per-Bolo56 &  03:29:23.77 &	+31:36:12.6& 294 & 1.02 & 0.37 & 9.8 \\
 Per6 &  398   & 93 & Per-Bolo61 &  03:30:25.08 & +30:27:42.6 & 290 & 0.82 & 0.30 & 10.5 \\
B1 &  413  & 96 & Per-Bolo63 &  03:30:46.74 &	+30:52:44.8 & 297 & 1.17 & 0.42 & 10.5\\
&   414  & 97 & Per-Bolo64 &  03:30:50.76 &	+30:49:21.6& 297 & 0.48 & 0.17 & 9.8\\
 & 479 &  108 & Per-Bolo70 & 03:32:43.72 & +30:59:48.5& 297 & 2.56 & 0.93 & 10.2 \\
&  491  & 117 & Per-Bolo76 &  03:33:10.91 &	+31:21:44.7& 297 & 1.00 & 0.36 & 10.8$^{*}$ \\
&   504  & 124 & Per-Bolo82 &  03:33:25.31 &	+31:05:37.5 & 297 & 1.80 & 0.65 & 9.7 \\
B1-E &   543  & 132 & Per-Bolo87 &  03:35:23.02 &	+31:06:50.6 & 301 & 0.67 & 0.24 & 11.7$^{*}$\\
IC348 &    615 & 141 & Per-Bolo88 & 03:40:14.92 & +32:01:40.8& 314 & 1.03 & 0.35 & 10.3$^{*}$\\
&    627 & 142 & Per-Bolo89 & 03:40:49.53 & +31:48:40.5 & 314 & 0.93 & 0.32 & 12.4\\
&   642 & 144 & Per-Bolo91 & 03:41:20.42 & +31:47:32.7& 314 & 0.69 & 0.24 & 13.3$^{*}$\\
&   656 & 145 & Per-Bolo92 & 03:41:40.64 & +31:58:05.4 & 314 & 0.97 & 0.33 & 9.5\\
&    657 & 146 & Per-Bolo93 & 03:41:45.86 & +31:48:10.8& 314 & 0.87 & 0.30 & 10.3$^{*}$\\
&   658 & 147 & Per-Bolo94 & 03:41:46.68 & +31:57:29.4 & 314 & 0.81 & 0.28 & 9.6\\
&   709 & 156 & Per-Bolo99 & 03:43:38.06 & +32:03:07.4 & 314 & 1.48 & 0.51 & 13.5 \\
&   715 & 158 & Per-Bolo101 & 03:43:46.34 & +32:01:43.5 &314  & 1.45 & 0.49 & 14.8\\
&   739 & 168 & Per-Bolo109 & 03:44:05.28 & +32:00:38.8 &314 & 1.53 & 0.52 & 9.2$^{*}$\\
&   746 & 170 & Per-Bolo111 & 03:44:14.38 & +31:58:00.7 & 314& 1.42 & 0.49 & 10.6\\
&   747 & 171 & Per-Bolo112 & 03:44:15.08 & +32:09:13.1 & 314& 0.71 & 0.24 & 12.9\\
&   752 & 174 & Per-Bolo114 & 03:44:23.10 & +32:10:01.0 & 314& 0.63 & 0.22 & 14.7\\
&   768 & 180 & Per-Bolo117 & 03:44:48.83 & +32:00:31.6 & 314& 1.43 & 0.49 & 10.8\\
HPZ6 &   780 & 183 & Per-Bolo119 & 03:45:16.47 & +32:04:47.6& 318 & 0.98 & 0.33 & 10.7 \\ 
B5 &   799 & 188 & Per-Bolo121 & 03:47:31.31 & +32:50:56.9& 325 & 1.01 & 0.33 & 10.4 \\
&   800 & 192 & Per-Bolo122 & 03:47:38.97 & +32:52:16.6& 325 & 1.93 & 0.64 & 11.7 \\
		\hline
    \end{tabular}
      \begin{description} 
     \item $^{1}$ Catalog of \textit{Herschel} cores for the Perseus region (\citealt{2021A&A...645A..55P}). $^{2}$ \cite{2008ApJS..175..509R} provided a list of NH$_3$ clumps for the Perseus region. $^{3}$ Bolocam naming convention for the Perseus cores is from \cite{2008ApJ...684.1240E}. 
$^{4}$ RA and DEC values in this table are those that correspond with the most recent \textit{Herschel} data. $^{5}$ Distance, $D$, is the mean distance for that particular region as listed in Table 5 in  \citealt{2021A&A...645A..55P}. $^{6}$ Median value from \textit{Herschel} maps within the ARO 12m 62$''$ beam. $^{7}$ Kinetic temperature values for RADEX calculations derived from NH$_3$ as in \cite{2008ApJS..175..509R}, except for cores marked by $^{*}$ where only upper limits are presented and instead the \textit{Herschel} $T_\mathrm{dust}$ value is used (Table A.2. in \cite{2021A&A...645A..55P}).
      \end{description}
\end{table*}

In Table\,\ref{physparams} the core numbers for each catalog are listed, and we refer to each core by their \textit{Herschel} number and corresponding coordinates throughout the remainder of the paper. It should be noted that in some cases the corresponding NH$_3$ peak is slightly offset from the \textit{Herschel} peak. Observations with the Green Bank 100m dish probing how NH$_3$ follows the dust (from SCUBA observations) also show that in a sample of these cores there are slight offsets in the peak positions \citep{2014MNRAS.440.1730M}, but negligible within our single-dish beam sizes. For consistency, it is the \textit{Herschel} dust peak position that is used when carrying out our observations. 
The coordinates of each of the starless and prestellar cores were also checked to make sure there was no overlap (within an ARO 12m 62 arcsec beam) with any other core, including protostellar sources, that were listed in \citealt{2021A&A...645A..55P} and this narrowed our catalog down to the final 35 cores. The \textit{Herschel} data provides us with corresponding H$_2$ column density, \textit{N}(H$_2$), information as well as dust temperatures, $T_\mathrm{dust}$, for each core, with $T_\mathrm{dust}$ ranging from 8 -- 15 K, peak \textit{N}(H$_2$) ranging from 0.45 -- 23 $\times$ 10$^{22}$ cm$^{-2}$ and core masses ranging from 0.3 -- 5.5 M$_{\odot}$ \citep{2021A&A...645A..55P}. In the cases where $T_{k}$ is uncertain, and only an upper limit is provided in \citealt{2008ApJS..175..509R}, we use the $T_\mathrm{dust}$ value (see Table\,\ref{physparams}). In general, $T_{k}$ and $T_\mathrm{dust}$ do not differ by more than a factor of 1.6. 

The starless and prestellar cores targeted in this survey span across the entire $\sim$10\,pc of Perseus and thus different distances need to be adopted for each individual sub-region. For example, the eastern region of IC348 and the western region of NGC1333 are at different distances of 321$\pm$10\,pc and 293$\pm$22\,pc, respectively, according to \cite{2018ApJ...865...73O}, who use VLBA and \textit{Gaia} data. \cite{2019ApJ...879..125Z} went on to derive a distance map of Perseus (along with other clouds) using `per-star distance–extinction estimates' along with stellar distances from \textit{Gaia} Data Release 2 (DR2) parallax measurements \citep{2016A&A...595A...1G, 2018A&A...616A...1G}, and find similarly distances in Perseus range from 234 to 331 pc. For this paper we adopt the mean distance for each sub-region based on more recent distances also derived from \textit{Gaia} DR2 results as described in \cite{2021A&A...645A..55P}, and listed here in column 7 of Table\,\ref{physparams}. These distances agree well within $< 10\%$ to those derived in both \cite{2018ApJ...865...73O} and \cite{2019ApJ...879..125Z}.

\begin{figure*}
\includegraphics[width=178mm]{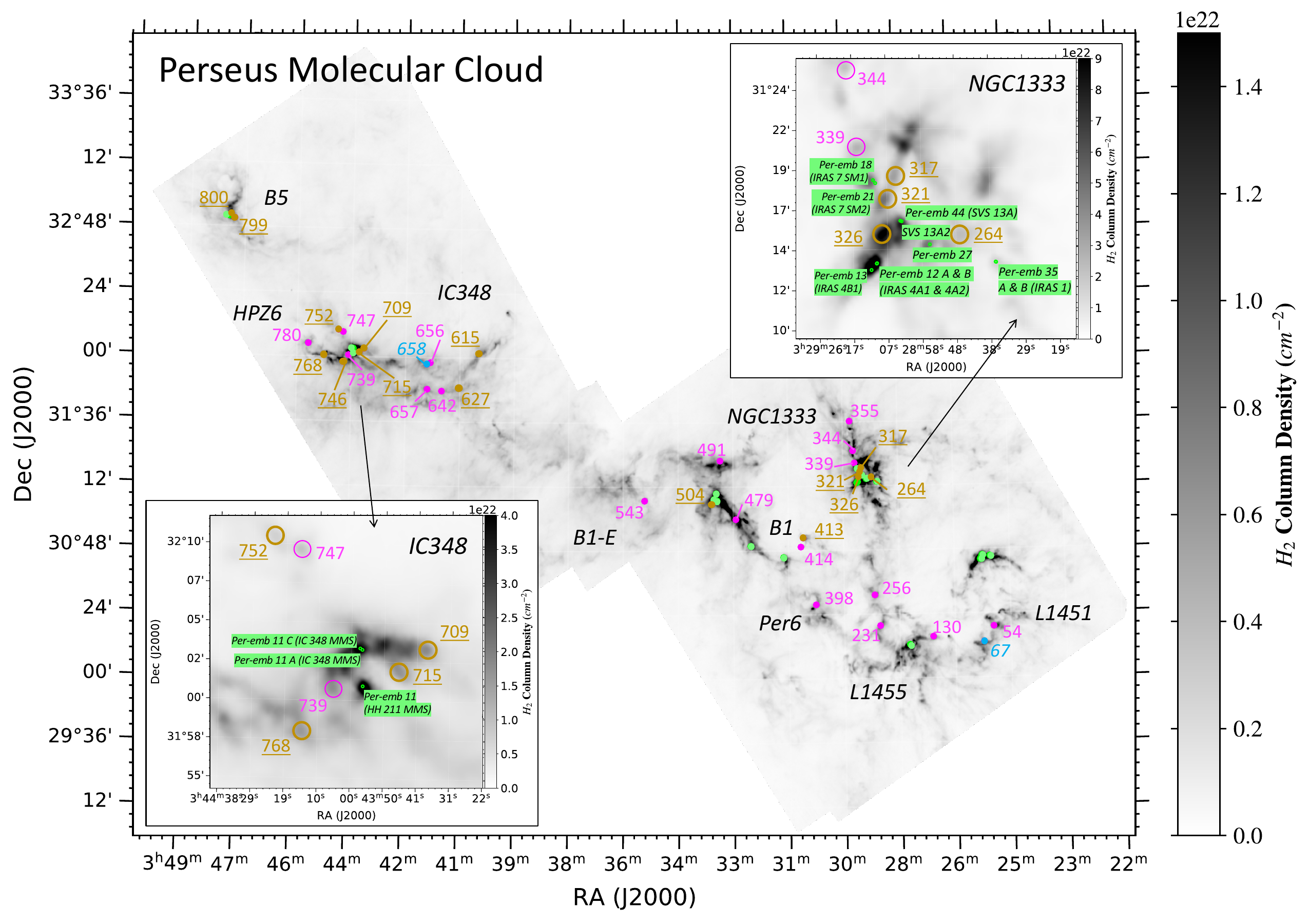} 
\caption{ \label{fig1} Perseus Molecular Cloud H$_2$ column density map from \textit{Herschel} \citep{2012A&A...540A..10S, 2012A&A...547A..54P, 2014ApJ...787L..18S}. Overplotted are the numbered starless and prestellar cores observed (pink, gold and orange points) and those underlined in gold have had successful ARO 12m CH$_3$CHO detections at the $\sigma_{T_\mathrm{mb}} = 6\,\mathrm{mK}$ level and were followed up with the Yebes 40m telescope. Cores 67 and 658 are italicized and labeled in blue, denoting these are the cores that were followed up with additional ARO 12m observations probing down to  $\sigma_{T_\mathrm{mb}} = 2.5\,\mathrm{mK}$. The remaining starless and prestellar cores are in magenta. In all cores CH$_3$OH has been detected. For comparison, in green are the locations of the 27 protostars from the PEACHES sample with CH$_3$OH detections (see their Table 5 in \citealt{2021ApJ...910...20Y}). The apparent overlap in sources is a projection effect, as seen from the zoom-in insert panels for regions IC348 and NGC1333. The size of the starless and prestellar core circles are at a FWHM of 62 arcsec. }
\end{figure*}

Lastly, we note that each core has been labeled `prestellar' in the \textit{Herschel} catalog, based firstly on the fact that no internal source of energy (e.g., protostar) is present and on if the core's self-gravity exceeds its pressure support. Because there is increasing evidence to suggest that gravity is not the only term that should be considered when determining if a core is likely to form a star \citep{2022MNRAS.515.5219G, 2022MNRAS.517..885O,2023ASPC..534..193P, 2023MNRAS.521.4579S}, we continue to refer to these objects as a whole as both `starless and prestellar' cores. In Figure\,\ref{fig1} the location of each starless and prestellar core is labeled.

\begin{table*}
	\caption{COM Line List}
 \setlength{\tabcolsep}{12pt}
	\label{LineList}
	\begin{tabular}{llllllll} 
    \tablecolumns{8}
     \tablewidth{0pt}
     \tabcolsep=0.4cm

       Telescope: Receiver &  Molecule & Transition & Rest Frequency, $\nu$ & $E_u/k$ & g$_u$  & $A_{ul}$ & $\theta_\mathrm{beam}^{1}$  \\ 
     &   &   &  (GHz)  & (K)  &  & (s$^{-1}$) & arcsec \\
        \hline  
         ARO 12m: 3mm  &  &  &  & &   &  \\ 
    & CH$_3$OH   &  2$_{-1,2} - 1_{-1,1}$  E& 96.739363 & 12.5 & 20 & 2.6E-06 & 62.3 \\
   &  &  2$_{0,2} - 1_{0,1}$ A & 96.741377 & 7.0 & 20 & 3.4E-06  & 62.3 \\
  &  &  2$_{0,2} - 1_{0,1}$ E & 96.744549  & 20.1  & 20 & 3.4E-06 & 62.3 \\
 & CH$_3$CHO & $5_{0,5}-4_{0,4}$ E & 95.947439 & 13.9  &22 & 3.0E-05  & 62.8 \\
 &  & $5_{0,5}-4_{0,4}$ A & 95.963465 & 13.8   & 22& 3.0E-05 & 62.8 \\
& CH$_2$CHCN &	$10_{0,10} - 9_{0,9}$&	94.276641	& 24.9  &63	&  6.2E-05	& 64.0 \\
& &	$10_{1,9} - 9_{1,8}$&	96.982446	& 27.8  &63	&  7.2E-05	& 62.2\\
 Yebes 40m: Q-band &    &  &  & &   &  \\ 
 & t-HCOOH & 2$_{1,2} - 1_{1,1}$ & 43.303709  & 6.2  & 5 &  5.7E-07 & 41.8 \\
 &  & 2$_{0,2} - 1_{0,1}$ & 44.911750 & 3.2 & 5 & 8.2E-07 & 40.3\\
 &   & 2$_{1,1} - 1_{1,0}$ & 46.581226  &  6.5 & 5 &  7.1E-07 &  38.8 \\
 &H$_2$CCO & 2$_{1,2} - 1_{1,1}$ o- & 40.039022 & 15.9 & 15 & 4.5E-07 & 45.2 \\
 & & 2$_{0,2} - 1_{0,1}$ p- & 40.417950 & 2.9 & 5 & 6.2E-07 &  44.8 \\
 & & 2$_{1,1} - 1_{1,0}$ o- & 40.793832 & 16.0 & 15 & 4.7E-07 & 44.3\\
 & CH$_3$OH   &  1$_{0,1} - 0_{0,0}$  A & 48.372460  & 2.3 & 12  & 3.6E-07  & 37.4\\
&    &  1$_{-0,1} - 0_{-0,0}$  E &  48.376887 & 15.4 &  12 & 3.6E-07 & 37.4\\
&CH$_3$CN & $2_1 - 1_1$  & 36.794765   & 9.8 & 10 & 1.8E-06  & 49.2 \\
& &  $2_0 - 1_0$   &  36.795474 &   2.6 & 10 & 2.5E-06 &  49.2 \\
&  CH$_3$CHO & 2$_{1,2} - 1_{1,1}$ A  & 37.464204  & 4.9 & 10  & 1.2E-06 & 48.3\\
&       &  2$_{1,2} - 1_{1,1}$ E  & 37.686932 & 5.0 & 10 & 1.1E-06 & 48.0 \\
&       & 2$_{0,2} - 1_{0,1}$ E & 38.506035  &  2.8 & 10 & 1.7E-06 & 47.0 \\
&      & 2$_{0,2} - 1_{0,1}$ A & 38.512079 &  2.8 & 10 & 1.7E-06 & 47.0  \\
&     & 2$_{1,1} - 1_{1,0}$ E & 39.362537  & 5.2 & 10  &  1.3E-06  & 46.0 \\
&     & 2$_{1,1} - 1_{1,0}$ A & 39.594289  & 5.1 & 10 & 1.4E-06 & 45.7 \\
& CH$_2$CHCN & 4$_{1,4} - 3_{1,3}$ & 37.018922 & 6.6 & 27 & 3.6E-06 & 48.9 \\
&   & 4$_{0,4} - 3_{0,3}$&  37.904849 & 4.5 & 27 & 4.1E-06 & 47.7\\
&   & 4$_{2, 3}- 3_{2, 2}$ & 37.939620 & 13.2  & 27 &  3.1E-06 & 47.7 \\
& HCOOCH$_3$  & 3$_{0, 3}- 2_{0, 2}$ E & 36.102224 & 3.5 & 14 & 6.2E-07 & 50.1 \\
&            & 3$_{0, 3}- 2_{0, 2}$ A & 36.104793 & 3.5 & 14 & 6.2E-07 & 50.1 \\
& & 3$_{2, 2}- 2_{2, 1}$ E & 36.678607 & 6.2 & 14 &  3.3E-07 &  49.3 \\
&             & 3$_{2, 2}- 2_{2, 1}$ A & 36.657467 & 6.2 & 14 &  3.6E-07 & 49.3 \\
& & 4$_{1,4} - 3_{1,3}$ E &  45.395795 & 6.1 & 18 & 1.2E-06   & 39.8 \\
& & 4$_{1,4} - 3_{1,3}$ A &  45.397380 & 6.1 & 18 & 1.2E-06   & 39.8 \\
& & 4$_{0,4} - 3_{0,3}$ E &  47.534093 & 5.8 & 18 & 1.5E-06  & 38.1 \\
& & 4$_{0,4} - 3_{0,3}$ A &  47.536915 & 5.8 & 18 & 1.5E-06 &  38.0 \\
& CH$_3$OCH$_3$ & $3_{1, 2} - 3_{0, 3}$ AE+EA & 32.977274  & 7.0 & 70 & 3.4E-07 & 54.8 \\
& & $3_{1, 2} - 3_{0, 3}$ EE & 32.978232 & 7.0 & 112 & 3.4E-07 & 54.8 \\
& & $3_{1, 2} - 3_{0, 3}$ AA & 32.979187  & 7.0 & 70 & 3.4E-07 & 54.8 \\
& & $4_{1, 3} - 4_{0, 4}$ AE+EA & 35.592414  & 10.8 & 54 &  4.0E-07 & 50.8 \\
& & $4_{1, 3} - 4_{0, 4}$ EE & 35.593408 & 10.8 & 144 &  4.0E-07 & 50.8\\
& & $4_{1, 3} - 4_{0, 4}$ AA & 35.594402  & 10.8 & 54 &  4.0E-07 & 50.8\\
		\hline
    \end{tabular}
      \begin{description} 
     \item  Values for CH$_3$CN, CH$_3$CHO and HCOOCH$_3$ from JPL catalog\footnote{\url{https://spec.jpl.nasa.gov/}} (\citealt{1998JQSRT..60..883P}) and for the remaining transitions from CDMS database\footnote{\url{https://cdms.astro.uni-koeln.de}} (\cite{2001A&A...370L..49M}, \cite{2005JMoSt.742..215M}, \cite{2016JMoSp.327...95E}). ${^1}$The beam size corresponding to the selected molecular transition. 
      \end{description}
\end{table*} 

\section{Observations and Data Reduction} \label{sec:obs}

Molecular line observations of COMs were taken with the Arizona Radio Observatory (ARO) Telescope 12m dish on Kitt Peak outside of Tucson, Arizona as well as with the 40m radio telescope of the Yebes Observatory (Guadalajara, Spain), known as the Yebes 40m. Lines targeted and analyzed in this paper are listed in Table\,\ref{LineList}.

\subsection{ARO 12m} \label{subsec:arored}

Single pointing observations on the ARO 12m for all 35 sources (Table\,\ref{physparams} and Figure\,\ref{fig1}) were carried out from October 2021 to April 2022 and again from April to May of 2023 with the 3mm sideband separating dual polarization receiver. Each scan was 5 minutes using absolute position switching (APS) between the source and the off position void of emission every 30 seconds. 
Pointing was checked every $\sim$1-2 hours on a nearby quasar or planet. The AROWS spectrometer, with a resolution of 39\,kHz, was used for all observations with the two polarizations (vertical and horizontal) tuned to simultaneously observe favorable transitions of CH$_3$OH, CH$_3$CHO and CH$_2$CHCN (see Table\,\ref{LineList}). An error/noise level of at least 10$\%$ in brightness temperature was adopted based on the monitoring of standard sources (see Figure\,\ref{fig:standardsource} in Appendix\,\ref{appendixARO}). 

\begin{figure*}
\includegraphics[width=172mm]{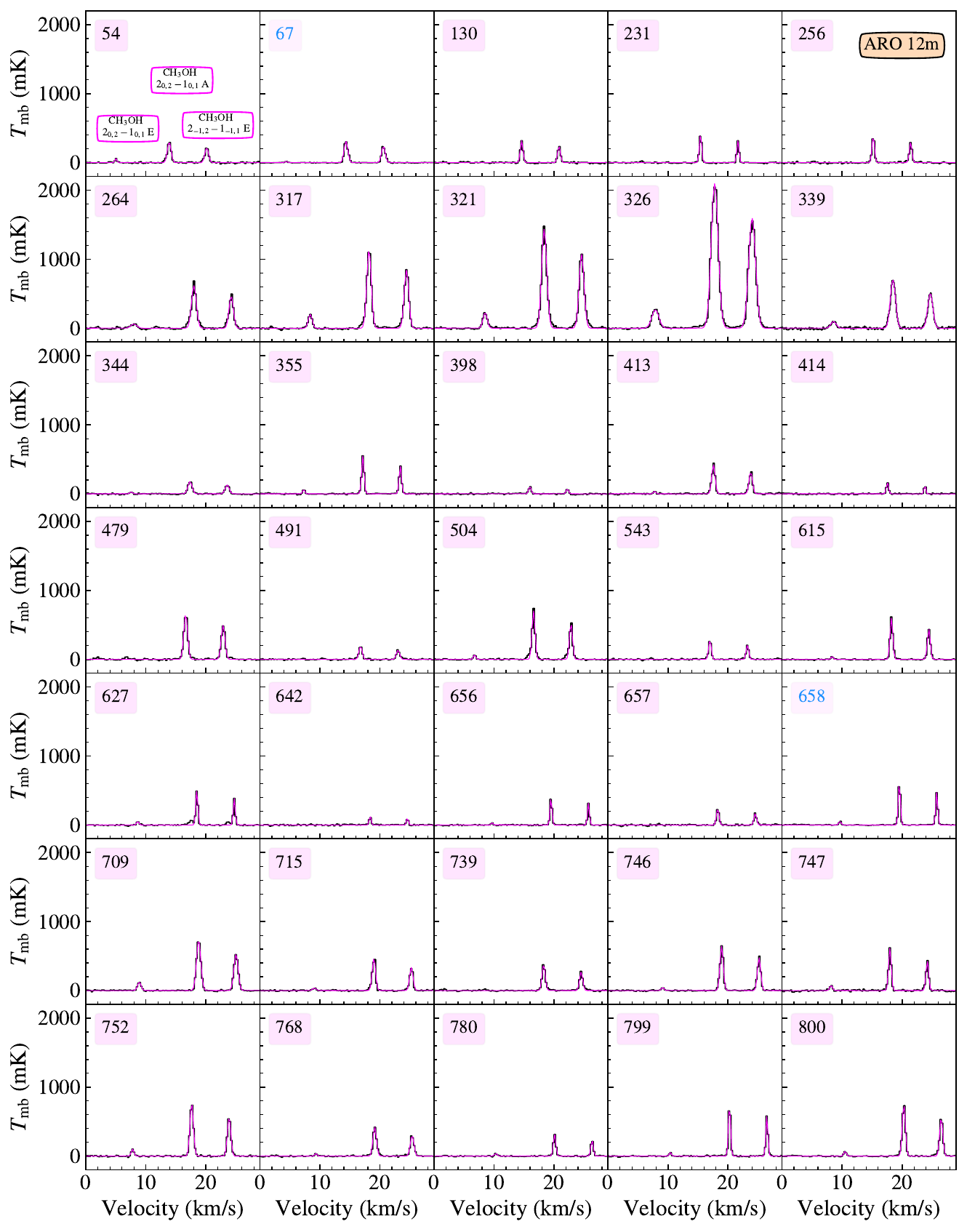} 
\caption{ \label{meth_spec_plot} Methanol, CH$_3$OH, spectrum (in black) in units of $T_\mathrm{mb}$ (K) versus velocity (km/s) from the ARO 12m for each of the 35 starless and prestellar cores targeted in this Perseus survey. Core numbers are labeled in the upper left of each panel. Gaussian fits are plotted in magenta. There are three 2-1 transitions observable in the spectral window, labeled in the uppermost left panel. Longer integration time on cores 67 and 658 (labeled in blue), resulted in RMS values around $\sigma_{T_\mathrm{mb}} = 3\,\mathrm{mK}$. For the remaining cores, the average RMS value is $\sigma_{T_\mathrm{mb}} = 7\,\mathrm{mK}$. }
\end{figure*}

\begin{figure*}
\includegraphics[width=172mm]{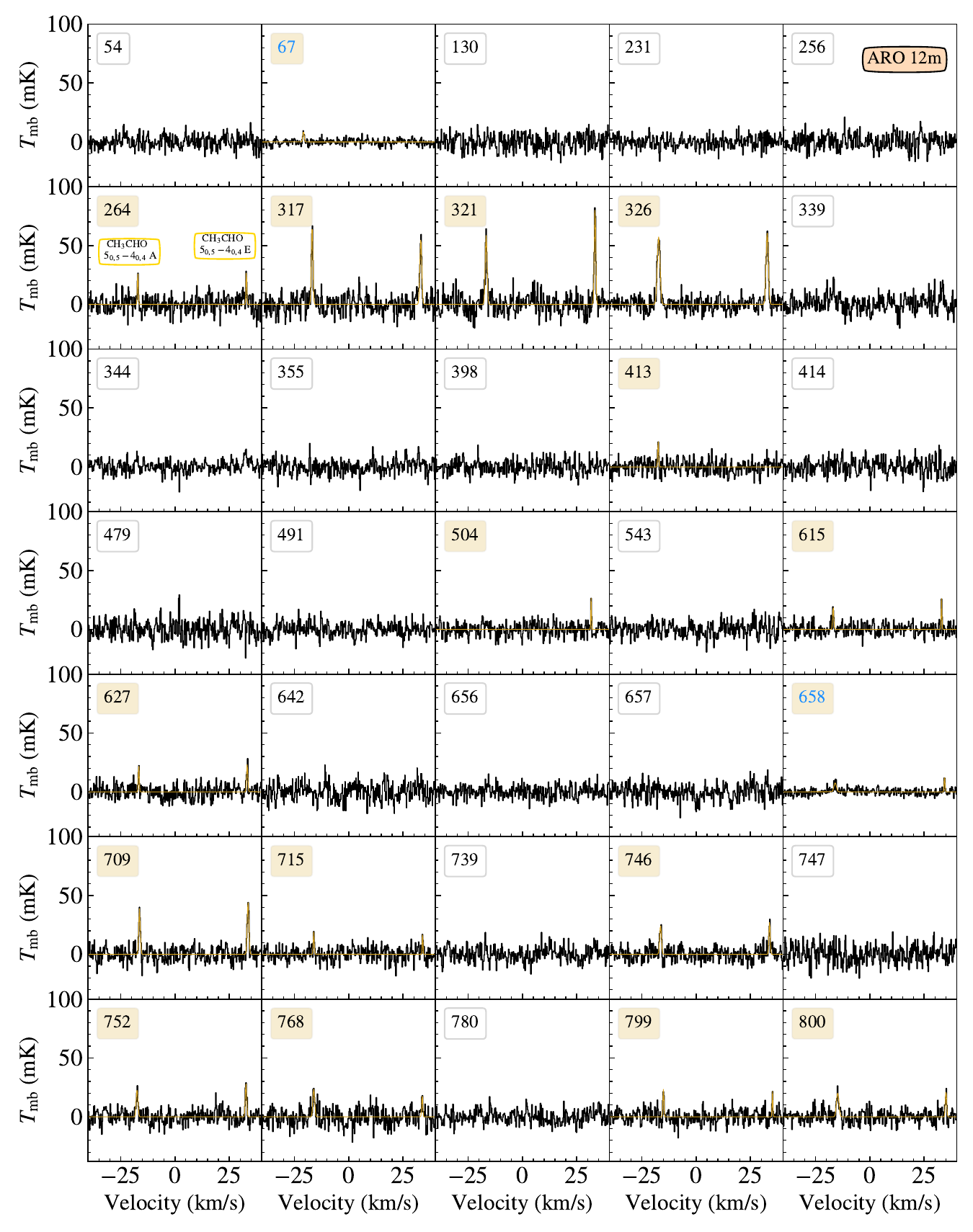} 
\caption{ \label{acet_spec_plot} Acetaldehyde, CH$_3$CHO, spectrum (in black) in units of $T_\mathrm{mb}$ (K) versus velocity (km/s) from the ARO 12m for each of the 35 starless and prestellar cores targeted in this Perseus survey. Core numbers are labeled in the upper left of each panel. Gaussian fits are plotted in gold. There are two 5-4 transitions observable in the spectral window, labeled in the first column of the second row panel. Longer integration time on cores 67 and 658 (labeled in blue), resulted in RMS values around $\sigma_{T_\mathrm{mb}} = 2.5\,\mathrm{mK}$. For the remaining cores, the average RMS value is $\sigma_{T_\mathrm{mb}} = 6\,\mathrm{mK}$.  }
\end{figure*}

Data reduction was performed using the CLASS program of the GILDAS package (\citealt{2005sf2a.conf..721P}, \citealt{2013ascl.soft05010G})\,\footnote{\url{http://iram.fr/IRAMFR/GILDAS/}}. For each polarization, a median efficiency percentage for each planet was calculated, along with estimated errors, during the different observing season and ranged from $\sim80-89\%$ (see Appendix\,\ref{appendixARO}). A main beam temperature was set by this beam efficiency, $\eta$, where $T_\mathrm{mb}$ = $T_A^*$/$\eta$ \citep{1993PASP..105..117M}. The baselined, hanning smoothed (by 2 channels), and scaled spectra were then fit with Gaussian line profiles within the CLASS routine and plotted using the Matplotlib package \citep{2007CSE.....9...90H} in Python. The reduced spectra and Gaussian fits for the detected transitions of CH$_3$OH and CH$_3$CHO are plotted in Figures\,\ref{meth_spec_plot} and\,\ref{acet_spec_plot}, respectively (detection statistics detailed in section\,\ref{sec:results:detectionstats12}). In Figure\,\ref{fig1} the locations of each core are labeled based on CH$_3$CHO detections and, for additional reference, the location of the protostellar sources with at least CH$_3$OH detected from the PEACHES sample is also shown \citep{2021ApJ...910...20Y}. 

\subsection{Yebes 40m} \label{subsec:yebesred}

Follow-up observations for 15 of the 35 cores with initial ARO 12m CH$_3$CHO detections ($\sigma_{T_\mathrm{mb}} = 6\,\mathrm{mK}$; see section\,\ref{sec:results:detectionstats12}) were done with the Yebes 40m telescope during the spring 2022 and spring 2023 seasons (projects 22A022 and 23A025; PI: Scibelli). Each core was observed with the dual (horizontal and vertical) linear polarization Q-band receiver \citep{2021A&A...645A..37T} using the frequency switching technique with a standard throw of 10.52 MHz. The wide-band nature of the receiver allows for a total bandwidth of 18.5 GHz spanning from 31.5 -- 50 GHz (6 -- 9mm) with a resolution of 38.0kHz (0.38 km/s -- 0.23 km/s). Pointing corrections were obtained by observing strong nearby quasars or SiO masers during each observing run.

Once the data were obtained, they were inspected, reduced, and put on the main beam temperature, $T_\mathrm{mb}$, scale using publicly available Python-based scripts\footnote{\url{https://github.com/andresmegias/gildas-class-python/}} developed by \cite{2023MNRAS.519.1601M}, which invokes the CLASS program of the GILDAS package for the combining of the spectra (see Appendix\,\ref{yebesappendix} for more detail). Then, additional CLASS scripts were run to search for the following COMs (Table\,\ref{LineList} and Table\,\ref{linelistnondec}): the five-atom precursor COMs formic acid; t-HCOOH, and ketene; H$_2$CCO, which for simplicity we will continue to label as `COMs', as well as the six-atom COMs CH$_3$OH, methyl cyanide; CH$_3$CN, and the even larger species CH$_3$CHO, CH$_2$CHCN, HCOOCH$_3$, and CH$_3$OCH$_3$, in order to compare to literature values \citep{2014ApJ...795L...2V, 2016ApJ...830L...6J, 2020ApJ...891...73S, 2021MNRAS.504.5754S, 2021ApJ...917...44J, 2023MNRAS.519.1601M}. The online tool SPLATALOGUE\footnote{\url{https://splatalogue.online}} was used to identify COM transitions, where the molecular data come from the Cologne Database of Molecule Spectroscopy (CDMS; \citealt{2001A&A...370L..49M, 2005JMoSt.742..215M, 2016JMoSp.327...95E}) and the Jet Propulsion Laboratory Millimeter and Submillimeter Spectral Line catalog (JPL; \citealt{1998JQSRT..60..883P}). We note that while additional complex molecules lie within our data set, such as large carbon-chains and cyanopolyynes, these species will be the subject of subsequent papers. In general, the $\sigma_{T_\mathrm{mb}}$ values range from $\sim 2 - 9$\,mK. 

\begin{figure}
\centering
\begin{center}$
\begin{array}{c}
\includegraphics[width=80mm]{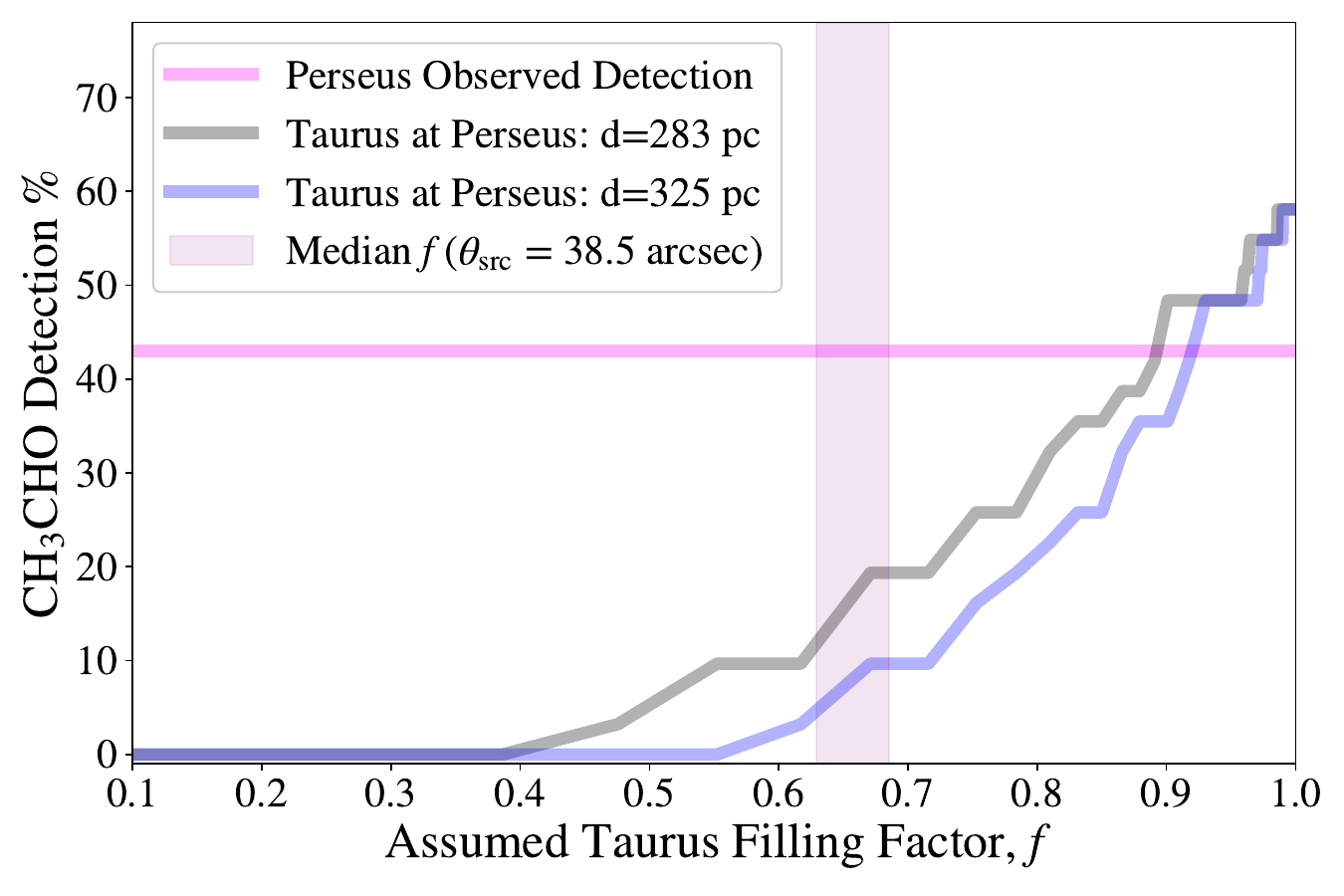} 
\end{array}$
\end{center}
\caption{\label{detect_frac} Expected detection percentages versus assumed filling factor, $f$, if the Taurus sample of CH$_3$CHO-detected cores from \citealt{2020ApJ...891...73S} were put at Perseus distances. For reference, we also plot as a horizontal magenta line the true observed detection percentage, 43\%, for the Perseus sample at the $\sigma_{T_\mathrm{mb}} = 5-8\,\mathrm{mk}$ RMS limit. If the Taurus cores are similar in size as the Perseus cores, i.e. $\theta_\mathrm{src} = 38.5$\,arcsec plotted at the range of distances as the light pink vertical band, one would have only expected a $\sim 10-20\%$ detection rate. 
}
\end{figure} 

\begin{figure*}
\includegraphics[width=155mm]{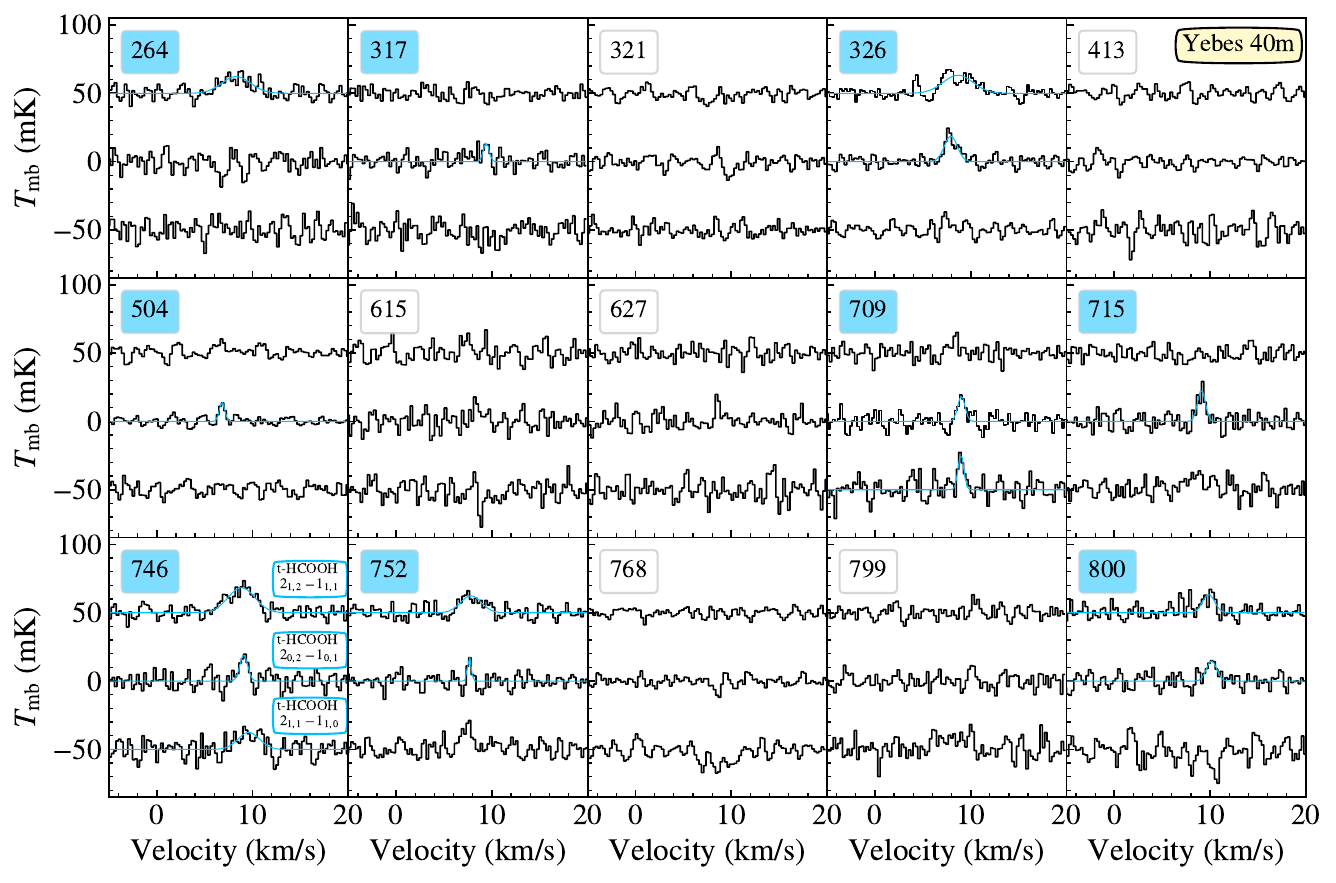} 
\caption{ \label{formic_spec_yebes40m} Formic acid, t-HCOOH, spectrum (in black) in units of $T_\mathrm{mb}$ (K) versus velocity (km/s) from the Yebes 40m for the 15 core sub-sample. Gaussian fits are plotted in blue. There are three $2-1$ transition observable and they are centered on the $v_\mathrm{lsr}$ of the core. Spectra are offset by intervals of 50\,mK.}
\end{figure*}

\begin{figure*}
\includegraphics[width=155mm]{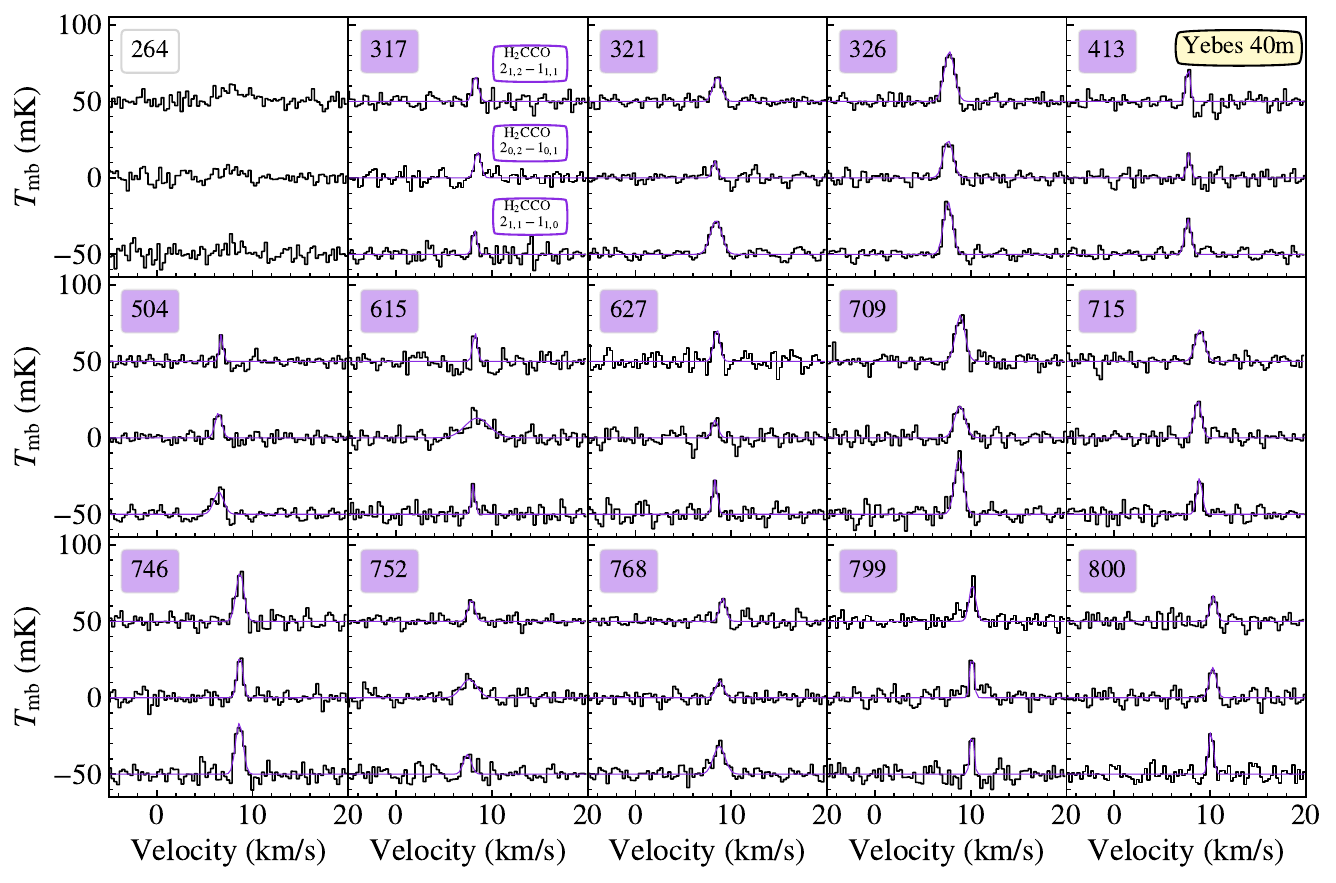} 
\caption{ \label{keetene_spec_yebes40m} Ketene, H$_2$CCO, spectrum (in black) in units of $T_\mathrm{mb}$ (K) versus velocity (km/s) from the Yebes 40m for the 15 core sub-sample. Gaussian fits are plotted in purple. There are three separate $2-1$ transitions observable, and all transitions are centered on the $v_\mathrm{lsr}$ of the core. Spectra are offset by intervals of 50\,mK.  }
\end{figure*}
\section{Results} \label{sec:results}

The main result from this survey is that gas-phase COMs are detectable and prevalent in cold ($\sim 10$\,K) starless and prestellar cores across the Perseus Molecular Cloud. In most of the cores the smaller COMs ($<$7-atom and 7-atom O-bearing) are detected, and in only a handful are higher complexity COMs (7-atom N-bearing and $>8$-atom O-bearing) also seen. Below we detail the detection statistics and calculate column densities, for both the full ARO 12m sample and the Yebes 40m sub-sample. 

\subsection{Detection Statistics: ARO 12m}\label{sec:results:detectionstats12}

From our preliminary ARO 12m survey, which had baseline RMS levels of $\sigma_{T_\mathrm{mb}} = 5-8\,\mathrm{mk}$, we found in 100\% (35/35) of the cores CH$_3$OH was detected, in $43\%$ (15/35) CH$_3$CHO was detected, and in 0\% of the cores (0/35) was CH$_2$CHCN detected. In general, the cores with the strongest detections (i.e., the brightest $T_\mathrm{mb}$ values) of CH$_3$OH had the strongest CH$_3$CHO detections (see Figures\,\ref{meth_spec_plot}\,\&\,\ref{acet_spec_plot}). A deeper search towards two of the cores (67 and 658) that both had no initial CH$_3$CHO detection was then carried out to achieve RMS values of $\sigma_{T_\mathrm{mb}} = 2-3\,\mathrm{mK}$. In both cores CH$_3$CHO was detected bringing the detection statistic up to $49\%$. It is important to note that for each of these deeper integration's on core 67 and core 658, it took $> 60$ hours and therefore additional integration on the 18 remaining sources with no CH$_3$CHO detections at the $\sigma_{T_\mathrm{mb}} = 6\,\mathrm{mK}$ level was not possible.

We also report that across the sample, in 31\% (11/35) of the cores the weakest CH$_3$OH line ($2_{0,2}-1_{0,1}$\,E) was not detected above the $3\sigma$ limit. And, for core 504 only the $5_{0,5}-4_{0,4}$\,E state of CH$_3$CHO was detected above the $3\sigma$ limit, while for cores 67 and 413 we only report the $5_{0,5}-4_{0,4}$\,A state with $>3\sigma$ confidence (note: similar to cores 11, and 20, 21 in \cite{2020ApJ...891...73S}). 

\begin{figure*}
\includegraphics[width=155mm]{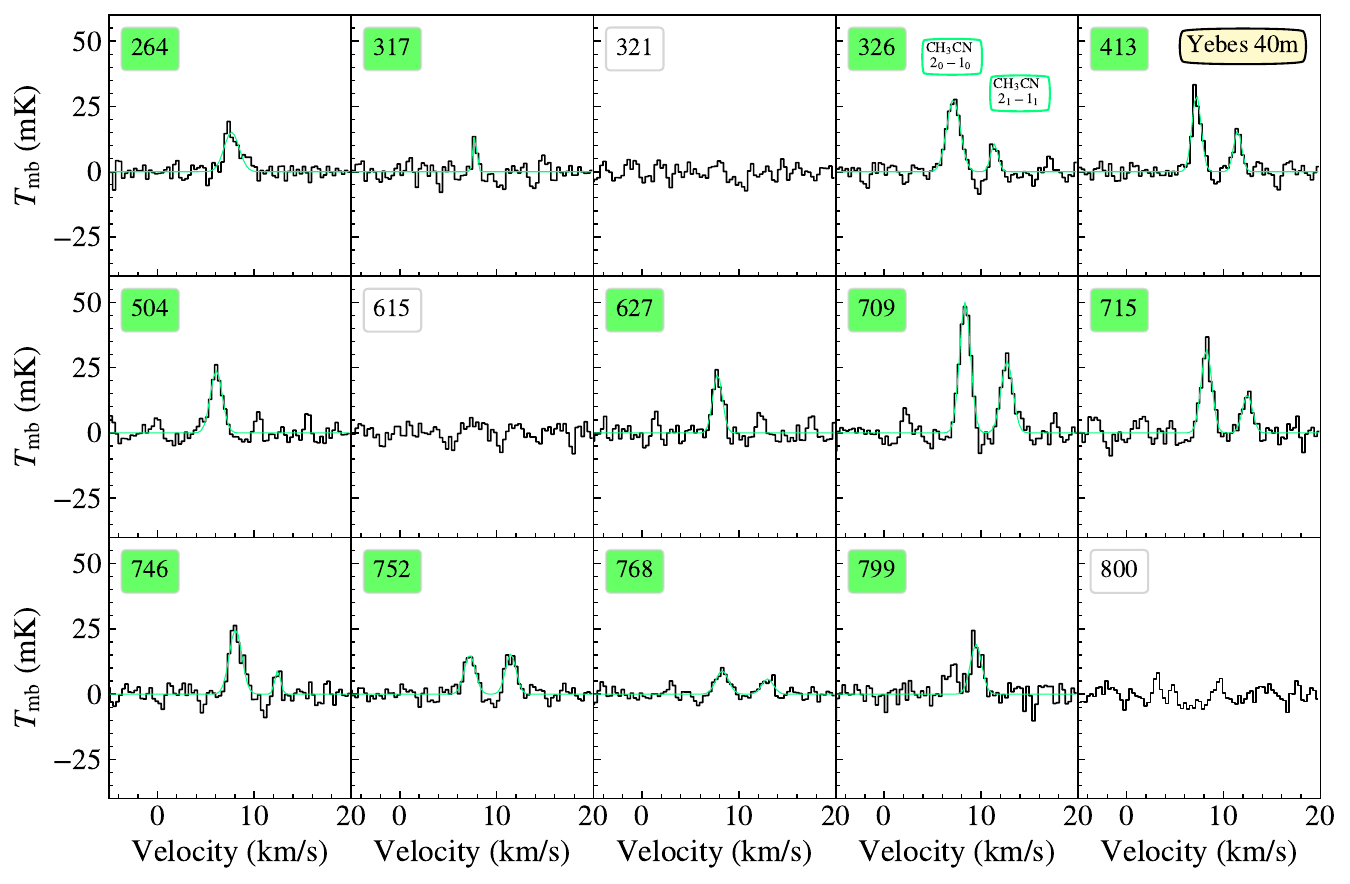} 
\caption{ \label{methcyn_spec_yebes40m} Methyl Cyanide, CH$_3$CN, spectrum (in black) in units of $T_\mathrm{mb}$ (K) versus velocity (km/s) from the Yebes 40m for the 15 core sub-sample. Gaussian fits are plotted in light green. There are two $2-1$ transitions that can be observable, and the $2_0 - 1_0$ state transitions is centered on the $v_\mathrm{lsr}$ of the core. }
\end{figure*}

\begin{figure*}
\includegraphics[width=155mm]{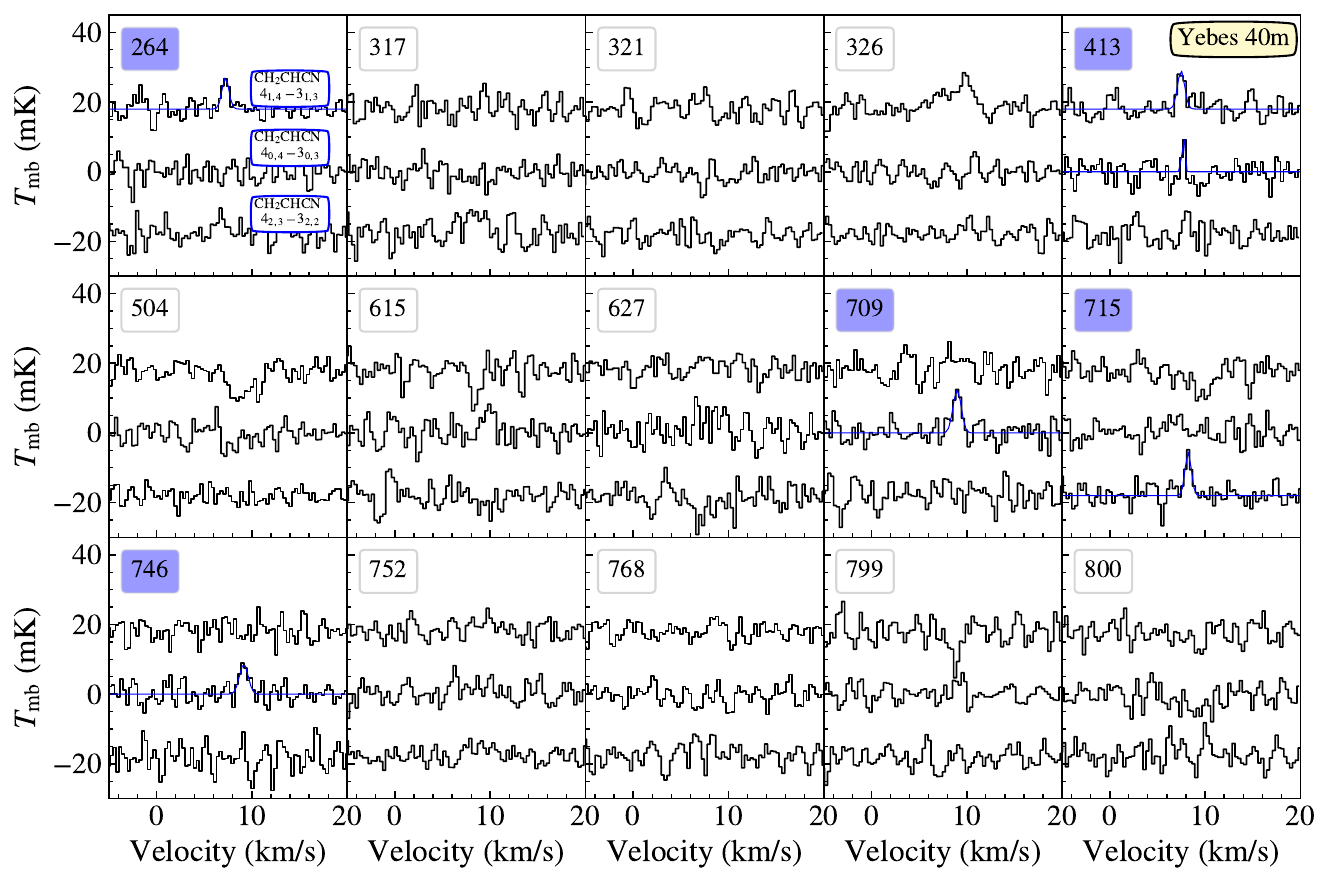} 
\caption{ \label{vyncyn_spec_yebes40m} Vinyl cyanide, CH$_2$CHCN, spectrum (in black) in units of $T_\mathrm{mb}$ (K) versus velocity (km/s) from the Yebes 40m for the 15 core sub-sample. Gaussian fits are plotted in blue. There are three separate $4-3$ transitions that could be observable, and all transitions are centered on the $v_\mathrm{lsr}$ of the core. Spectra are offset by intervals of 18\,mK for easier viewing.}
\end{figure*}

\begin{figure*}
\includegraphics[width=155mm]{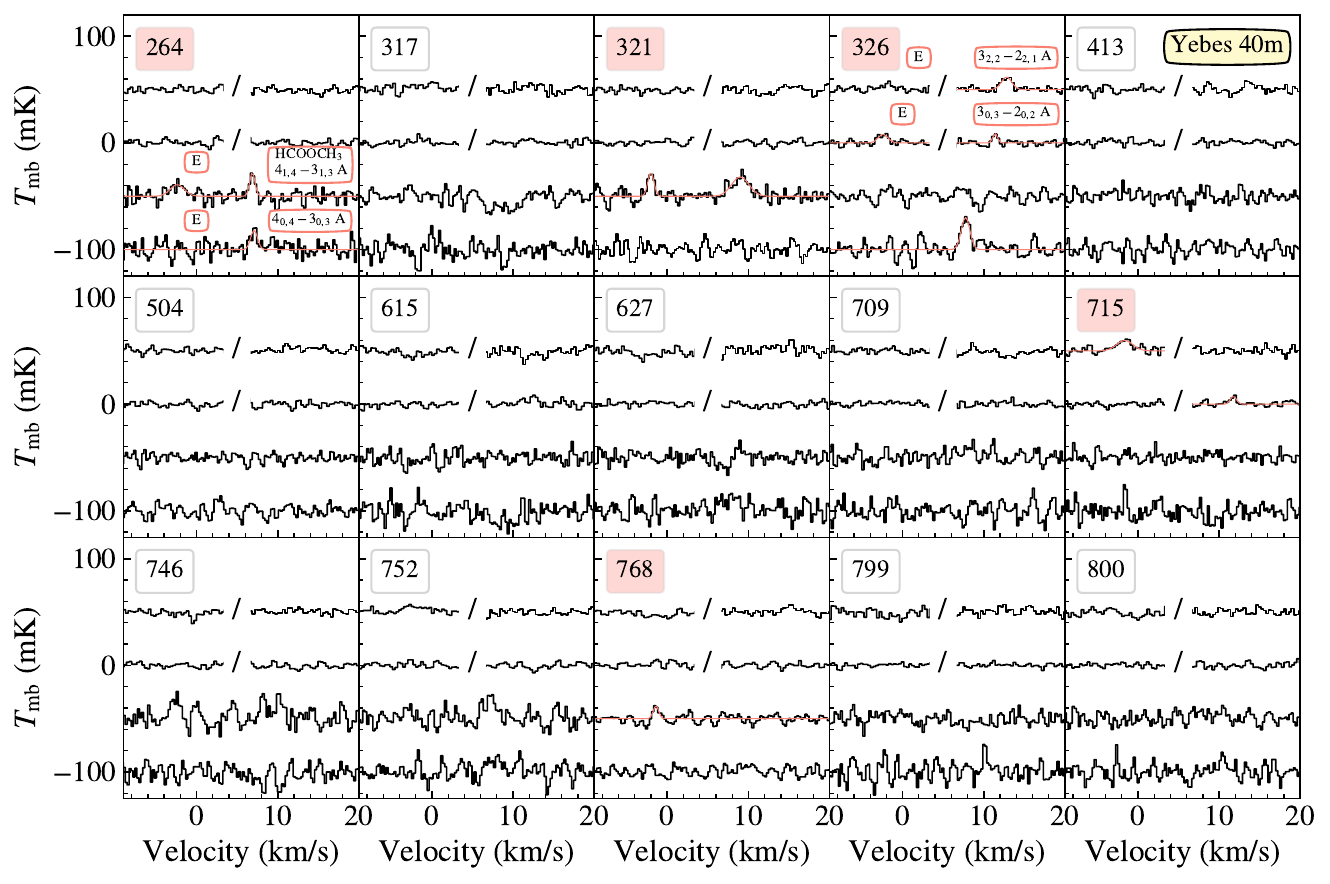} 
\caption{ \label{mf_spec_yebes40m} Methyl Formate, HCOOCH$_3$, spectrum (in black) in units of $T_\mathrm{mb}$ (K) versus velocity (km/s) from the Yebes 40m for the 15 core sub-sample. Gaussian fits are plotted in orange. Spectra are offset by intervals of 50\,mK. The top two rows show the $3 - 2$ E \& A states, respectively, which are not visible together within the plotted frequency range (denoted by the `/') and instead the E line is shifted by - 10 km/s and the A line by + 5 km/s from the $v_\mathrm{lsr}$ of the core.
}
\end{figure*}

\begin{figure*}
\includegraphics[width=155mm]{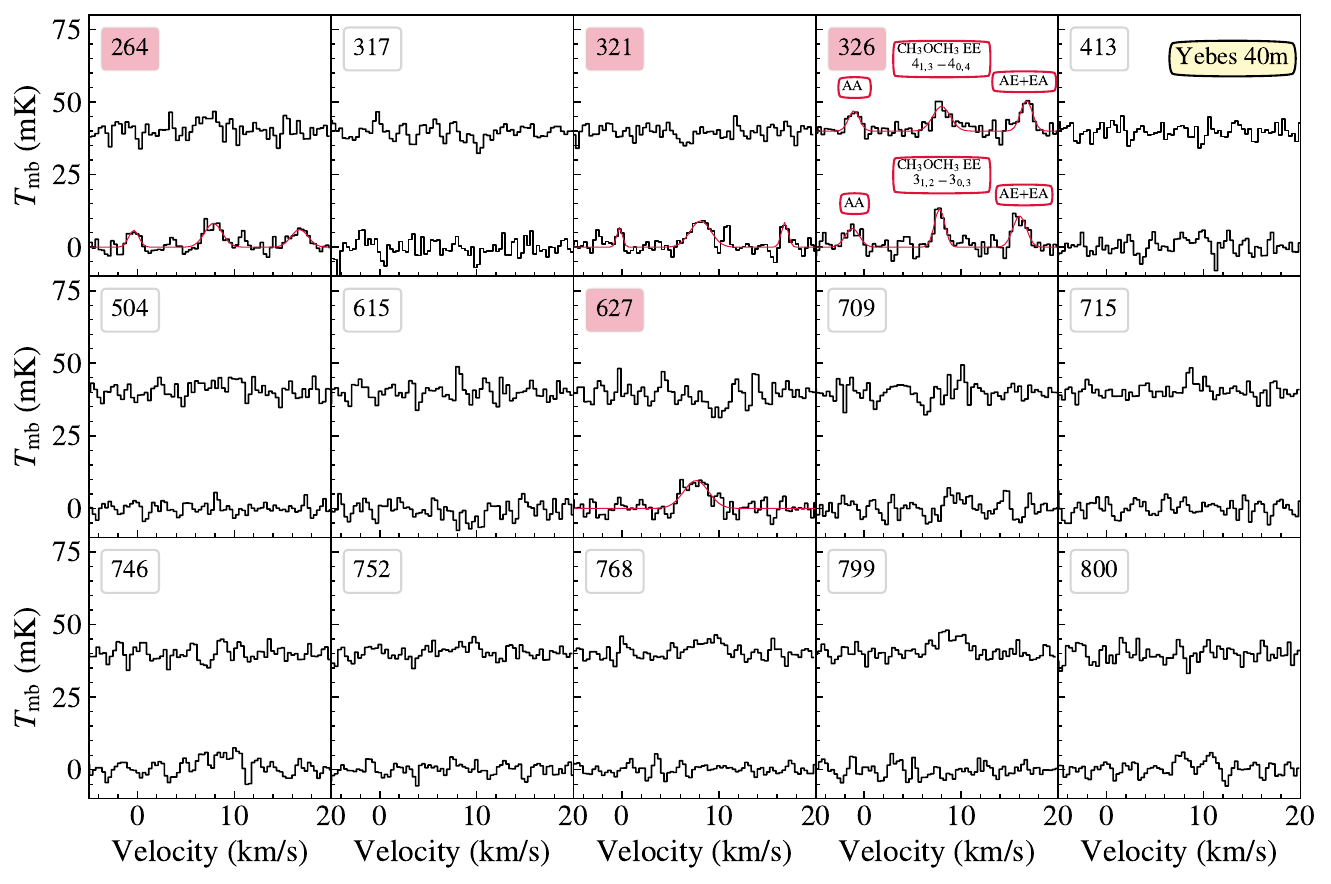} 
\caption{ \label{dme_spec_yebes40m} Dimethyl ether, CH$_3$OCH$_3$, spectrum (in black) in units of $T_\mathrm{mb}$ (K) versus velocity (km/s) from the Yebes 40m for the 15 core sub-sample. Gaussian fits are plotted in red. The EE transitions are centered on the $v_\mathrm{lsr}$ of the core. The $4-4$ lines are offset by 40\,mK for easier viewing. }
\end{figure*}

At a distance of 135\,pc \citep{2014ApJ...786...29S} the Taurus Molecular Cloud survey detected CH$_3$OH in 100\% and CH$_3$CHO in 70\% of the 31-core sample \citep{2020ApJ...891...73S}. Both the Taurus and Perseus surveys looked at these same lines with the ARO 12m at similar $\sigma_{T_\mathrm{mb}}$ levels $\sim$7\,mK, however, the distances to each molecular cloud region must be taken into account for accurate comparison as this affects the physical scales seen by the telescope. For example, assuming a fixed size of 7020\,au (0.03\,pc) for the cores in Taurus and Perseus the angular source size, $\theta_\mathrm{src}$, in Taurus will be 52\,arcsec while in Perseus, whose distances range from 283 to 325\,pc (Table\,\ref{physparams}), $\theta_\mathrm{src}$ can range from $\sim25-22$\,arcsec. To account for this, we consider the filling factor, \textit{f}, which measures the coupling between the source integrated intensity emission, $I(\theta, \phi)$, and the telescope beam, $P_n(\theta)$. By assuming $I(\theta, \phi)$ on the sky can be approximated by a Gaussian with FWHM $\theta_\mathrm{src}$, then the filling factor is given by, \begin{equation}\label{fillingfrac}
    f = \frac{\int I(\theta,\phi) P_n(\theta) \mathrm{d}\Omega}{I(\theta=0) \int P_n (\theta) \mathrm{d}\Omega } = \frac{\theta_\mathrm{src}^2}{\theta_\mathrm{src}^2 + \theta_\mathrm{beam}^2}. 
\end{equation} Given that the beam size, $\theta_\mathrm{beam}$, of the ARO 12m is 63 arcsec at the CH$_3$CHO frequencies, we calculate a range of $f$ values and scale $\theta_\mathrm{src}$ from the distance of Taurus to the range of observed distances in Perseus (e.g., at the Taurus distance a 7020\,au sized core has $f=0.4$ but in Perseus $f \sim 0.1$). Given our new constraints, we re-calculate the RMS level and detection percentages, assuming at least a $3\sigma$ detection, from the \cite{2020ApJ...891...73S} Taurus survey. In Figure\,\ref{detect_frac} we show how the detection percentage drops with decreasing $f$ and increasing distance. If the Taurus cores have similar physical source sizes as calculated for the Perseus cores (at $38.5$\,arcsec this is $\sim$10,900-12,500\,au; see section\,\ref{sec:resultsN}), the expected CH$_3$CHO detection rates at Perseus distances in Taurus are expected to be $\sim10-20\%$, which is $\sim 2-4\times$ lower than what has been observed in the Perseus sample. Thus, considering CH$_3$OH and CH$_3$CHO are the precursors to higher complexity COMs, the starless and prestellar cores in the Perseus Molecular Cloud are just as prevalent in COMs, if not more so, than the Taurus Molecular Cloud. 

\subsection{Detection Statistics: Yebes 40m}\label{sec:results:detectionstats40}

The 15 cores with CH$_3$CHO detections (as well as CH$_3$OH detections) at the $\sigma_{T_\mathrm{mb}} = 6$\,mK limit with the ARO 12 were observed with the Yebes 40m. Within the $31.5-50$\,GHz range of the data, we searched for energetically favorable COM transitions (listed in Table\,\ref{LineList}) and claim a detection if both the Gaussian fitted peak was at least $3\times$ brighter than the noise, i.e., $>3\sigma$, and the Gaussian fitted integrated intensity, or area under the line, was $3\times$ brighter than the associated error. While not the focus of this paper, all line-of-sight velocities ($v_\mathrm{lsr}$) and line-widths (FWHM) are reported for each detected line in Table\,\ref{GaussFits_Yebes} in Appendix\,\ref{yebesappendix}. The $v_\mathrm{lsr}$ ranges from $\sim 6-10$\,km/s depending on the source and FWHM values typically range from $\sim 0.4-2$\,km/s.

There were 2 additional transitions of the CH$_3$OH molecule and 6 additional transitions of the CH$_3$CHO molecule (see Table\,\ref{LineList}) that were observed in each core. In the case of core 264 the $1_{-0,1} - 0_{-0,0}$\,E state transition of CH$_3$OH was not detected above a $3\sigma$ limit and for core 715 the $2_{1,2} - 2_{1,1}$\,E transition of CH$_3$CHO was not detected above the $3\sigma$ limit (see in Appendix\,\ref{yebesappendix} Figures\,\ref{meth_spec_yebes40m},\,\ref{acet_spec_yebes40m} and\,\ref{acetaddl_spec_yebes40m}).

In the search for new molecules, we find for the smaller (5-atom) COMs a detection rate of 9/15 (60\%) for t-HCOOH (Figure\,\ref{formic_spec_yebes40m}) and 14/15 (93\%) for H$_2$CCO (Figure\,\ref{keetene_spec_yebes40m}). 
For the 6-atom N-bearing molecule CH$_3$CN in 12/15 (80\%) of the cores at least the brightest $2_0-1_0$ transitions was detected and in 7/12 of these cores the nearby $2_1 - 1_1$ transitions was also detected (Figure\,\ref{methcyn_spec_yebes40m}). 

The 7-atom N-bearing molecule CH$_2$CHCN, while not seen in the ARO 12m survey, was detected in $< 34\%$ of the 15 core sub-sample with the Yebes 40m data (Figure\,\ref{vyncyn_spec_yebes40m}). We report the most robust detection, with more than one transition detected above $3\sigma$, for one core only, 413.
For the remaining 4 cores (264, 709 and 746) with CH$_2$CHCN detected, only one transition is observed at $> 3\sigma$. Beyond the three CH$_2$CHCN transitions detected (listed in Table\,\ref{LineList}), other energetically favorable $a$-type transitions were searched for (listed in Table\,\ref{linelistnondec}), and no significant detections were made. 

Next, we searched for and detected the 8-atom molecule HCOOCH$_3$ in $<34\,\%$ of the cores. Multiple transitions are observed for cores 264, 321, 326 and 715, yet in core 768 only one transition was detected at $>3\sigma$ (Figure\,\ref{mf_spec_yebes40m}). In core 326 the highest number of transitions were detected (4 total), followed by core 264 (3 total).
Only one transition is detected for 768, the $4_{1,4} - 3_{1,3}$ E line, but it is also seen in cores 264 and 321.

Lastly, for the 9-atom molecule CH$_3$OCH$_3$ we find significant detections in $<27\%$ of the cores (Figure\,\ref{dme_spec_yebes40m}). Confident detections are made for cores 264, 321, and 326 (20\%) as they show all three nearby $3_{1,2} -3_{0,2}$ AE+EA, EE and AA state transitions, and for core 326 the additional three $4_{1,3} -4_{0,4}$ transitions. For core 627, only the brighter  $3_{1,2} -3_{0,2}$ EE transition is seen above $3\sigma$. 

While each molecule searched for was detected in at least a handful of cores, in no single core are all of the COMs targeted detected. Core 326 has the most detections, with CH$_3$OH, CH$_3$CHO, t-HCOOH, H$_2$CCO, CH$_3$CN, HCOOCH$_3$, and CH$_3$OCH$_3$ all detected, and CH$_2$CHCN being the only non-detection. Interestingly, in core 264 the smaller 5-atom COM H$_2$CCO is not detected in our survey limit, yet it is the only core for which the species CH$_2$CHCN, HCOOCH$_3$ and CH$_3$OCH$_3$ are all detected. Both of these cores are located in the active and shocked NGC1333 region (see Figure\,\ref{fig1}), and below we discuss the COM detections in context of the location within the Perseus cloud.

\begin{figure}
\centering
\begin{center}$
\begin{array}{c}
\includegraphics[width=80mm]{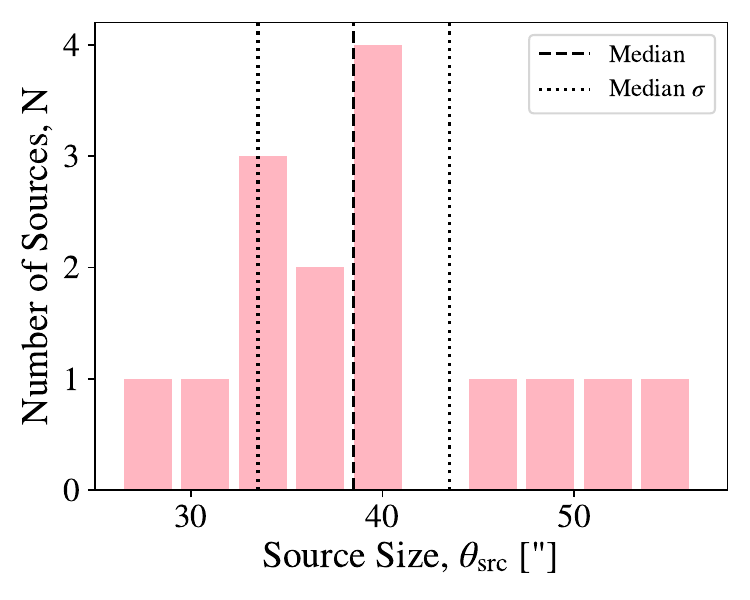} 
\end{array}$
\end{center}
\caption{\label{fig:sourceize} The distribution of source size, $\theta_\mathrm{src}$, for the Yebes 40m sub-sample of cores constrained from RADEX calculations for CH$_3$OH transitions observed at two different beam sizes, $\theta_\mathrm{beam}$. 
}
\end{figure}

\subsection{Detection Statistics: Trends in Spatial Distribution}  

Overall, from this survey the detection statistics for COMs in starless and prestellar cores has at least doubled. 
The cores with COM detections  reside in the two active Perseus clusters, NGC\,1333 and IC\,348, now thought to be of similar age \citep{2016ApJ...827...52L}. NGC\,1333 in particular is known to host many protostars and active outflows \citep{2000A&A...361..671K, 2013ApJ...774...22P}, which could help to get COMs off the grains in the cold starless and prestellar cores we've targeted in this survey. It is in the NGC\,1333 region where the majority of the protostars from the PEACHES survey reside with detections of at least CH$_3$OH (\citealt{2021ApJ...910...20Y}; see Figure\,\ref{fig1}). In particular, cores 264, 317, 321 and 326 are all located within $\sim$8\,arcmin box that also contains 10 protostars; Per-emb 18, Per-emb 21, Per-emb 44, Per-emb 27, Per-emb 35 A \& B, Per-emb 13, Per-emb 12 A \& B, and SVS 13A2. It is in core 326 where we detect all of the O-bearing COMs larger than CH$_3$OH. Core 326 is not only the densest core (\textit{n}($\mathrm{H}_2$) = 2.87 $\times 10^{5}$\,cm$^{-3}$; Table\,\ref{physparams}), but it is also $\sim$ a few beams away from protostars Per-emb 44, Per-emb 12 A \& B and Per-emb 13 which have warm gas-phase detections of CH$_3$CN, CH$_3$OH, CH$_3$CHO, HCOOCH$_3$ and CH$_3$OCH$_3$ \citep{2021ApJ...910...20Y}. 

Toward the IC\,348 cluster, which exhibits a lower abundance of objects at lower masses than NGC\,1333 \citep{2016ApJ...827...52L}, studies have been done to detect large complex molecules such as fullerenes, PAHs, and numerous hydrocarbon species (i.e., HCN, C$_2$H$_2$, C$_4$H$_2$, HC$_3$N, HC$_5$N, C$_2$H6, C$_6$H$_2$, C$_6$H$_6$) using mid-IR spectra from Spitzer \citep{2019MNRAS.489.1509I, 2023MNRAS.521.2248I}. Interestingly, only three nearby protostars in this region from the PEACHES sample had at least CH$_3$OH detected, i.e., Per-emb 11 A, C and Per-emb 1 \citep{2021ApJ...910...20Y}. In close proximity to these protostars ($< 5$\,arcmin) are the starless and prestellar cores 709 and 715 from this survey each with CH$_3$CHO detections, additional bright CH$_3$CN lines (see Figure\,\ref{methcyn_spec_yebes40m}), as well as detections of CH$_2$CHCN, t-HCOOH, H$_2$CCO and (for core 715 only) HCOOCH$_3$.

As for the other regions in Perseus, the star formation activity and evolutionary state (i.e., presence or absence of protostars) also seem to correlate with the presence of COMs in the starless and prestellar stage, though we note that the number of cores sampled in the regions outside the two main clusters (NGC\,1333 and IC\,348) is lower (see Figure\,\ref{fig1} and Table\,\ref{physparams}). It is in the B5 region where a dominant embedded protostar, B5-IRS1, with three gravitationally bound dense-gas fragments resides \citep{2015Natur.518..213P}, and hosts the CH$_3$CHO-abundant cores 799 and 800 (see Figure\,\ref{fig1}). Core 800, in fact, is spatially coincident with the `B5-Cond1' gas fragment as described in \cite{2015Natur.518..213P}. In contrast, in L1455 there is a low abundance of Class 0 protostars \citep{2007A&A...472..187H}, and no starless or prestellar cores from our sample with CH$_3$CHO detections at our $\sigma_{T_\mathrm{mb}} = 6$\,mK limit (Figure\,\ref{fig1}). Only one protostar from the PEACHES sample had at least CH$_3$OH detected in the L1455 region, Per-emb 20, and for this source no higher complexity COMs were detected (i.e., no t-HCOOH, CH$_3$CN, CH$_3$CHO, HCOOCH$_3$, or CH$_3$OCH$_3$). Lastly, we note that compared to the cores within NGC\,1333 and IC\,348 (a total of 20), the cores in these other regions (a total of 15) on average have lower dust and kinetic temperatures (e.g., median $T_\mathrm{k}$ differences of 12.1 and 10.5, respectively). Thus, regions within the Perseus Molecular Cloud with increased COM detections in the warm protostellar stage also exhibit an increased COM detection in the cold prestellar stage.

\subsection{Column Densities}\label{sec:resultsN}

\begin{figure}
\centering
\begin{center}$
\begin{array}{c}
\includegraphics[width=80mm]{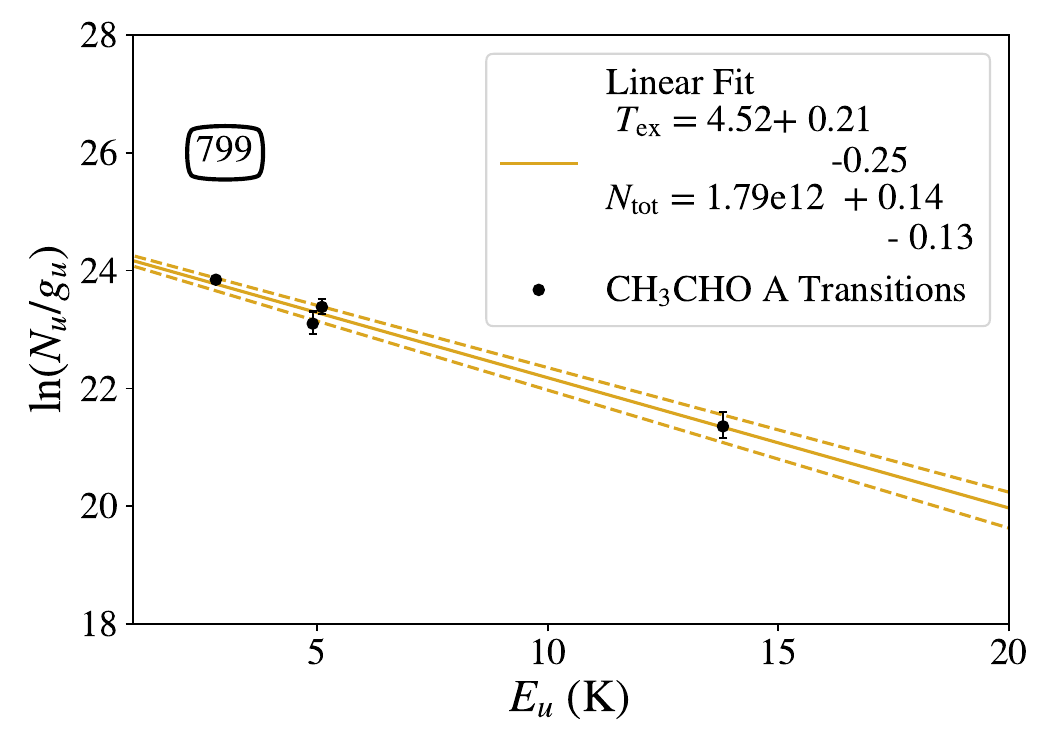} 
\end{array}$
\end{center}
\caption{\label{acet_rotdiagram} The rotation diagram method is used to constrain the column densities, $N$, and excitation temperatures, $T_\mathrm{ex}$, for each of the 15 cores with multiple A-state CH$_3$CHO transitions. Here we show the fit for core 799. 
}
\end{figure}

\begin{figure}
\centering
\begin{center}$
\begin{array}{c}
\includegraphics[width=80mm]{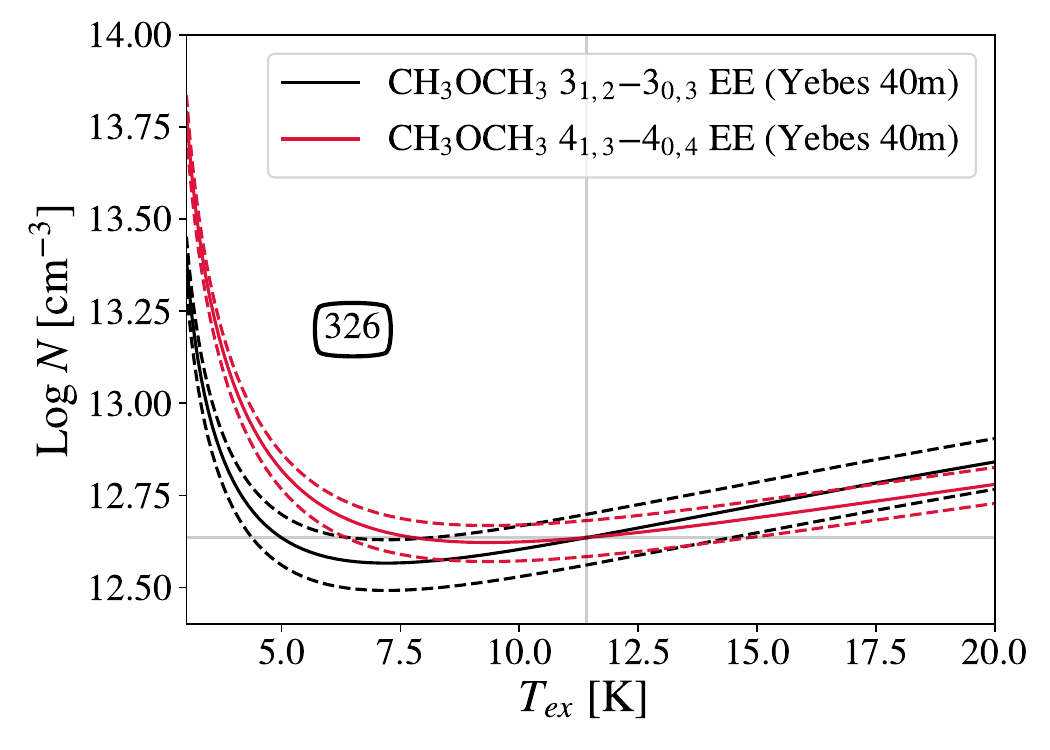} 
\end{array}$
\end{center}
\caption{\label{dme_ctex} In the CTEX radiative transfer method column density, $N$, in log-scale is plotted against the excitation temperature $T_\mathrm{ex}$ for two transitions of CH$_3$OCH$_3$ EE, as red and black curves, that were detected towards core 326. The dashed lines show the spread in error from the intensity measurements and are used to calculate the spread in error for $N$ and $T_\mathrm{ex}$. The grey horizontal and vertical lines show where the two transitions cross to find the best constraint. 
}
\end{figure}

Depending on the molecule, and how many available transitions were observed and detected for that molecule, four different methods were used to calculate column densities; the RADEX code method \citep{2007A&A...468..627V}, the rotation diagram (RD) method, the constant excitation (or CTEX) method, and the Local Thermodynamical Equilibrium (LTE) method. RADEX is the only non-LTE method that, while one-dimensional, is able to calculate an excitation temperature, $T_\mathrm{ex}$, a column density, $N$, and an opacity, $\tau$, for each transition separately. This method is limited, however, to the few molecules with which collisional rate coefficients have been calculated, which in our case are CH$_3$OH, CH$_3$CN and HCOOCH$_3$.

For CH$_3$OH we are able to utilize the difference in beam sizes of the ARO 12m and Yebes 40m CH$_3$OH transitions (62 arcsec and 37 arcsec, respectively; see Table\,\ref{LineList}) in order to constrain the source size, $\theta_\mathrm{src}$, and account for the true filling factor, $f$, when calculating the column density for our 15 core sub-sample (see Appendix\,\ref{N_detail_appendix} for more detail). We find $\theta_\mathrm{src}$ ranges from $26.5-56.5$ arcsec where the median $\theta_\mathrm{src}$ is 38.5 arcsec (5 arcsec median deviation; Figure\,\ref{fig:sourceize}), which was then used as the `typical' value when calculating $N$(CH$_3$OH) for the remaining cores with no Yebes 40m data. These source sizes are utilized in the column density calculations for CH$_3$CHO, as this molecule is believed to trace CH$_3$OH spatially \citep{2017ApJ...842...33V, 2021MNRAS.504.5754S}, but for the remaining molecules the source size is unknown so $f = 1$ is assumed.

The total CH$_3$OH column densities (A+E states) calculated from RADEX, $N_\mathrm{sum}$, range from $0.87 - 50.27 \times 10^{13}$\,cm$^{-2}$, the excitation temperatures range from $T_\mathrm{ex} \sim 5 -
13$\,K, and the optical depths are generally consistent with optically thin emission, yet for several cores the $2_{0,2} - 1_{0,1}$ A transition gets close to $\tau \sim 1$ (Table\,\ref{tab:colden_met} in Appendix\,\ref{N_detail_appendix}). These $N_\mathrm{sum}$ values are roughly a factor $2-8\times$ higher than for the Taurus sample, which ranged from $0.42-3.4 \times 10^{13}$\,cm$^{-2}$ \citep{2020ApJ...891...73S}, however no filling factor was applied to the Taurus sample and could contribute to this offset in $N_\mathrm{sum}$. If we assume the median emitted source size for the Perseus sample is similar to the Taurus sample and scale this to the appropriate distance, the $N$(CH$_3$OH) values in \cite{2020ApJ...891...73S} increase by factors of 1.6 -- 1.8. Thus, the CH$_3$OH column densities in Perseus are in fact elevated compared to Taurus.

The RADEX derived column densities for CH$_3$CN range between $0.81 - 5.8 \times 10^{11}$\,cm$^{-2}$ and the excitation temperatures range from $5.57 - 13.2$\,K, with a median value of 7.39\,K, where the $\tau$ calculated is consistent with optically thin with $\tau_\mathrm{max} <0.009$. Toward the dust peak of core L1517B a CH$_3$CN column density value is found within this range, at $2.1 \pm 0.6 \times 10^{11}$\,cm$^{2}$ \citep{2023MNRAS.519.1601M}. For HCOOCH$_3$, the average $N$ (and standard deviation) value of the detected transitions is $ 2.97 \pm 2.18 \times 10^{12}$\,cm$^{-2}$, where $T_\mathrm{ex}$ ranges from $10.5  - 16.5$\,K and for all cases $\tau_\mathrm{max} < 0.003$ is also consistent with optically thin emission.

\begin{figure*}
\centering
\begin{center}$
\begin{array}{cc}
\includegraphics[width=170mm]{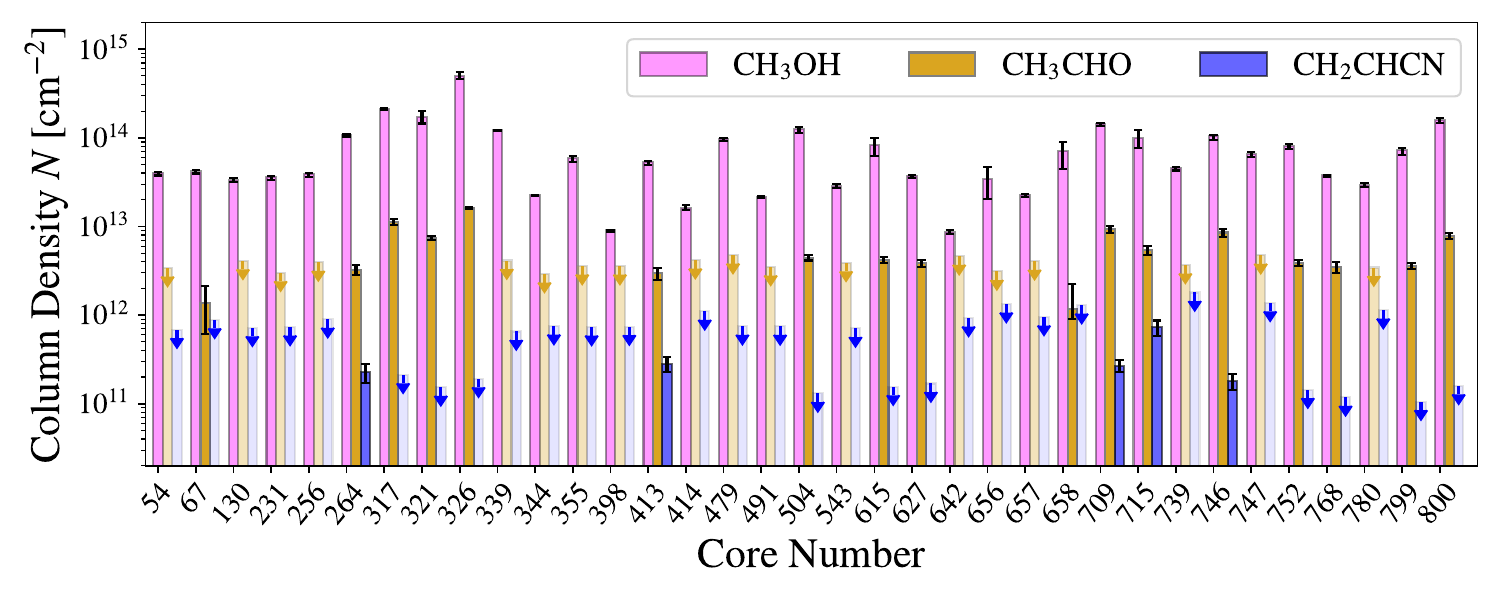} \\
\includegraphics[width=170mm]{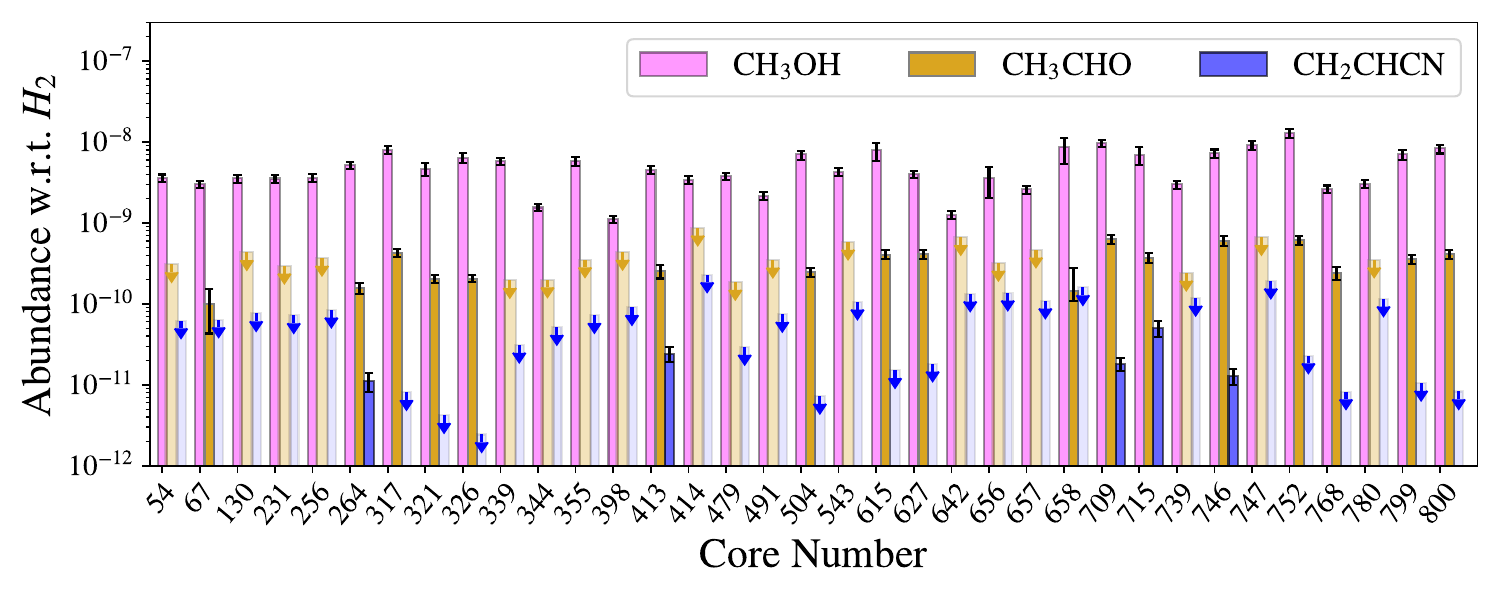} \\
\end{array}$
\end{center}
\caption{ \label{fig:disc_12mcompare} Comparison of COM column density values (top) and abundances (bottom) within the full 35 core ARO 12m sample. Plotted in magenta are values calculated for CH$_3$OH, in gold for CH$_3$CHO and in blue for CH$_2$CHCN. Error bars are in black and upper limits are shown as lighter shaded bars with downward arrows. } 
\end{figure*}

\begin{figure*}
\centering
\begin{center}$
\begin{array}{cc}
\includegraphics[width=170mm]{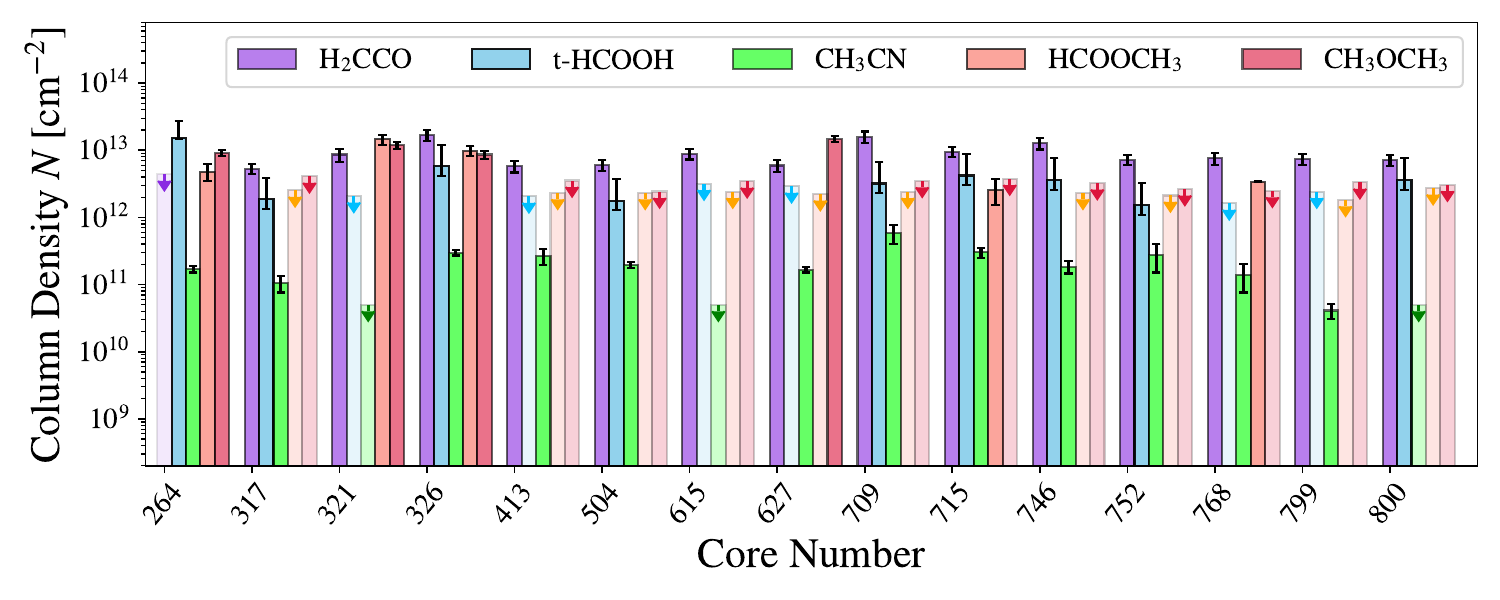} \\
\includegraphics[width=170mm]{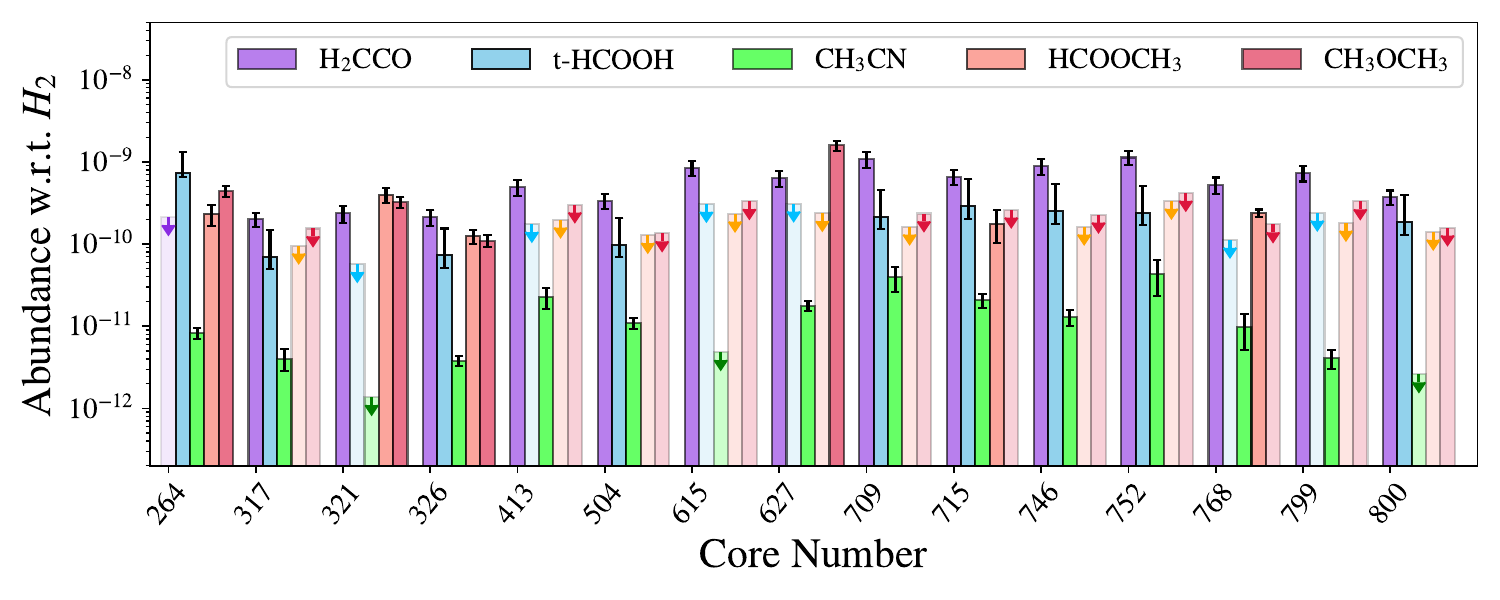} \\
\end{array}$
\end{center}
\caption{ \label{fig:disc_40mcompare} Comparison of COM abundances within the 15 core Yebes 40m sub-sample. Plotted in purple are values calculated for H$_2$CCO, in blue for t-HCOOH, in lime for CH$_3$CN, in orange for HCOOCH$_3$ and in red for CH$_3$OCH$_3$. Error bars are in black and upper limits are shown as lighter shaded bars with downward arrows.}
\end{figure*}

For the remaining molecules, the other three methods (RD, CTEX, or LTE) are used, which assume a fixed $T_\mathrm{ex}$ for all transitions. The upper state column density (assuming optically thin, $\tau_\nu \ll 1$, emission) can then be calculated where, \begin{equation}
    {N}_u = \frac{I}{h A_{ul} f} \frac{u_\nu({T}_\mathrm{ex})}{[J_\nu({T}_\mathrm{ex}) - J_\nu({T}_\mathrm{cmb})]},
\end{equation} and $h$ is the Planck constant, $f$ is the (frequency dependent) filling factor (equation\,\ref{fillingfrac}), $I$ is the integrated intensity of the line, $A_{ul}$ is the spontaneous emission coefficient (or `Einstein A'), $T_\mathrm{cmb}$ is the background temperature of 2.73\,K, and $u_\nu$ (Planck energy density) and $J_\nu$ (Planck function in temperature units) are defined as, \begin{equation}
    u_\nu \equiv \frac{8 \pi h \nu^3 }{c^3} \frac{1}{\exp{(h\nu/kT - 1)}}, 
\end{equation} \begin{equation}
    J_\nu \equiv \frac{h \nu }{k} \frac{1}{\exp{(h\nu/kT - 1)}}.
\end{equation} In these equations, $c$ is the speed of light, $k$ is the Boltzmann constant and $\nu$ is our line frequency. For a total column density, $N_{\mathrm{tot}}$, \begin{equation} \label{eq:N}
    \frac{{N}_u}{g_u} = \frac{N_\mathrm{tot}}{{Q(T_\mathrm{ex})}} \exp{(-E_u / k {T_\mathrm{ex}})},
\end{equation} where $g_u$ is the upper state degeneracy, $E_u$ is the upper state energy, and ${Q(T_\mathrm{ex})}$ is the partition function dependent on the excitation temperature of the molecule. 

In a standard RD method, the log normal of the left side of equation\,\ref{eq:N}, ln(${N}_u/g_u$), is plotted vs $E_u$ so that the excitation temperature, ${T_\mathrm{ex}}$, is the inverse of the slope of the linear fit and the y-intercept is used to find ${N_\mathrm{tot}}$ \citep{1999ApJ...517..209G}. This method is used to calculate the column density for the A state of CH$_3$CHO (Figure\,\ref{acet_rotdiagram}), with more than three transitions ranging from $E_u/k \sim 3-14$\,K for cores with both ARO 12m and Yebes 40m observations. The ${T_\mathrm{ex}}$ calculated from our RD method range from $ 4.27 - 7.07$ with a median value of  5.55\,K and median standard deviation of 0.9\,K. This median value of 5.55\,K is then used to calculate the $N$ of CH$_3$CHO A for the remaining cores. The full sample ranges from $0.59 - 8.08\,\times\,10^{12}$\,cm$^{-2}$ with a median value (and median standard deviation) of $2.10 \pm 0.61 \times 10^{12}$\,cm$^{-2}$.

Calculated with the same set of equations, the CTEX method also uses at least two transitions with different $E_u$ values to simultaneously constrain ${T_\mathrm{ex}}$ as well as the column density (see details in Appendix of \cite{2002ApJ...565..344C} and equation 80 of \cite{2015PASP..127..266M}), but instead finds the intersection of the plotted $N$ vs. ${T_\mathrm{ex}}$ curves (e.g., \cite{2021MNRAS.501..347A, 2021MNRAS.504.5754S, 2022MNRAS.515.5219G}). In our sample, for core 326 there are two bright CH$_3$OCH$_3$ EE transitions, $3_{1,2} - 3_{0,3}$ and $4_{1,3} - 4_{0,4}$, with a large enough $E_u$ gap, at 7\,K and 11\,K, respectively, that the CTEX method can constrain $N$ and $T_\mathrm{ex}$. 
We find for a $T_\mathrm{ex} = 11.4\pm5.1$\,K the column density is constrained to $N$(CH$_3$OCH$_3$)$ = 4.32\pm0.57 \times 10^{12}$\,cm$^{-2}$ (see Figure\,\ref{dme_ctex}). 

Lastly, for the LTE method, rather than calculating a single ${T_\mathrm{ex}}$, one needs to be assumed in order to calculate ${N_\mathrm{tot}}$. For the remaining molecules, H$_2$CCO, t-HCOOH, and CH$_2$CHCN, we use this LTE method. We calculate for a $T_\mathrm{ex}$ of 10\,K, $N$(o-H$_2$CCO) ranging from $1.8 - 7.6 \times 10^{12}$\,cm$^{-2}$ and $N$(p-H$_2$CCO) ranging from $0.8 - 4.4 \times 10^{12}$\,cm$^{-2}$. In the case of t-HCOOH assuming $T_\mathrm{ex}$ of 10\,K, the $N$ values range from $1.5 - 15 \times 10^{12}$\,cm$^{-2}$. For the N-bearing species CH$_2$CHCN we assume the $T_\mathrm{ex}$ is comparable to those found in the RADEX calculations for CH$_3$CN to estimate $N$, calculating values ranging from $0.18 - 0.72 \times 10^{12}$\,cm$^{-2}$. More detail on the exact analysis methods on a molecule-by-molecule and core-by-core basis, along with all listed column density values, can be found in Appendix\,\ref{N_detail_appendix}. 

\section{Discussion} \label{sec:discussion}

The abundances of COMs in the starless and prestellar cores in Perseus provide key insights into formation histories at the earliest phases of low-mass star formation, where initial chemical conditions are set. Here we calculate and compare across our samples the COM abundances with respect to molecular hydrogen (H$_2$), calculated from the \textit{Herschel} column density values listed in column 8 of Table\,\ref{physparams} and the COM column densities, $N$, from section\,\ref{sec:resultsN}, including 3$\sigma$ upper limits. In the case of the asymmetric rotors the $N$ derived from a single state is doubled (e.g., A+E) to find $N_\mathrm{sum}$, which assumes a 1:1 ratio (or 1:2:1 in the case of CH$_3$OCH$_3$) that may not always be the case. We do find the mean A:E ratio for CH$_3$OH across all cores (using the brightest 96\,GHz transitions) is 1.1 and conclude doubling the A state column density is a reasonable approximation to make. For H$_2$CCO, a total $N$ is found first by adding both para and ortho states (Table\,\ref{tab:colden_5atom}). In Figure\,\ref{fig:disc_12mcompare} we plot both $N$ and abundance with respect to H$_2$ values across the full 35-core sample, for which we have measured constraints for CH$_3$OH, CH$_3$CHO and CH$_2$CHCN. 

In Figure\,\ref{fig:disc_40mcompare} a similar plot is shown for the Yebes 40m sub-sample of 15 cores for which we have measured constraints for H$_2$CCO, t-HCOOH, CH$_3$CN, HCOOCH$_3$, and CH$_3$OCH$_3$. Noticeable is the higher abundance of CH$_3$OH compared to all other molecules targeted (ranging from 1.12 to 12.9 $\times 10^{-9}$), and the lower abundances of the N-bearing species CH$_3$CN and CH$_2$CHCN, which are well correlated (i.e., increased CH$_3$CN leads to CH$_2$CHCN) with values $\sim$ a few $\times 10^{-11}$ (see also Table\,\ref{tab:abundances}). Specifically, we find for the cores with detections that the abundance of CH$_2$CHCN ranges from $1.10 - 5.03 \times 10^{-11}$ and that the abundance of CH$_3$CN ranges from $0.38 - 4.34 \times 10^{-11}$. For the remaining O-bearing COMs abundances range from $\sim 10^{-10}-10^{-9}$, i.e., detected molecular abundances $X$(CH$_3$CHO) from $0.99 - 6.27 \times 10^{-10}$, $X$(t-HCOOH) from $0.71 - 7.34 \times 10^{-10}$, $X$(H$_2$CCO) from $2.01 - 11.3 \times 10^{-10}$, $X$(HCOOCH$_3$) from $1.25 - 4.00 \times 10^{-10}$ and $X$(CH$_3$OCH$_3$) from $1.10 - 16.0 \times 10^{-10}$ (Table\,\ref{tab:abundances}). 
In the subsequent subsections, we perform detailed comparisons of these abundances and discuss implications for COM formation in cold cores.

\begin{figure*}
\centering
\begin{center}$
\begin{array}{c}
\includegraphics[width=172mm]{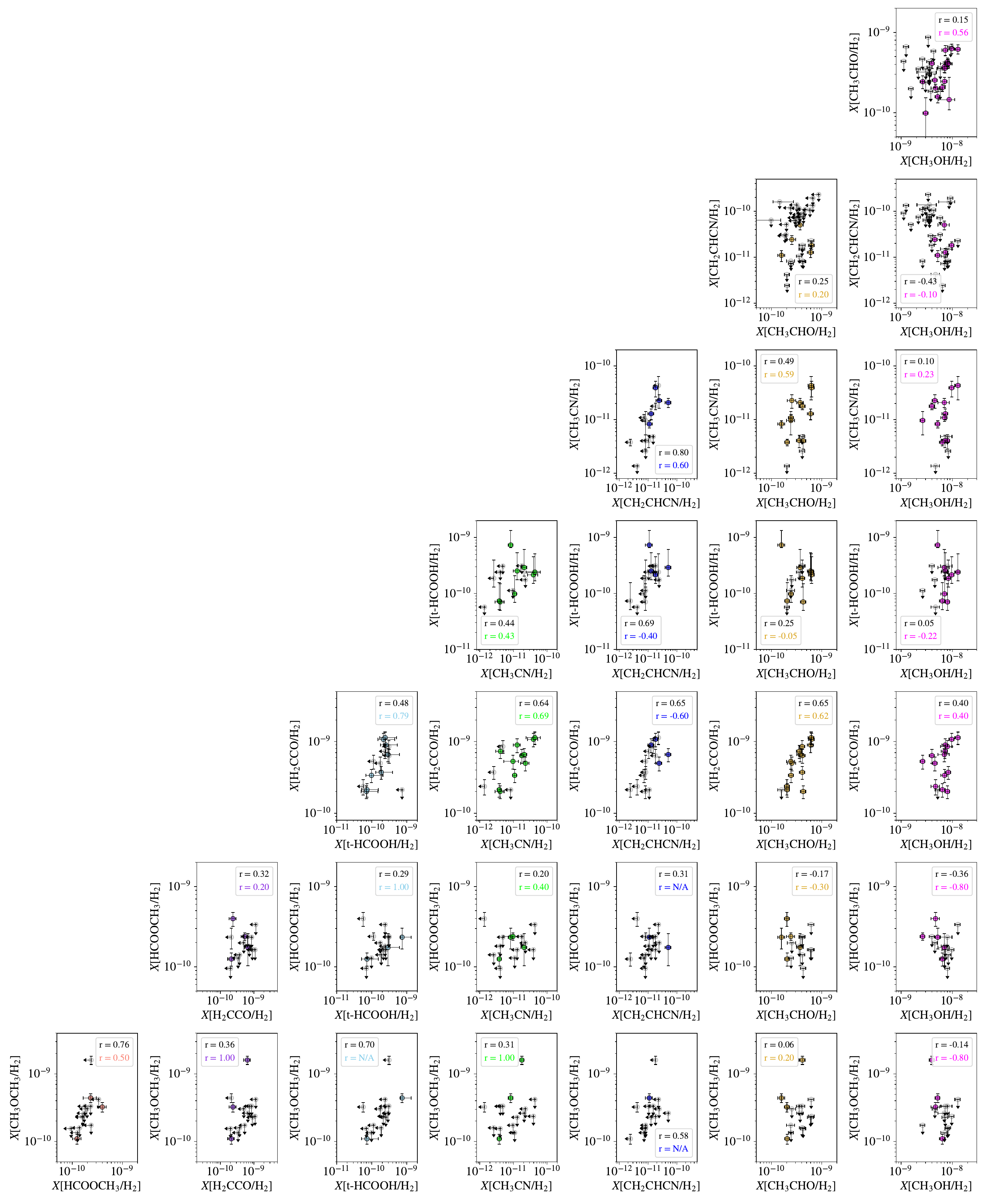} \\
\end{array}$
\end{center}
\caption{ \label{fig:correlation_cornerplot} Correlation corner plot for the eight species of COMs in this study plotted as abundances with respect to H$_2$. In the rightmost column are CH$_3$OH abundances (magenta points) with the following columns plotting abundances of CH$_3$CHO (gold points), CH$_2$CHCN (blue points), CH$_3$CN (green points), t-HCOOH (light blue points), H$_2$CCO (purple points) and HCOOCH$_3$ (orange points). In the case where either or both correlated molecules are not detected, upper limits are plotted in grey. In each panel the Spearman rank correlation coefficient `r' is reported for the full sample of cores, including upper limits, in black text as well as the `r' value for only the sample cores with detections of both molecules being compared in colored text (note: if there are $<3$ points `r' is reported as `N/A').}
\end{figure*}

\begin{figure*}
\centering
\begin{center}$
\begin{array}{c}
\includegraphics[width=170mm]{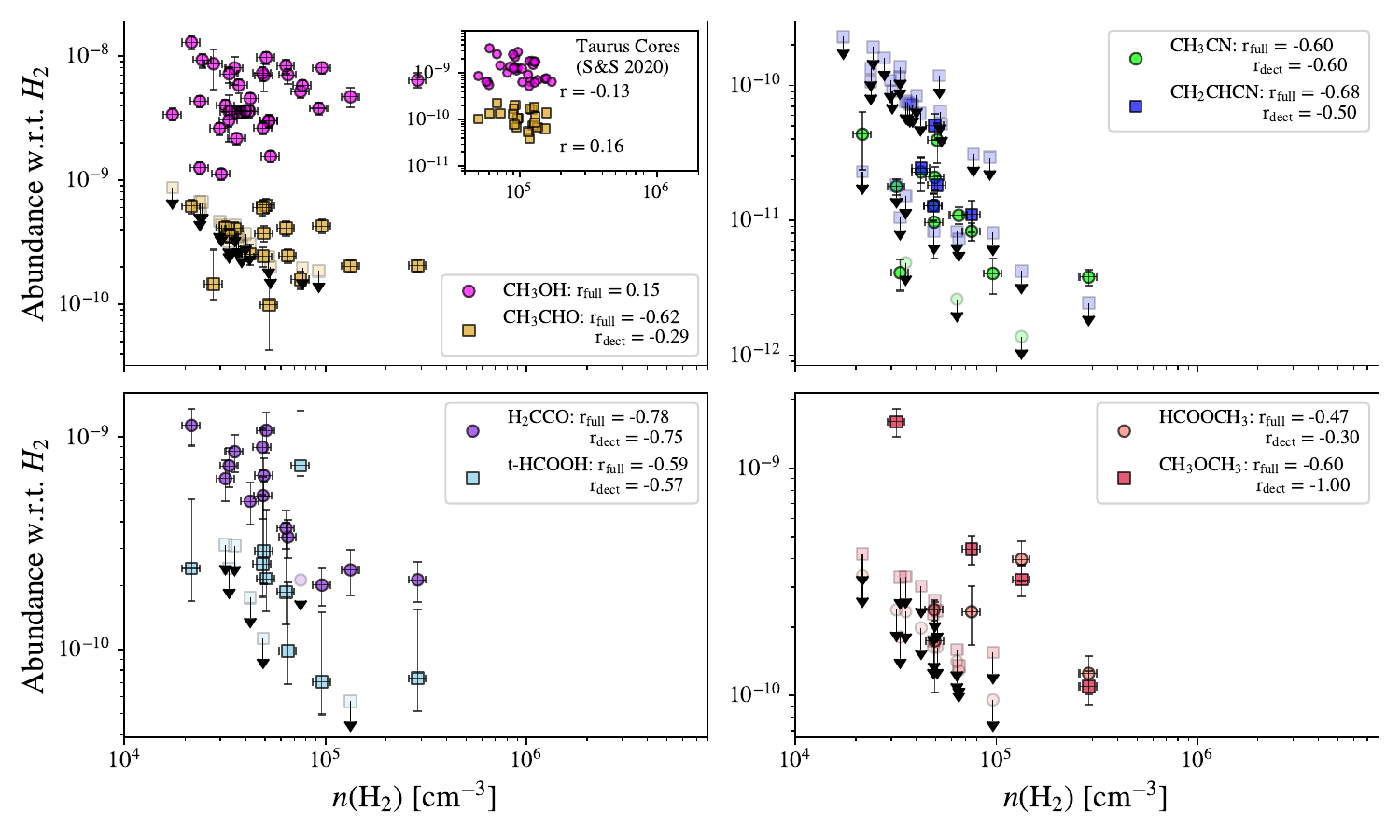} \\
\end{array}$
\end{center}
\caption{ \label{fig:phys_compare} Comparison of COM abundances with respect to H$_2$ to the average H$_2$ volume density, $n$(H$_2$) in units of cm$^{-3}$ for each of the cores in our sample. The Spearman rank correlation coefficient, noted here as r, is included in the legend in each panel. 
In most cases a negative r value is found, suggesting the COM abundances drop with increasing volume density. Error bars, including a 10\% error in $n$(H$_2$), are plotted in black. Upper limits are shown as slightly more translucent points with black downward facing arrows. We note `r' is calculated both for the full sample of cores (including upper limits) as well as only the cores with detections. An insert plot in the top left panel compares the abundance and density range for the sample of Taurus cores presented in \citealt{2020ApJ...891...73S}, where error-bars have been removed for clarity.
}
\end{figure*}

\begin{table*}
	\centering
	\caption{COM abundances with respect to H$_2$ for starless and prestellar cores in Perseus}
 \setlength{\tabcolsep}{10pt}
	\label{tab:abundances}
	\begin{tabular}{cllllllll} 
    \tablecolumns{10}
     \tablewidth{0pt}
     \tabcolsep=0.1cm
Core \#$^{1}$ &  t-HCOOH & H$_2$CCO &  CH$_3$CN & CH$_3$OH & CH$_3$CHO & CH$_2$CHCN & HCOOCH$_3$ & CH$_3$OCH$_3$ \\  
(\textit{Herschel}) & $\times 10^{-10}$  & $\times 10^{-10}$ &  $\times 10^{-11}$ & $\times 10^{-9}$ & $\times 10^{-10}$ & $\times 10^{-11}$ & $\times 10^{-10}$ & $\times 10^{-10}$ \\
\hline
54 & -- & -- & -- &3.59$\SPSB{+0.41}{-0.37 }$ &$<3.07$ &$<6.21$ & -- & -- \\ [1pt]
\textit{67} & -- & -- & -- &3.01$\SPSB{+0.34}{-0.32 }$ &0.99$\SPSB{+0.56}{-0.55 }$  &$<6.39$  & -- & -- \\ [1pt]
130 & -- & -- & -- &3.56$\SPSB{+0.42}{-0.38 }$ &$<4.33$ &$<7.62$ & -- & -- \\[1pt]
231 & -- & -- & -- &3.54$\SPSB{+0.43}{-0.37 }$ &$<2.95$ &$<7.20$ & -- & --   \\[1pt]
256 & -- & -- & -- &3.63$\SPSB{+0.41}{-0.39 }$ &$<3.69$ &$<8.43$& -- & --  \\[1pt]
\textbf{264} &7.34$\SPSB{+0.76}{-6.02 }$ &$<2.12$ & 0.83$\SPSB{+0.13}{-0.70 }$ & 5.17$\SPSB{+0.57}{-0.53 }$ &1.58$\SPSB{+0.25}{-0.25 }$ &1.10$\SPSB{+0.29}{-0.29 }$ &2.34$\SPSB{+0.67}{-0.64 }$ &4.40$\SPSB{+0.64}{-0.64 }$ \\ [1pt]
\textbf{317} &0.71$\SPSB{+0.21}{-0.79 }$ &2.01$\SPSB{+0.41}{-0.41 }$ &0.40$\SPSB{+0.12}{-0.00 }$ & 8.02$\SPSB{+0.85}{-0.81 }$ &4.28$\SPSB{+0.52}{-0.51 }$&$<0.81$  & $<0.96$ &$<1.55$ \\ [1pt]
\textbf{321} &$<0.57$ &2.37$\SPSB{+0.57}{-0.57 }$ &$<0.14 $ & 4.69$\SPSB{+0.86}{-0.89 }$ &2.03$\SPSB{+0.23}{-0.23 }$ &$<0.42$ &3.98$\SPSB{+0.79}{-0.51 }$ &3.24$\SPSB{+0.51}{-0.51 }$ \\[1pt]
\textbf{326} &0.73$\SPSB{+0.22}{-0.82 }$ &2.13$\SPSB{+0.46}{-0.46 }$ &0.38$\SPSB{+0.05}{-0.24 }$ & 6.38$\SPSB{+0.84}{-0.91 }$ &2.05$\SPSB{+0.21}{-0.21 }$ &$<0.24$&1.25$\SPSB{+0.23}{-0.18 }$ &1.10$\SPSB{+0.18}{-0.18 }$ \\ [1pt]
339 & -- & -- & --  &5.78$\SPSB{+0.58}{-0.58 }$ &$<1.96$ &$<3.11$ & -- & --    \\[1pt]
344 & -- & -- & --  & 1.55$\SPSB{+0.16}{-0.16 }$ &$<1.99$ &$<5.15 $& -- & --   \\[1pt]
355  &  -- & -- & --  &5.83$\SPSB{+0.81}{-0.68 }$ &$<3.53$ &$<7.22$& -- & --   \\[1pt]
398  &  -- & -- & --  &1.12$\SPSB{+0.13}{-0.11 }$ &$<4.36$ &$<9.05$& -- & -- \\[1pt]
\textbf{413} &$<1.76$ &4.99$\SPSB{+1.11}{-1.11 }$ &2.28$\SPSB{+0.64}{-0.00 }$ & 4.55$\SPSB{+0.55}{-0.48 }$ &2.54$\SPSB{+0.48}{-0.48 }$ &2.42$\SPSB{+0.52}{-0.52 }$ &$<1.98$ & $<3.03 $  \\ [1pt]
414 &  -- & -- & --  &3.38$\SPSB{+0.37}{-0.41 }$ &$<8.70$ &$<23.0$ & -- & -- \\[1pt]
479 &  -- & -- & -- &3.79$\SPSB{+0.42}{-0.39 }$ &$<1.86$ &$<2.92$ & -- & -- \\[1pt]
491 & -- & -- & --  &2.17$\SPSB{+0.23}{-0.22 }$ &$<3.50$ &$<7.46$& -- & -- \\[1pt]
\textbf{504}  &0.98$\SPSB{+0.29}{-1.10 }$ &3.38$\SPSB{+0.70}{-0.70 }$ &1.09$\SPSB{+0.16}{-0.00 }$ & 7.04$\SPSB{+1.04}{-0.81 }$ &2.45$\SPSB{+0.29}{-0.31 }$  &$<0.73$ &$<1.29$ &$<1.35$ \\ [1pt]
 543 &  -- & -- & --  &4.31$\SPSB{+0.48}{-0.46 }$ &$<5.83 $&$<10.6$ & -- & --  \\[1pt]
\textbf{615} &$<3.08$ &8.54$\SPSB{+1.72}{-1.72 }$  &$<0.48$ &8.00$\SPSB{+2.14}{-1.83 }$ &4.07$\SPSB{+0.52}{-0.53 }$  &$<1.51 $ &$<2.34$ &$<3.33$\\[1pt]
\textbf{627}&$<3.12$&6.37$\SPSB{+1.40}{-1.40 }$ &1.78$\SPSB{+0.24}{-0.00 }$ & 3.99$\SPSB{+0.44}{-0.42 }$ &4.11$\SPSB{+0.57}{-0.58 }$ &$<1.82$ &$<2.38$ &16.02$\SPSB{+2.23}{-2.23 }$ \\[1pt]
 642  &  -- & -- & --  & 1.25$\SPSB{+0.14}{-0.14 }$ &$<6.65 $&$<13.4 $& -- & -- \\[1pt]
656  &   -- & -- & --  &3.57$\SPSB{+1.53}{-1.29 }$ &$<3.20$ &$<13.7$& -- & --  \\[1pt]
 657  &  -- & -- & --  &2.61$\SPSB{+0.31}{-0.27 }$ &$<4.66$ &$<10.9$& -- & --  \\[1pt]
\textit{658} & -- & -- & --  &8.65$\SPSB{+3.25}{-2.55 }$ &1.45$\SPSB{+0.37}{-1.30 }$ &$<16.0$& -- & -- \\[1pt]
\textbf{709} &2.15$\SPSB{+0.64}{-2.40 }$ &10.76$\SPSB{+2.33}{-2.33 }$ &3.93$\SPSB{+1.29}{-0.00 }$ &9.72$\SPSB{+1.11}{-1.01 }$ &6.27$\SPSB{+0.84}{-0.84 }$ &1.81$\SPSB{+0.33}{-0.33 }$ &$<1.63$ &$<2.36$\\[1pt]
 \textbf{715}&2.91$\SPSB{+0.87}{-3.24 }$ &6.59$\SPSB{+1.36}{-1.36 }$  &2.08$\SPSB{+0.40}{-0.85 }$  & 6.97$\SPSB{+1.76}{-1.72 }$ &3.72$\SPSB{+0.55}{-0.56 }$ &5.03$\SPSB{+1.11}{-1.11 }$ & 1.74$\SPSB{+0.71}{-0.00 }$ &$<2.62$ \\[2pt]
739  & -- & -- & --  & 3.00$\SPSB{+0.36}{-0.32 }$ &$<2.41$ &$<11.8$ & -- & --  \\[1pt]
 \textbf{746} &2.53$\SPSB{+0.76}{-2.82 }$ &8.97$\SPSB{+1.97}{-1.97 }$  &1.29$\SPSB{+0.29}{-0.00 }$&7.29$\SPSB{+0.98}{-0.81 }$ &6.00$\SPSB{+0.87}{-0.86 }$&1.27$\SPSB{+0.29}{-0.29 }$ &$<1.63$ &$<2.28$ \\[1pt]
 747  &  -- & -- & --  & 9.28$\SPSB{+1.22}{-1.03 }$ &$<6.65$ &$<19.3$ & -- & -- \\[1pt]
 \textbf{752} &2.41$\SPSB{+0.72}{-2.69 }$ &11.35$\SPSB{+2.24}{-2.24 }$ &4.34$\SPSB{+1.99}{-0.00 }$ &  12.9$\SPSB{+1.7}{-1.4 }$ &6.16$\SPSB{+0.80}{-0.80 }$ &$<2.29$ &$<3.36$ &$<4.19$\\[1pt]
\textbf{768} &$<1.13$ &5.30$\SPSB{+1.18}{-1.18 }$ &0.97$\SPSB{+0.45}{-0.25 }$ & 2.63$\SPSB{+0.29}{-0.27 }$ &2.43$\SPSB{+0.44}{-0.43 }$ &$<0.82$ & 2.39$\SPSB{+0.25}{-0.00 }$ &$<1.74$ \\[1pt]
780  &  -- & -- & --  & 3.03$\SPSB{+0.35}{-0.34 }$ &$<3.51$ &$<11.6$ & -- & -- \\ [1pt]
\textbf{799} &$< 2.41$  &7.33$\SPSB{+1.54}{-1.54 }$ &0.41$\SPSB{+0.11}{-0.00 }$ & 7.18$\SPSB{+1.11}{-0.87 }$ &3.57$\SPSB{+0.45}{-0.45 }$  &$<1.05$ &$<1.81$ &$<3.31$\\[1pt]
\textbf{800} &1.87$\SPSB{+0.56}{-2.08 }$ &3.73$\SPSB{+0.76}{-0.76 }$ &$<0.26$ &8.34$\SPSB{+1.12}{-0.92 }$ &4.10$\SPSB{+0.53}{-0.53 }$ &$<0.83$ &$<1.41$ &$<1.59$ \\
		\hline
    \end{tabular}
      \begin{description} 
     \item COMs listed by increasing number of atoms. $^{1}$ The cores observed with both the ARO 12m and Yebes 40m are bolded. For the italicized cores 67 and 658, the ARO 12m RMS is lower, $\sim 2$\,mK, than for the rest of the sample at $\sim 6$\,mK. In the case of t-HCOOH and H$_2$CCO, errors encompass full range of $N$ from the full range of assumed $T_\mathrm{ex}$ values, i.e., at 5\,K, 10\,K, and 20\,K. Errors quoted as `$0.00$' are $<0.005$. 
      \end{description}
\end{table*}

\subsection{Abundance Correlations} \label{sec:abund_corr}

We first explore how the abundances of each of the eight species of COMs focused on in this survey correlate with each other (Figure\,\ref{fig:correlation_cornerplot}). To quantitatively compare the relationship, we calculate the Spearman rank correlation coefficient `$r$' which measures the uniformity of the relationship between two datsets (part of the `scipy.stats' package: \cite{2020SciPy-NMeth}). A value of -1 or +1 imply an exact negative or positive monotonic relationship, respectively, and a value of 0 implies no correlation. Strong positive trends are typically expected for molecules that are chemically related or for molecules that depend on similar physical conditions, such as temperature. 

Firstly, we find a positive trend ($r$\,$=0.56$, considering only the detections) when CH$_3$CHO is compared to CH$_3$OH. This positive trend is consistent with the correlation found for the sample of 31 cores in Taurus, where $r$\,$=0.54$ \citep{2020ApJ...891...73S}. Additionally, the positive correlation we find for our full sample that considers the upper limits from the full Perseus sample (35 total cores) is compatible with this overall trend. These upper limits have limited constraints on CH$_3$CHO, thus deeper searches for CH$_3$CHO in these cores would likely cause the full trend to remain positively correlated.

Notably, there is not a significant positive correlation when the  CH$_3$CN abundance is compared against CH$_3$OH (r\,$=0.10/0.23$; for the full sample and detected sample, respectively). This is unlike what is seen during warm-phase chemistry from observations of CH$_3$CN and CH$_3$OH towards the neighboring protostars in Perseus \citep{2020A&A...635A.198B, 2021ApJ...910...20Y}. The PEACHES survey in particular find a tight correlation (r=$0.87$) between CH$_3$CN and CH$_3$OH that spans more than two order of magnitude in column densities and suggest a possible chemical relation between these two species and a large chemical diversity among the protostars \citep{2021ApJ...910...20Y}. Though, the CALYPSO survey cautioned that a strong correlation may not imply a chemical link \citep{2020A&A...635A.198B}. Similar to the PEACHES sample, we do see a positive correlation with CH$_3$OCH$_3$ and HCOOCH$_3$ (r\,$=0.76/0.50$) like they see for their sample of protostars. It should still be noted that the sample size of PEACHES is larger (factor of 3) when compared to our sub-sample of starless and prestellar cores in Perseus with CH$_3$CN, CH$_3$OCH$_3$ and HCOOCH$_3$ constraints. 

Other species with correlated abundances (r values $>0.5$ for both the full sample and detected sample; Figure\,\ref{fig:correlation_cornerplot}) include the N-bearing COMs CH$_3$CN and CH$_2$CHCN (r\,$=0.80/0.60$), H$_2$CCO and CH$_3$CN (r\,$=0.64/0.69$), as well as H$_2$CCO and CH$_3$CHO (r\,$= 0.65/0.62$). In \cite{2017ApJ...841..120B} they also find positive correlations between CH$_3$CHO and CH$_3$OH (r\,$=0.9$) and H$_2$CCO and CH$_3$CN (r\,$=0.99$) from single-dish data for small samples of low-mass protostars ($<6$), the majority of which are in Perseus. 

Next, we compare COM abundances with respect to H$_2$ to the average H$_2$ volume density, $n$(H$_2$), in units of cm$^{-3}$ for our sample of cores (Figure\,\ref{fig:phys_compare}). This $n$(H$_2$) value comes from Table\,\ref{physparams}, column 9, and is the median value calculated from the \textit{Herschel} maps, within the ARO 12m 62\,arcsec beam, where an error of 10\% is assumed. This is also the same beam size used to calculate the average H$_2$ volume density, $n$(H$_2$), for the Taurus cores in \cite{2020ApJ...891...73S}, though we stress that due to the difference in distances we are biased to lower $n$(H$_2$) values on average for the Perseus sample. 

There is significant scatter for CH$_3$OH, both in Perseus (r\,$=0.15$) and for the sample of cores in Taurus cores (r\,$= -0.13$; \citealt{2020ApJ...891...73S}), suggesting that CH$_3$OH does not correlate with $n$(H$_2$). As for CH$_3$CHO, a general negative correlation is found except when the densest core 326 is considered (r\,$=-0.62/-0.29$; for the full sample and detected sample, respectively) compared to no correlation found (r\,$=0.16$) for the full sample of Taurus cores with CH$_3$CHO detections. Compared to Perseus, the majority of the Taurus cores lie within the same $4\times 10^{4} - 2\times 10^{5}$\,cm$^{-3}$ range (see insert plot in top left panel of Figure\,\ref{fig:phys_compare}), though again we are biased to lower $n$(H$_2$) values due to the further distance of Perseus. The abundances with respect to H$_2$, however, do differ with both CH$_3$OH and CH$_3$CHO abundances in Taurus averaging an order of magnitude lower than those in Perseus, which could explain why similar trends in volume density are not seen. For the CH$_3$CHO abundances in Taurus specifically, the scatter is also due to the fact that a limited number (two or less) of molecular transitions were available to carry out the column density analysis, which also needed to assume a filling factor of one, $f=1$ (i.e., no source size was taken into account; see \cite{2020ApJ...891...73S}).

Looking at the other molecules, there does appear to be negative correlations for $n$(H$_2$) with the 5-atom species H$_2$CCO (r\,$=-0.78/-0.75$), t-HCOOH (r\,$=-0.59/-0.57$), the N-bearing species CH$_3$CN (r\,$=-0.60/-0.60$), CH$_2$CHCN (r\,$=-0.68/-0.50$), as well as CH$_3$OCH$_3$ (r\,$=-0.60/-1.0$; note only four data points for the detected sample). As for HCOOCH$_3$, no significant correlation is found, but it trends in the negative direction as well (r\,$=-0.47/-0.30$; see Figure\,\ref{fig:phys_compare}). 

The negative trend with respect to $n$(H$_2$) confirms that COM abundances (excluding CH$_3$OH) decrease with increasing volume density towards the dust peak of the starless and prestellar cores in Perseus. It has been seen with spatially resolved maps of the very dense and highly evolved prestellar core L1544 that the smaller COMs H$_2$CCO and CH$_3$CN are peaking away from the dust peak of the core \citep{2017A&A...606A..82S} and larger COMs have been found to be enhanced away of the dust peak toward the methanol peak \citep{2016ApJ...830L...6J}. It is therefore also probable the COMs in our Perseus sample are also peaking away from the dust peak in the higher density cores. This effect is due to depletion effects where in the densest cores molecules will freeze-out onto the grains more efficiently and thus their COM abundances are expected to be lower. While CH$_3$OH is also known to be chemically differentiated spatially in evolved cores like L1544 \citep{2014A&A...569A..27B} and `typical' cores in Taurus' L1495 filament \citep{2022ApJ...927..213P}, the abundance is likely high enough across larger spatial scales that we can not differentiate the CH$_3$OH depletion toward the dust peak of the Perseus cores within our observed beam sizes. It could also be that unlike CH$_3$OH, known to form on grains \citep{1992ApJ...399L..71C}, the other COMs in the Perseus cores could be forming predominantly in the gas-phase (e.g., \citealt{2015MNRAS.449L..16B, 2020MNRAS.499.5547V, 2023MNRAS.526.4535G}) and are more likely to experience depletion at high densities (see section\,\ref{sec:discussion_implications} for more discussion on formation routes). Future work to better characterize the spatial distribution of COMs towards these Perseus cores is needed. 

Outside of $n$(H$_2$), we also explore how abundances with respect to H$_2$ correlate with kinetic temperature, $T_\mathrm{k}$, and find no significant correlations (r\,$ < 0.3$). Additionally, 14 cores in our sample overlap with the magnetic field alignment catalog reported in \cite{2023MNRAS.525..364P}, and therefore we explore possible correlations with COM abundance for the larger sample of cores with CH$_3$OH and CH$_3$CHO detections. We find no significant trend (r\,$ > -0.4$) in CH$_3$OH abundance with velocity gradient or magnetic field alignment. For CH$_3$CHO, however, a negative correlation in the velocity gradient is found for the seven overlapping catalog matched cores  (r\,=\,-0.67), but no strong correlation is found for the magnetic field alignment (r\,=\,-0.30). 

\subsection{COM Abundance Across Low-Mass Star Formation}

\begin{figure*}
\centering
\begin{center}$
\begin{array}{c}
\includegraphics[width=170mm]{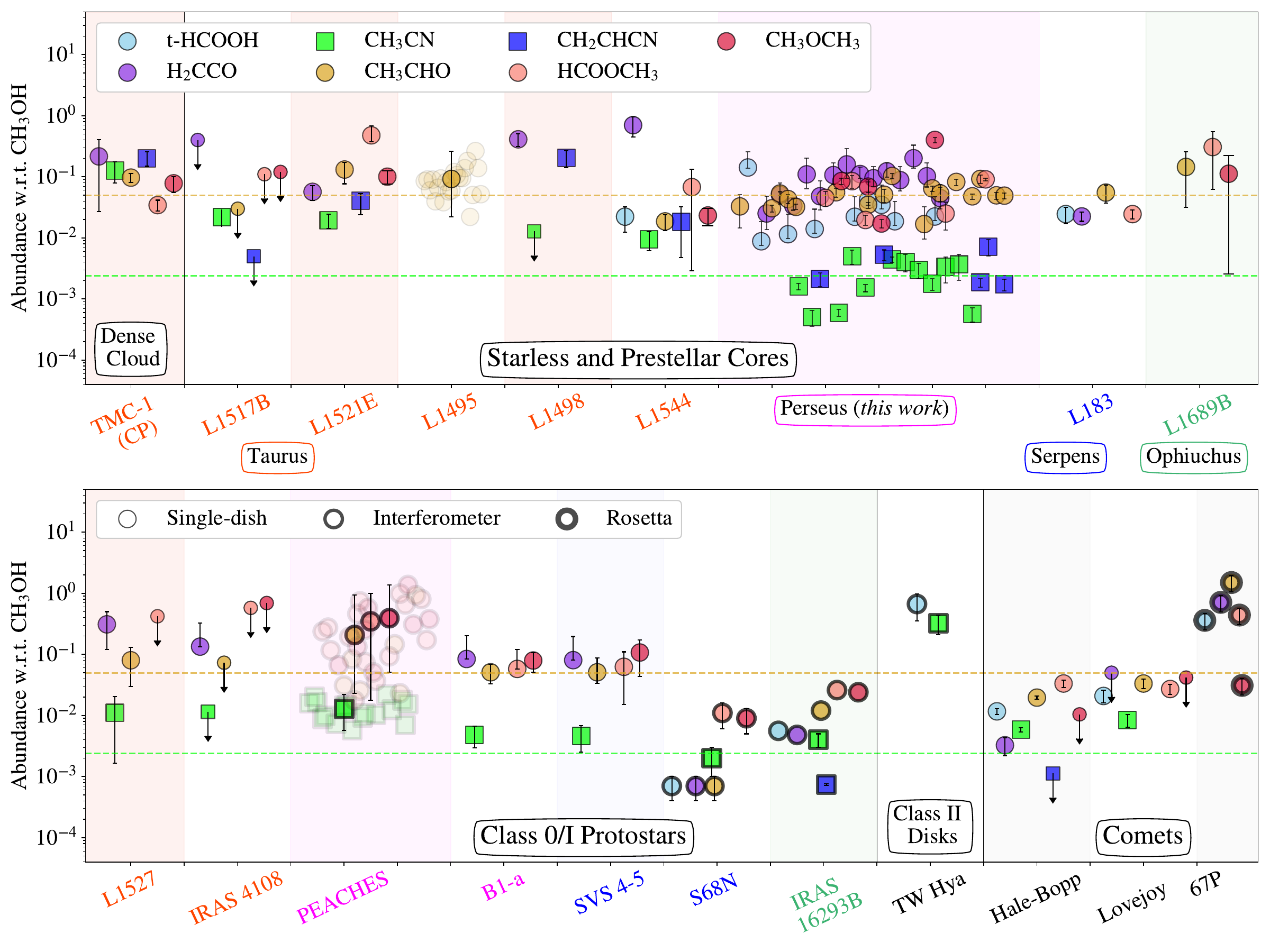} \\
\end{array}$
\end{center}
\caption{ \label{fig:meth_compare} A comparison of CH$_3$OH abundances across the stages of low-mass star formation, including a large sample of starless and prestellar cores (upper panel), as well as for a selection of protostars, disks and comets (lower panel). 
A gold and green vertical dashed line represents the median CH$_3$CHO and CH$_3$CN values (Table\,\ref{methabund}), respectively, from the Perseus starless and prestellar core sample presented in this work. Values for the dense cloud TMC-1 are at the cyanopolyyne peak, or CP \citep{2015ApJ...802...74S, 2018ApJ...854..116S, 2016ApJS..225...25G, 2021A&A...649L...4A}. For the starless and prestellar cores, we compare only to dust peak measurements (L1517B; \citealt{2023MNRAS.519.1601M}, L1512E; \citealt{2019A&A...630A.136N, 2021MNRAS.504.5754S}, L1495; \citealt{2020ApJ...891...73S}, L1498; \citealt{2021ApJ...917...44J}, L1544; \citealt{2014ApJ...795L...2V, 2016ApJ...830L...6J, 2021ApJ...917...44J}, L183; \citealt{2020A&A...633A.118L}, and L1689B; \citealt{2012A&A...541L..12B}). Protostars are further distinguished by whether observations where made with a single-dish or interferometer by the thickness of the marker (L1527; \citealt{2019PASJ...71S..18Y}, PEACHES; \citealt{2021ApJ...910...20Y}, IRAS 4108, B1-a, SVS 4-5; \citealt{2016ApJ...819..140G, 2017ApJ...841..120B}, S68N; \citealt{2020A&A...639A..87V, 2021A&A...650A.150N}, and IRAS 16293B; \citealt{2018A&A...616A..90C, 2018A&A...620A.170J}). For the sources that are made up of a large samples of cores, i.e., `L1495' and `PEACHES', we plot the average values and the spread in the sample as error bars, where the sample is also shown by more translucent markers. The Class II disk, TW Hya, is plotted from interferometer measurements detailed in \citealt{2016ApJ...823L..10W, 2018ApJ...859..131L, 2018ApJ...862L...2F}. Lastly, we also include radio observations of the long-range comets, Hale-Bop and Lovejoy \citep{2019ECS.....3.1550B}, as well as Jupiter-family comet, 67P/Churyumov–Gerasimenko or 67P, from measurements taken with the mass spectroscopy instrument aboard the Rosetta mission \citep{2023A&A...678A..22H}.
}
\end{figure*}

CH$_3$OH is a `mother' molecule to COM formation that survives the disk formation process and is used to normalize other COM abundances in order to compare across the various stages of low-mass star formation, from cores to comets (e.g., \citealt{2019MNRAS.490...50D, 2020A&A...639A..87V, 2021MNRAS.504.5754S, 2021ApJ...910...20Y}). In Figure\,\ref{fig:meth_compare}, we plot the abundances of COMs with respect to CH$_3$OH for not only the sample of starless and prestellar cores in Perseus focused on in this study, but the abundance values for other cold cores as well as a selection of Class 0/I protostars, a Class II disk, and three comets.  

\subsubsection{Starless and Prestellar Cores}

We first compare in the top panel of Figure\,\ref{fig:meth_compare} the CH$_3$OH abundances of our sample of cores in Perseus to the cyanopolyyne peak (CP) of the dense cloud TMC-1 \citep{2015ApJ...802...74S, 2018ApJ...854..116S, 2016ApJS..225...25G, 2021A&A...649L...4A}, as well as the dust peak of a sample of other starless and prestellar cores with COM detections and/or upper limits, i.e., L1517B; \cite{2023MNRAS.519.1601M}, L1512E; \cite{2019A&A...630A.136N, 2021MNRAS.504.5754S}, the 22 cores in the L1495 filament; \cite{2020ApJ...891...73S}, L1498; \cite{2021ApJ...917...44J}, L1544; \cite{2014ApJ...795L...2V, 2016ApJ...830L...6J, 2021ApJ...917...44J}, L183; \cite{2020A&A...633A.118L}, and L1689B; \cite{2012A&A...541L..12B}.

We find for O-bearing species abundances are similar, i.e., within roughly one order of magnitude, as seen in \cite{2021MNRAS.504.5754S}. However, more than one order of magnitude drop in abundance is seen for the N-bearing species CH$_3$CN and CH$_2$CHCN when comparing between cores with detections in Taurus compared to the Perseus cores. Though, the upper limit for the young Taurus core L1517B does show a similar drop in CH$_2$CHCN/CH$_3$OH abundances. 

In Table\,\ref{methabund} we list all median values, along with standard deviations, for COM abundances with respect to CH$_3$OH (as well as H$_2$ and CH$_3$CN). Representative of the O-bearing and N-bearing species, respectively, the median CH$_3$CHO/CH$_3$OH and CH$_3$CN/CH$_3$OH ratios in the Perseus sample (plotted as gold and green dashed lines in Figure\,\ref{fig:meth_compare}) are offset by a factor of 25. If we compare this to the Taurus cores with detections of both CH$_3$CHO and CH$_3$CN, the offset in this methanol abundances ranges from factors of 1.3 to  6.7. This gap in the abundance of N-bearing COMs with respect to CH$_3$OH compared to O-bearing COMs in Perseus that is not seen in the sample of cores we have in the Taurus Molecular Cloud could suggest different O/N ratios toward these two clouds. 

In order to test this, we first look at the median abundance ratio of our two N-bearing molecules, CH$_2$CHCN/CH$_3$CN, which is 1.06$\pm$0.27 for our Perseus sample of cores (Table\,\ref{methabund}). Compared to the Taurus cores with detections of both molecules, the ratio is slightly enhanced where CH$_2$CHCN/CH$_3$CN for TMC-1 is $\sim 1.58$, for L1521E $\sim 2.1$ and L1544 $\sim 1.9$. For those cores with upper limits, interestingly CH$_2$CHCN/CH$_3$CN in core L1498 is $>16$ and in core L1517B $<0.22$ (see Figure\,\ref{fig:meth_compare}). Still, the median value for all Taurus cores is enhanced at 1.94$\pm$0.36, which suggests a different O/N ratio is likely not the reason for the N-bearing and O-bearing COM abundance discrepancy, i.e., rather than a standardized inventory of enhanced `N', the formation efficiency of CH$_2$CHCN may instead be diminished in Perseus.
And, we know from observations across the molecular filaments in Taurus and Perseus (that avoid protostars, HII regions and outflows) the ratio of the simpler O- to N- species HCO$^+$/HCN and H$^{13}$CO$^+$/H$^{13}$CN have been found to be within a factor of $\sim1$ \citep{2021A&A...648A.120R}.

It may also be that N-bearing COMs are probing the chemical evolution in starless and prestellar cores. As discussed by \cite{2023MNRAS.519.1601M}, it is possible that N-bearing and O-bearing COMs form at different dynamical timescales with N-bearing species forming first and bigger molecules forming later on, based on the detections in L1517B, L1498 and L1544. Though, this scenario breaks down for the chemically young core L1521E and chemically rich TMC-1, also believed to be young (see Figure\,\ref{fig:meth_compare}). 

If we look at our sample of Perseus cores, we find for core 413, which has the most confident CH$_2$CHCN detection (i.e., two transitions at $>3\sigma$), there are no HCOOCH$_3$ or CH$_3$OCH$_3$ detections. While core 264 is the only core where we detect all three molecules, CH$_2$CHCN, HCOOCH$_3$ and CH$_3$OCH$_3$ (Figures\,\ref{vyncyn_spec_yebes40m},\,\ref{mf_spec_yebes40m},\,and\,\ref{dme_spec_yebes40m}). Using density as a simple evolutionary tracer, it may be that core 413, with an average H$_2$ abundance of $n$(H$_2$)=0.42\,$\times$\,10$^{5}$\,cm$^{-3}$ is less evolved than core 264, with $n$(H$_2$)=0.75\,$\times$\,10$^{5}$\,cm$^{-3}$ (Table\,\ref{physparams}). 
It should be noted that for the densest core 326, at $n$(H$_2$)=2.87\,$\times$\,10$^{5}$\,cm$^{-3}$, there are strong detections of both HCOOCH$_3$ and CH$_3$OCH$_3$, yet we do not see CH$_2$CHCN. In this case, environmental effects may play a bigger role in the chemistry, as it sits directly along the path of the CO outflows from nearby protostar SVS 13 \citep{2013ApJ...774...22P}. Interestingly, looking at Figure\,\ref{fig:phys_compare}, CH$_2$CHCN is only detected in a narrower range of densities, from $n$(H$_2$)=$0.42-0.75\times$\,10$^{5}$\,cm$^{-3}$, when compared to the full sample of cores observed. As for CH$_3$CN, this COM is seen over much wider range of densities and in cores both rich in the higher complexity O-bearing species (HCOOCH$_3$ and CH$_3$OCH$_3$) as well as CH$_2$CHCN. 

So, while the exact evolutionary stage of a core may be difficult to probe by just looking at the presence of N-bearing COMs,  what could be different in the Perseus and Taurus cores is the rate at which the cores are evolving, or rather the timescale on which the core is collapsing. If the N-bearing COMs are forming first, as proposed in \cite{2023MNRAS.519.1601M}, this process may be happening more quickly in Perseus and CH$_3$CN and CH$_2$CHCN may experience freeze-out (as discussed in section\,\ref{sec:abund_corr}) earlier and thus we see a drop in the gas-phase abundances compared to the O-bearing species that have formed later. There is contradictory evidence, however, for this proposed scenario, e.g., the prevalence of CH$_3$CN across a variety of cores at different densities in Perseus and the lack of correlation between the cores with detection of both CH$_3$CN and CH$_2$CHCN (see Figure\,\ref{fig:meth_compare}).

It is more likely there are compounding effects, both environmental and evolutionary, that influence COM abundances with respect to CH$_3$OH for starless and prestellar cores in the localized regions within both Taurus and Perseus. A more thorough analysis should be done to differentiate the evolutionary stages of each of the Perseus cores (e.g., such as measuring CO depletion factors or isotopologue fractions) and to characterize the local environment in order to further explore this difference in the N-bearing and O-bearing COM abundances with respect to methanol.

\begin{table}
	\centering
	\caption{Median Abundances for Perseus Starless and Prestellar Cores}
 \setlength{\tabcolsep}{7pt}
	\label{methabund}
	\begin{tabular}{llll} 
    \tablecolumns{10}
     \tablewidth{0pt}
     \tabcolsep=0.1cm
Species & X/H$_2$ $\times 10^{-10}$ & X/CH$_3$OH &  X/CH$_3$CN \\  
\hline
t-HCOOH & 2.15(0.76) &  0.022(0.011)   &  15.7(5.28)\\ [1pt]
H$_2$CCO & 5.83(2.58) & 0.098(0.036)   & 35.8(13.8) \\ [1pt]
CH$_3$CN & 0.119(0.078)  &  0.002(0.001)  &  1\\ [1pt]
CH$_3$OH & 43.1(16.74) &  1 & 450(200) \\ [1pt]
CH$_3$CHO &  3.56(1.14) & 0.050(0.015)  &  22.8(7.73)\\ [1pt]
CH$_3$CHCN &  0.181(0.060) & 0.002(0.004) & 1.06(0.27) \\ [1pt]
HCOOCH$_3$ &  2.34(0.59) & 0.045(0.026)  & 26.4(4.16) \\ [1pt]
CH$_3$OCH$_3$ & 3.82(1.65) &   0.077(0.034)  & 53.1(24.2) \\ 
\hline 
\hline 
O-bearing & 3.69(1.45) & 0.050(0.025) & 31.1(11.8)\\ 
N-bearing & 0.150(0.031) & 0.002(0.001) & 1.06(0.27) \\ 
		\hline
    \end{tabular}
      \begin{description} 
     \item Note: values reported are for the subset(s) of detected cores. The standard deviation of each median value is reported in parentheses.
      \end{description}
\end{table}

\subsubsection{Protostars}

Next, we look at the Class 0/I protostars in the bottom panel of Figure\,\ref{fig:meth_compare} to find the O-bearing COMs are generally consistent within an order of magnitude, suggesting at least some of the chemical inventory in the cold core stage is being inherited to the protostellar stage. For the N-bearing COMs (specifically CH$_3$CN), lower abundances are found with respect to CH$_3$OH than the O-bearing COMs across not only the sample protostars, but also the sample of starless and prestellar cores (excluding TMC-1; Figure\,\ref{fig:meth_compare}). The singular CH$_2$CHCN data point also falls below the O-bearing COM abundances from ALMA observations of IRAS\,16293B \citep{2018A&A...616A..90C}. An outlier, the S68N protostar does show t-HCOOH, H$_2$CCO and CH$_3$CHO abundances falling below CH$_3$CN in Figure\,\ref{fig:meth_compare}, yet \citealt{2021A&A...650A.150N} find that when comparing to a larger variety of N-bearing COMs, they are roughly an order of magnitude lower in abundance than the O-bearing COMs, which is not only true for S68N but for several other protostars as well (i.e., compare Figure 7 in \citealt{2020A&A...639A..87V} and Figure 7 in \citealt{2021A&A...650A.150N}). 
The consistency in this general trend of N-bearing COMs being lower in abundance than O-bearing species in cold cores and protostars further suggests that there is a limited reprocessing of material as cores evolve. 

Looking only in Perseus, we find that when compared to the cold cores in our sample (i.e., the median CH$_3$CN/CH$_3$OH line in Figure\,\ref{fig:meth_compare}) the abundance for the N-bearing species CH$_3$CN is $\sim$2$\times$ higher in the bright Perseus B1-a source taken with single-dish data \citep{2017ApJ...841..120B} and $\sim$6$\times$ higher when compared the abundances from the full PEACHES sample derived from ALMA observations \citep{2021ApJ...910...20Y}. While \cite{2021ApJ...910...20Y} speculated seeing a decrease of CH$_3$CN/CH$_3$OH from the prestellar to the protostellar phase in Perseus, here we see the opposite, potentially evidence that warm-phase grain-surface pathways may release back into the gas-phase the CH$_3$CN and CH$_2$CHCN species and regulate the abundances seen in later stages in Perseus. 

The lack of statistics of both O-bearing and N-bearing species in protostars in Taurus make it more difficult to draw robust conclusions about COM inheritance in comparison to Perseus. The CALYPSO survey, originally meant for studying the angular momentum problem in protostars, was not sensitive enough to detect COM emission towards the three protostars in their sample located in Taurus (IRAM04191, L1521F and L1527; \citealt{2020A&A...635A.198B}). Despite limited detections, it is evident that the Taurus protostars follow the same general trend as the protostars in Perseus (as well as for Serpens and Ophiuchus), i.e., N-bearing COMs are in lower abundances than O-bearing COMs, and thus it is unclear how the different clouds may affect COM abundances in this stage.  We note that there has been a direct comparison ALMA study (matched to the same sensitivity and resolution as PEACHES) in the massive star forming region of Orion (named `ORANGES'), which does find that a lower fraction, 29$\%$, of the 19 solar-mass Class 0/I protostars targeted show CH$_3$OH emission when compared to the PEACHES survey \citep{2022ApJ...929...10B}. More analysis of abundance ratios need to be done, however, in order to conclude if environmental factors indeed play a large role in the complex chemical evolution at the protostellar stage. 

We also point out, as \cite{2021ApJ...910...20Y} show (their Figure 17), that the abundances of both the N-bearing species CH$_3$CN and O-bearing species CH$_3$CHO, HCOOCH$_3$ and CH$_3$OCH$_3$ in the PEACHES sample are systematically higher than for other isolated protostars in different environments also observed with ALMA (as we show for S68N and IRAS\,16293B in Figure\,\ref{fig:meth_compare}), though, again, in both single-dish and interferometric measurements O-bearing species are generally enhanced compared to N-bearing species. A proper study of the statistics and sampling biases in these literature values of protostars should be done in order to say more. 

\subsubsection{Protoplanetary disks}

Constraining the COM inventory in the next stage of low-mass star formation, Class II disks, remains a challenge. In Figure\,\ref{fig:meth_compare} we plot two abundances constraints from the T-Tauri disk TW HYa using column density values for CH$_3$OH \citep{2016ApJ...823L..10W}, t-HCOOH \citep{2018ApJ...862L...2F}, and CH$_3$CN \citep{2018ApJ...859..131L}. The t-HCOOH/CH$_3$OH and CH$_3$CN/CH$_3$OH ratio in TW\,HYa is $\sim$20$\times$ and $\sim$160$\times$ higher, respectively, than the median values found in our Perseus sample of starless and prestellar cores (Table\,\ref{methabund}). Still, the lower abundance of CH$_3$CN compared to t-HCOOH may suggest even at this later stage there is a limited reprocessing of the chemical inventory inherited from the prestellar phase.

Beyond TW\,HYa, the detection of higher complexity COMs in T-Tauri disks remains elusive. There are constraints for disks with higher mass stars at their centers, such as the Herbig disks IRS\,48 and HD\,100546, where CH$_3$OH, HCOOCH$_3$ and CH$_3$OCH$_3$ (in IRS\,48 only) have been detected \citep{2024arXiv240204001B, 2024arXiv240204002B}. The abundances of this warm reservoir of gas interestingly show even higher abundances of HCOOCH$_3$ and CH$_3$OCH$_3$ when compared to CH$_3$OH, i.e., for IRS\,48 peak values for HCOOCH$_3$/CH$_3$OH $\sim$20 and CH$_3$OCH$_3$/CH$_3$OH $\sim$6. Though, as the authors point out, these above unity ratios may be due to high optical depth of the CH$_3$OH emission in disks and future work studying isotopic species such as $^{13}$CH$_3$OH is needed to say more about whether differences in CH$_3$OH abundances in protoplanetary disks is really due to a potential re-processing of CH$_3$OH. 

\subsubsection{Comets}

Lastly, in Figure\,\ref{fig:meth_compare} we plot abundances of COMs with respect to CH$_3$OH for the long-range comets Hale-Bopp and Lovejoy \citep{2019ECS.....3.1550B}, which show good agreement with abundance values in starless cores and protostars. We also include measurements taken with the mass spectroscopy instrument aboard the Rosetta mission for the Jupiter-family comet 67P/Churyumov–Gerasimenko or 67P \citep{2023A&A...678A..22H}. For 67P, the data from a mass spectroscopy analysis should be taken lightly as it can be difficult to disentangle individual isomers (i.e., for HCOOCH$_3$ this includes methyl formate, acetic acid and glycolaldehyde), and because in this current study 67P was analyzed during a `very dusty' period close to the comet's perihelion when heavy organics are expected to be released. In fact, it is only in the recent analysis of 67P where CH$_3$CHO$_3$ has been claimed, as it was not seen in its previous analysis \citep{2019ECS.....3.1854S}. Irrespective of exact abundance values, the presence of these complex species throughout low-mass star formation, from cores to comets, warrants further investigation into their formation and possible inheritance across a larger range of environments. 

\subsection{Implications for COM Formation} \label{sec:discussion_implications}

In order to get COMs into the gas-phase in starless and prestellar cores, current models rely on the inclusion of chemical reactive desorption (CRD) and gas-phase radiative associations at often high chemical desorption efficiencies, such as the model of the evolved prestellar core L1544 detailed in \cite{2017ApJ...842...33V}, based on the work in \cite{2013ApJ...769...34V}. The basis of this work has then been applied to other cores with varying degrees of success. For example, \cite{2021MNRAS.504.5754S} find that CRD plus gas-phase chemistry alone is not sufficient to explain observed COMs in the chemically young starless core L1521E. Yet, \cite{2021ApJ...917...44J} find it does relatively well for the chemically evolved but dynamically young starless core L1498. Additional methods to increase COM gas-phase abundances include the incorporation of non-diffusive grain-surface processes \citep{2020ApJS..249...26J}, Eley-Rideal processes \citep{2015MNRAS.447.4004R}, cosmic-ray radiation chemistry \citep{2018PCCP...20.5359S, 2018ApJ...861...20S}, cosmic-ray sputtering \citep{2021A&A...652A..63W, 2022MNRAS.516.4097P}, and turbulence induced desorption due to collisions \citep{2022MNRAS.515..785K}. Below we give a general overview of predicted model abundances that are able to match (within order-of-magnitude agreement) our observations. 

\subsubsection{O-bearing COMs}

The formation of CH$_3$OH is known to proceed efficiently through the hydrogenation of CO on icy grains \citep{1992ApJ...399L..71C}. Relevant to cold ($\sim$ 10\,K) cores, there is additional motivation for the reaction with radicals CH$_3$ and H$_2$CO that leads to the formation of CH$_3$OH and the radical HCO that can go on to aid in the formation of additional COMs \citep{2022ApJ...931L..33S}. Within CRD models (e.g., in \citealt{2021MNRAS.504.5754S}), CH$_3$CHO is formed from the slow radiative association of CH$_3$ + HCO and the neutral-neutral reaction of CH + CH$_3$OH \citep{2000PCCP....2.2549J}. The models of \cite{2017ApJ...842...33V} and \cite{2021MNRAS.504.5754S} do predict abundances (with respect to H$_2$) of CH$_3$CHO at $\sim$10$^{-10}$, which is within order-of-magnitude agreement with the range of abundances for the Perseus cores (Table\,\ref{methabund}). For the 5-atom species H$_2$CCO and t-HCOOH, the CRD models of \cite{2021ApJ...917...44J} and \cite{2023MNRAS.519.1601M} are also able to produce abundances comparable to what we see in our Perseus cores, $\sim$ a few $\times 10^{-10}$.

For the higher complexity species HCOOCH$_3$ and CH$_3$OCH$_3$, explaining the presence of such high gas-phase abundances (a few $\times$ 10$^{-10}$) has not been as easy for the CRD models. For example, \cite{2021MNRAS.504.5754S} find no agreement between model ($\sim10^{-12}$) and observed ($\sim10^{-10}$) abundances of HCOOCH$_3$ and CH$_3$OCH$_3$ toward core L1521E. More recent analysis that includes the addition of cosmic ray sputtering, for example, can get abundances up to the levels we see in the Perseus cores (e.g., in \cite{2022MNRAS.516.4097P} their model 8HSC10). Turbulence-induced desorption from the models of \cite{2022MNRAS.515..785K} may also be able to increase the gas-phase abundances of HCOOCH$_3$ and CH$_3$OCH$_3$ in particular (see their Table 2 `max' model).
It should also be noted that for the handful of cores in our Perseus survey with detections of HCOOCH$_3$ and CH$_3$OCH$_3$ they mainly reside in the active and shocked region of NGC1333 (e.g., cores 264, 321, and 326) and thus there are likely compounded interactions towards some of our sources from thermal processes (e.g., shocks or outflow interactions) that may be able to more efficiently get these radical species and COMs off the grains. As evident for cores 264 and 326, independent RADEX fits of HCOOCH$_3$ find enhanced $T_\mathrm{k}$ values of $\sim$20\,K (see Appendix\,\ref{radexappendix}). 

Given the wide range of physical and environmental conditions of these Perseus cores, more detailed modeling on a case-by-case basis would be needed to say more about exact COM formation mechanisms. Still, despite variations, the range of calculated O-bearing COM abundances for these starless and prestellar cores in Perseus can in general be described by chemical models that include nonthermal desorption mechanisms. 

\subsubsection{N-bearing COMs}

The presence and formation of N-bearing COMs in prestellar cores remains more of a mystery. CH$_2$CHCN is of particular interest because it is thought to be one of the strongest candidates for forming membranes of potential astrobiological importance. \cite{2015Icar..256....1S} showed that through theoretical liquid-phase calculations, CH$_2$CHCN is one of the most favored to form these thermodynamically stable membranes in liquid methane at the surface temperature of Titan. Recent observations of Titan with ALMA have been successful in detecting CH$_2$CHCN \citep{2017SciA....3E0022P}, highlighting the molecules' significance in astrobiology. 

Toward the starless core L1498, CH$_2$CHCN is detected toward the dust peak, while CH$_3$CN is only detected toward the methanol peak \citep{2021ApJ...917...44J}. Conversely, toward core L1517B they do not detect CH$_2$CHCN at either the dust or methanol peak, yet do in fact detect CH$_3$CN at only the dust peak \citep{2023MNRAS.519.1601M}. In both studies, they use a CRD model to attempt to reproduced abundance constraints and are unable to do so for CH$_2$CHCN, which in their models reach abundances with respect to H$_2$ $\sim10^{-13}$ (note in \cite{2021MNRAS.504.5754S} they reproduce higher values $\sim10^{-12}$). Even for CH$_3$CN, the CRD models have trouble reproducing values above $\sim10^{-11}$, which would be consistent with our Perseus observations (see Table\,\ref{methabund}). 

In the case of CH$_2$CHCN, the gas-phase production routes are known to be inefficient (e.g., KIDA database; \cite{2012ApJS..199...21W}). Despite this, we have shown that similar to the O-bearing COMs (excluding grain-produced CH$_3$OH), the abundances with respect to H$_2$ for the N-bearing COMs may also deplete with increasing volume density (Figure\,\ref{fig:phys_compare}), suggesting gas-phase formation may still be important. \cite{2023MNRAS.526.4535G} show for CH$_3$CN there are efficient routes in the gas-phase that can produce rather high abundances, $\sim10^{-10}$, when modeled for the dark cloud TMC-1. Perhaps for CH$_2$CHCN other gas-phase formation routes are more efficient and/or the reactive desorption efficiency for this molecule is particularly high. The other desorption studies mentioned here unfortunately do not provide constraints for N-bearing COM abundances \citep{2015MNRAS.447.4004R,  2018PCCP...20.5359S, 2018ApJ...861...20S, 2020ApJS..249...26J, 2021A&A...652A..63W, 2022MNRAS.516.4097P, 2022MNRAS.515..785K}. We therefore stress that additional experimental and theoretical work on efficient gas and grain chemical pathways and desorption mechanisms of N-bearing COMs are still needed to explain the abundances (a few $\times \sim10^{-11}$) we find for CH$_2$CHCN and CH$_3$CN in the starless and prestellar cores in Perseus.

\section{Conclusions}\label{sec:conclusion}

We find a prevalence of COMs in starless and prestellar cores in the Perseus Molecular Cloud from a combined ARO 12m and Yebes 40m molecular line survey, more than doubling the COM abundance statistics for cold ($\sim$10\,K) cores. We detect CH$_3$OH (100\%: 35/35), CH$_3$CHO (49\%: 17/35), H$_2$CCO (93\%: 14/15), t-HCOOH (60\%: 9/15), CH$_3$CN (80\%: 12/15), CH$_2$CHCN (< 34\%), HCOOCH$_3$ (< 34\%) and CH$_3$OCH$_3$ (< 27\%) in multiple cores, but in no single core were all these molecules detected. Our main conclusions are:

\begin{enumerate}

    \item We find CH$_3$OH and CH$_3$CHO are just as prevalent, if not more so, in our sample of 35 cores in the roughly 2$\times$ farther Perseus Molecular Cloud when compared to the  COM survey of 31 starless and prestellar cores in the nearby (135\,pc) L1495/B218 filament of the Taurus Molecular Cloud \citep{2020ApJ...891...73S}. Moreover, CH$_3$CHO is likely widespread, as evident from detections of this molecule in the randomly selected cores 67 and 658 when higher sensitivity ($\sim 2$\,mK) observations were carried out.

    \item Most starless and prestellar cores in our survey reside in the active Perseus clusters NGC\,1333 and IC\,348, which are the regions that show increased COM complexity. It is within these regions where not only are there detections of higher complexity COMs in the prestellar stage, there is also an increase in COM detections in the warm protostellar stage when compared to the spatial location of the PEACHES sample \citep{2021ApJ...910...20Y}. 

   \item By combining multiple CH$_3$OH transitions with varying beam sizes, we account for the true emitted source size during our column density calculations. The total CH$_3$OH column densities, $N$(CH$_3$OH), for the 35-core sample range from $0.87 - 50.27 \times 10^{13}$\,cm$^{-2}$ with a median source size of 38.5$\pm$4 arcsec. 

   \item When normalized to molecular hydrogen, we find for the O-bearing COMs abundances that range from $2.15 - 43.1$\,$\times 10^{-10}$. For the N-bearing COMs we find roughly an order of magnitude lower abundances, with a values ranging from $ 0.118 - 0.181 \times 10^{-10}$. 

   \item From an abundance correlation analysis, we find no significant positive correlation (r\,$=0.10/0.23$, for the detected and full sample, respectively) with the CH$_3$CN/CH$_3$OH abundances , which is unlike what is seen toward the nearby protostars in Perseus \citep{2020A&A...635A.198B, 2021ApJ...910...20Y}. We do find significant (r$>0.5$) correlation for the following pairs of molecular abundances: CH$_3$OCH$_3$/HCOOCH$_3$ (r\,$=0.76/0.50$), CH$_3$CN/CH$_2$CHCN (r\,$=0.80/0.60$), H$_2$CCO/CH$_3$CN (r\,$=0.64/0.69$) and H$_2$CCO/CH$_3$CHO (r\,$= 0.65/0.62$).

   \item We find COM abundances (excluding CH$_3$OH) with respect to H$_2$ decrease with increasing average volume density, $n$(H$_2$), which is most likely due to depletion effects, where in the densest cores, molecules will freeze-out onto the grains. Additional observations that can map the spatial distribution of COMs in starless and prestellar cores are needed to better trace potential freeze-out and compare to chemical models. 

   \item By normalizing abundances to the grain-produced species CH$_3$OH, we also compare COM abundances in our Perseus sample to a large sample of other starless and prestellar cores, as well as to a sample of Class 0/I protostars, a Class II T-Tauri disk, and a handful of comets. In general, the similarity in O-bearing COM abundances throughout each stage suggests at least some of the COMs in the prestellar stage are being inherited to the later stages. Perhaps most notable, compared to starless and prestellar cores in Taurus, we find for our Perseus sample the abundances of N-bearing COMs are  $\sim$20 times lower, whereas the O-bearing COMs are in agreement ($\sim$ within an order of magnitude). This discrepancy in N-bearing abundances within the Perseus starless and prestellar cores suggests that perhaps different environmental effects are at play and/or these cores are evolving at a different rate compared to those sampled in Taurus. In the later stages, N-bearing species tend to also be lower in abundance with respect to methanol than O-bearing species, which further motivates that there is a limited amount of reprocessing of the complex chemical inventory throughout low-mass star formation.

   \item Despite variation in environment and evolutionary stage (i.e., varying densities) of the cores, the molecular hydrogen abundances for the O-bearing species are consistent with chemical models that rely on nonthermal desorption mechanisms to get COMs into the gas-phase. However, echoing the work of \cite{2021ApJ...917...44J} and \cite{2021MNRAS.504.5754S}, more experimental and theoretical work needs to be done in order to reproduce abundances of N-bearing COMs in the starless and prestellar core phase. A more detailed analysis for these individual cores in Perseus, taking into account their physical and environmental conditions, will be needed to say more about the dominate reaction and formation pathways for the COMs studied. 
   
\end{enumerate}

Overall, our synergistic observations of molecular line emission in both the 3mm and 7mm (Q-band) regime has proven to be a powerful mechanism for studying the complex chemistry of cold (10\,K) molecular gas in starless and prestellar cores, which represent one of the earliest stages in low-mass star formation that precedes the formation of Sun-like stars and potentially Earth-like planets. 

\section*{Acknowledgements}

We thank Yao-Lun Yang for providing comparison data from the PEACHES survey that improved this paper. Additionally, we thank our anonymous reviewer for their constructive comments. We are thankful that we have the opportunity to conduct astronomical research on Iolkam Du'ag (Kitt Peak) in Arizona and we recognize and acknowledge the very significant cultural role and reverence that these sites have to the Tohono O'odham Nation.
We sincerely thank the operators of the Arizona Radio Observatory (Michael Begam, Kevin Bays, Clayton Kyle, and Robert Thompson) for their assistance with the observations. 
The 12 m Telescope is operated by the Arizona Radio Observatory (ARO), Steward Observatory, University of Arizona, with funding from the State of Arizona, NSF MRI Grant AST-1531366 (PI Ziurys), NSF MSIP grant SV5-85009/AST- 1440254 (PI Marrone), NSF CAREER grant AST-1653228 (PI Marrone), and a PIRE grant OISE-1743747 (PI Psaltis). 

We also acknowledge observations carried out with the Yebes 40 m telescope (22A022 and 23A025). Paula Tarr\'io and Alba Vidal Garc\'ia carried out the observations and the first inspection of the data quality. The 40\,m radio telescope at Yebes Observatory is operated by the Spanish Geographic Institute (IGN; Ministerio de Transportes, Movilidad y Agenda Urbana). Yebes Observatory thanks the European Union’s Horizon 2020 research and innovation programme for funding support to ORP project under grant agreement No 101004719. 

Support for S.S. and Y.S. was provided by National Science Foundation Astronomy and Astrophysics Grant (AAG) AST-2205474 as well as the Universities Space Research Association (USRA) Stratospheric Observatory of Infrared Astronomy (SOFIA) grant 09-0155. S.S. acknowledges the National Radio Astronomy Observatory is a facility of the National Science Foundation operated under cooperative agreement by Associated Universities, Inc. A.M. has received support from grant number PRE2019-091471 funded by the MICIU/AEI/ 10.13039/501100011033 and by “ERDF/EU”. I.J.S and A.M. acknowledge financial support from grants No. PID2019-105552RB-C41 and PID2022-136814NB-I00 funded by MICIU/AEI/ 10.13039/501100011033 and by “ERDF/EU”.

\section*{Data Availability} \label{dataavailability}
 
The CLASS/Python reduction pipeline is available on github:{\url{https://github.com/andresmegias/gildas-class-python/}}. The reduced spectra (T$_\mathrm{mb}$ scale) presented in this paper is available on the Harvard Dataverse at the following link:{\url{https://doi.org/10.7910/DVN/H4YZ4U}}.



\bibliographystyle{mnras}
\bibliography{references.bib} 


\clearpage
\newpage

\appendix

\onecolumn

\section{Supplemental Material from ARO 12m Observations} \label{appendixARO}

\begin{figure*}
\includegraphics[width=150mm]{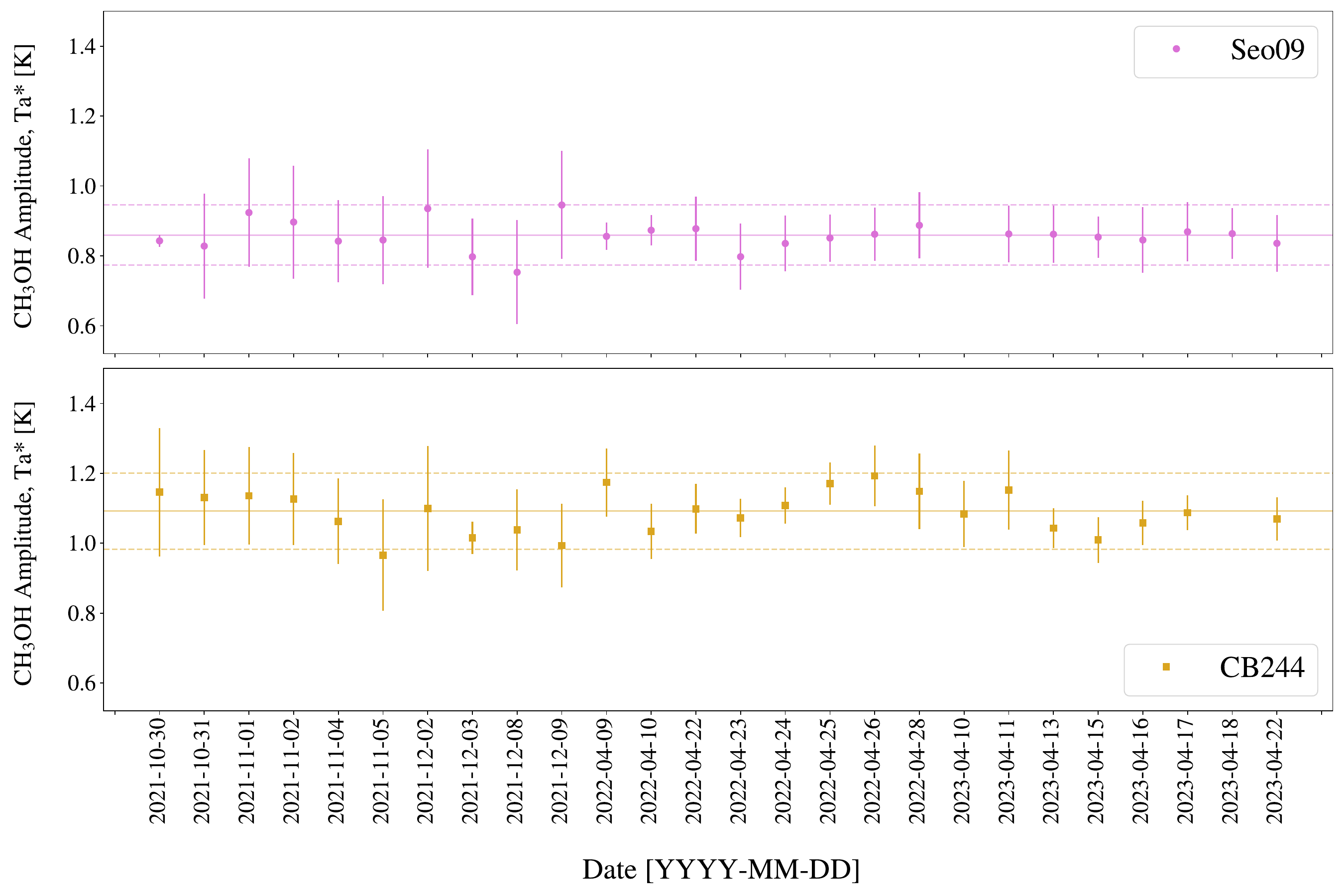} 
\caption{ \label{fig:standardsource} Standard sources Seo09 and CB244 were monitored during ARO 12m observations from 2021 to 2023. We plot the Gaussian fitted amplitude for the brightest CH$_3$OH $2_{0,2}-1_{0,1} \mathrm{A}$ transition (errorbars 3$\sigma_\mathrm{rms}$) versus the date observed. The amplitudes do not vary by more than $\pm10\%$ from the mean (dashed horizontal lines).}
\end{figure*}

Additional details regarding the ARO 12m observational setup (section\,\ref{subsec:arored}) are described here. Firstly, while observations were underway we monitored two standard sources over the roughly two year span of observations. Science scans were taken at the peak dust continuum position of nearby starless cores CB244 (23:25:27.1, +74:18:25.3, J2000.0) and Seo09 (04:18:07, +28:05:13, J2000.0; \citealt{2015ApJ...805..185S}) to monitor any changes in flux. 
In Figure\,\ref{fig:standardsource} the amplitudes (Ta* in units of Kelvin) of the brightest CH$_3$OH $2_{0,2}-1_{0,1} \mathrm{A}$ transition are plotted versus time, spanning from 2021 to the last observation date in 2023. The data collected for each session was baselined and hanning smoothed by 2 channels before the amplitude was measured with a Gaussian fit. Figure\,\ref{fig:standardsource} shows that the change in flux for each standard source did not vary by more than 10$\%$. 

To convert to the main beam temperature scale, where $T_\mathrm{mb}$ = $T_A^*$/$\eta$ \citep{1993PASP..105..117M}, different $\eta$ values were calculated for each observing season. During the fall of 2021 we used $\eta_\mathrm{ARO1} = 84.25\pm0.73\,\%$ and $\eta_\mathrm{ARO2} = 85.40\pm0.48\,\%$ by taking the median measurements from Jupiter, Uranus and Venus. During the Spring of 2022 we used $\eta_\mathrm{ARO1} = 86.60\pm1.36\,\%$ and $\eta_\mathrm{ARO2} = 87.96\pm1.53\,\%$ by taking the median measurements from Jupiter, Mars and Uranus. And, in the Spring of 2023 we used $\eta_\mathrm{ARO1} = 80.02\pm0.50\,\%$ and $\eta_\mathrm{ARO2} = 89.09\pm0.59\,\%$ by taking the median measurements from Mars and Venus.

As for our spectrometer setup with the AROWS backend, we tuned simultaneously to four lines: the $2_{0,2} - 1_{0,1}$\,E transition of CH$_3$OH at 96.74455 GHz, the center of the two $5_{0,4}-4_{0,4}$ transition of CH$_3$CHO at 95.9554 GHz, the $10_{1,9}-9_{1,8}$ transition of CH$_2$CHCN at 96.982446 GHz, and the $10_{0,10}-9_{0,9}$ transition of CH$_2$CHCN at 94.276641 GHz (see Table\,\ref{LineList}). The FWHM beam sizes are therefore between $\sim$ 62 -- 64 arcsec. This setup was achieved with the AROWS multi-window mode where each line was placed in four separate spectral windows within 4 GHz either in the Lower Side Band (LSB), as was done during spring 2022 observations, or in the Upper Side Band (USB), as was done during fall 2021 and spring 2023 observations. Note that the spring 2022 observations were done in this different sideband configuration in order to get rid of standing waves present in the USB at the time. 

\begin{figure*}
\includegraphics[width=175mm]{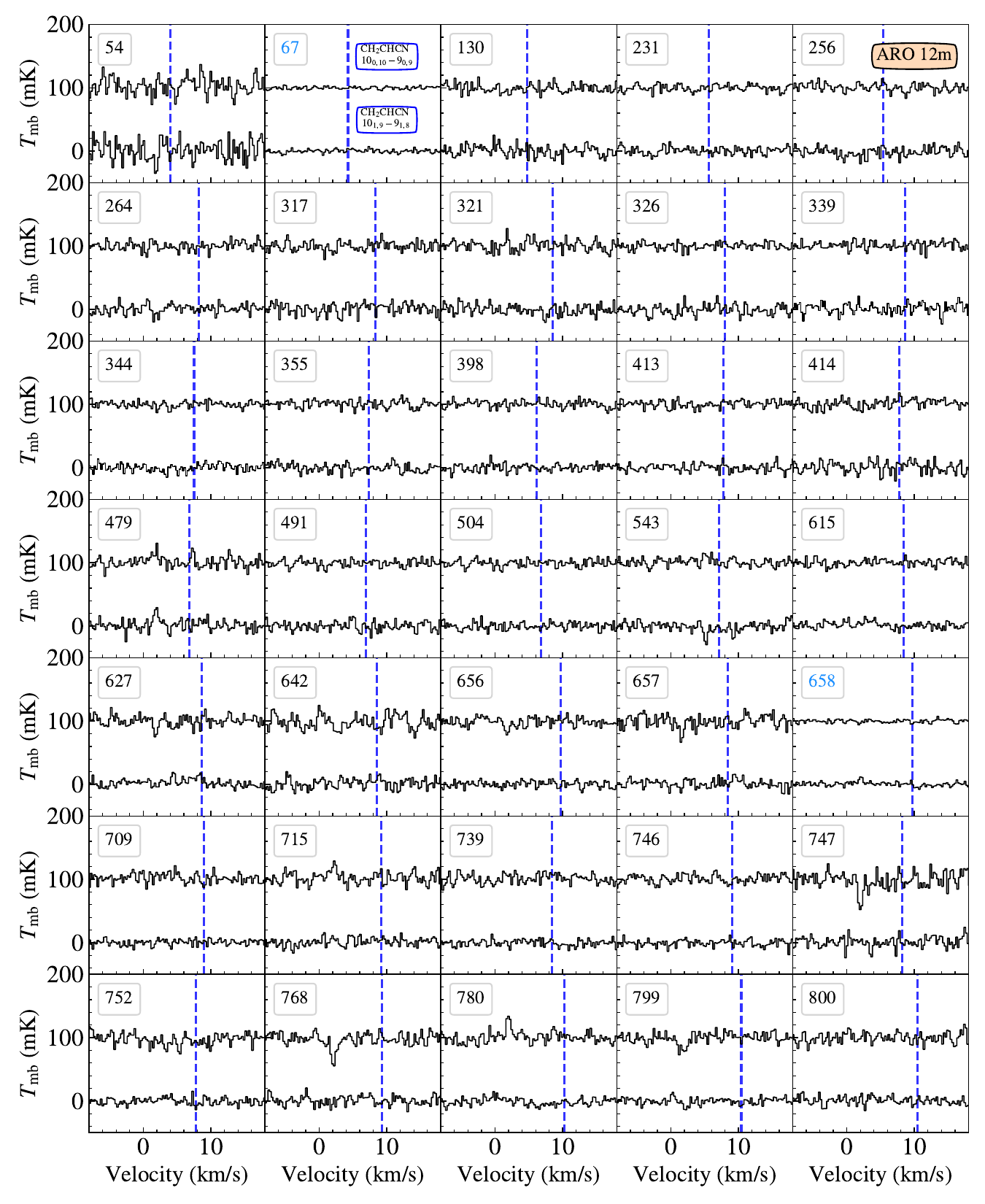} 
\caption{ \label{vincyn_spec_12m} Non-detections of vinyl cyanide, CH$_2$CHCN, the 94GHz and 96GHz line spectrum (in black) in units of $T_\mathrm{mb}$ (K) versus velocity (km/s) from the ARO 12m for each of the 35 starless and prestellar cores targeted in this Perseus survey. Core numbers are labeled in the upper left of each panel. Longer integration time on cores 67 and 658 (labeled in blue), resulted in RMS values around $\sigma_{T_\mathrm{mb}} = 2-3\,\mathrm{mK}$. For the remaining cores, the average RMS value is $\sigma_{T_\mathrm{mb}} = 6-7\,\mathrm{mK}$. The $10_{0,10} - 9_{0,9}$ line at 94GHz is shifted by 100\,mK for easier viewing. A vertical blue dashed line is overlaid on each panel to show the v$_{lsr}$ found from the brightest ARO 12m methanol line. }
\end{figure*}

The non-detected vinyl cyanide, CH$_2$CHCN, lines are plotted in Figure\,\ref{vincyn_spec_12m}. Both the $10_{0,10} - 9_{0,9}$ and $10_{1,9} - 9_{1,8}$ lines at 94.27 and 96.98 GHz, respectively, are shown with the 94.27\,GHz line shifted up by 100\,mK for easier viewing. For core 54, the vertical polarization was corrupted in the two CH$_2$CHCN windows, likely due to a standing wave interference, and therefore the noise level is higher. 

In Table\,\ref{GaussFits} we list the Gaussian fit line parameters, as well as the RMS values, calculated using the CLASS software and found for all of the ARO 12m spectral data towards each of the 35 cores. Line velocities have been re-shifted to the corresponding rest frequencies. 

\begin{deluxetable}{lllllll}
\tablecaption{ARO 12m Line Parameters \label{GaussFits}}
\tablewidth{-1pt}
\tabcolsep=0.4cm
\tablehead{ 
\colhead{Core} & \colhead{Line} & \colhead{Area} & \colhead{Vel } & \colhead{FWHM} & \colhead{$T_\mathrm{mb}$} & \colhead{RMS}  \\ 
\colhead{} & \colhead{} &\colhead{ (K-km s$^{-1}$)} & \colhead{(km s$^{-1}$)} & \colhead{(km s$^{-1}$)} & \colhead{(K)} & \colhead{(K)} }
\startdata
54 & CH$_3$OH 2$_{0,2} - 1_{0,1}$ E & 0.026(0.003) &5.009(0.026) &0.397(0.049) &0.061 &0.0067\\ & CH$_3$OH 2$_{0,2} - 1_{0,1}$ A & 0.233(0.004) &4.049(0.006) &0.724(0.016) &0.302 &0.0067\\ & CH$_3$OH 2$_{-1,2} - 1_{-1,1}$ E &0.168(0.004) &4.048(0.009) &0.727(0.024) &0.217 &0.0067\\
&CH$_3$CHO $5_{0,5} - 4_{0,4}$ A & -- &-- &-- &-- &0.0053\\ 
&CH$_3$CHO $5_{0,5} - 4_{0,4}$ E &-- &-- &-- &-- &0.0053\\
&CH$_2$CHCN $10_{0,10} - 9_{0,9}$ & -- &-- &-- &-- & 0.0131\\
&CH$_2$CHCN $10_{1,9} - 9_{1,8}$& -- &-- &-- &-- & 0.0150\\

67* & CH$_3$OH 2$_{0,2} - 1_{0,1}$ E &  0.0095(0.002) & 4.381(0.034)  &  0.463(0.096)  & 0.0193 & 0.0029 \\ 

& CH$_3$OH 2$_{0,2} - 1_{0,1}$ A&  0.24701(0.002) & 4.508(0.003) & 0.747(0.006) & 0.3106 & 0.0029\\ 
&CH$_3$OH 2$_{-1,2} - 1_{-1,1}$ E  &  0.18519(0.002) &   4.496(0.004) & 0.734(0.009) & 0.2370 & 0.0029 \\ 

&CH$_3$CHO $5_{0,5} - 4_{0,4}$ A & 0.0064(0.001) & 4.398(0.082) & 0.730(0.152) & 0.0083 & 0.0023\\ 
&CH$_3$CHO $5_{0,5} - 4_{0,4}$ E & -- &-- &-- &-- & 0.0023 \\
&CH$_2$CHCN $10_{0,10} - 9_{0,9}$ & -- &-- &-- &-- & 0.0023\\
&CH$_2$CHCN $10_{1,9} - 9_{1,8}$& -- &-- &-- &-- & 0.0027\\

130& CH$_3$OH 2$_{0,2} - 1_{0,1}$ E &-- &-- &-- &-- &0.0079\\ & CH$_3$OH 2$_{0,2} - 1_{0,1}$ A &0.181(0.004) &4.772(0.006) &0.531(0.014) &0.319 &0.0079\\ &CH$_3$OH 2$_{-1,2} - 1_{-1,1}$ E &0.142(0.004) &4.758(0.008) &0.555(0.02) &0.241 &0.0079\\ 
&CH$_3$CHO $5_{0,5} - 4_{0,4}$ A & -- &-- &-- &-- &0.0064\\ 
&CH$_3$CHO $5_{0,5} - 4_{0,4}$ E &-- &-- &-- &-- &0.0064\\
&CH$_2$CHCN $10_{0,10} - 9_{0,9}$ & -- &-- &-- &-- & 0.0064\\
&CH$_2$CHCN $10_{1,9} - 9_{1,8}$& -- &-- &-- &-- & 0.0077\\

231 & CH$_3$OH 2$_{0,2} - 1_{0,1}$ E &-- &-- &-- &-- &0.0064 \\ & CH$_3$OH 2$_{0,2} - 1_{0,1}$ A &0.204(0.003) &5.578(0.004) &0.458(0.009) &0.418 &0.0064 \\ &CH$_3$OH 2$_{-1,2} - 1_{-1,1}$ E  &0.146(0.003) &5.571(0.005) &0.429(0.009) &0.32 &0.0064\\
&CH$_3$CHO $5_{0,5} - 4_{0,4}$ A & -- &-- &-- &-- &0.0047\\ &CH$_3$CHO $5_{0,5} - 4_{0,4}$ E &-- &-- &-- &-- &0.0047\\
&CH$_2$CHCN $10_{0,10} - 9_{0,9}$ & -- &-- &-- &-- & 0.0052 \\
&CH$_2$CHCN $10_{1,9} - 9_{1,8}$& -- &-- &-- &-- & 0.0061 \\

256 & CH$_3$OH 2$_{0,2} - 1_{0,1}$ E &-- &-- &-- &--&0.0076\\ 
& CH$_3$OH 2$_{0,2} - 1_{0,1}$ A  &0.208(0.004) &5.361(0.005) &0.502(0.012) &0.389 &0.0076\\ 
& CH$_3$OH 2$_{-1,2} - 1_{-1,1}$ E  &0.163(0.004) &5.352(0.007) &0.518(0.016) &0.296 &0.0076 \\
&CH$_3$CHO $5_{0,5} - 4_{0,4}$ A & -- &-- &-- &-- &0.0062\\ 
&CH$_3$CHO $5_{0,5} - 4_{0,4}$ E &-- &-- &-- &-- &0.0062\\
&CH$_2$CHCN $10_{0,10} - 9_{0,9}$ & -- &-- &-- &-- & 0.0053\\
&CH$_2$CHCN $10_{1,9} - 9_{1,8}$& -- &-- &-- &-- & 0.0071 \\

264 & CH$_3$OH 2$_{0,2} - 1_{0,1}$ E & 0.0765(0.007) & 7.969(0.053) & 1.193(0.113) &  0.0627 &0.0085 \\ 
& CH$_3$OH 2$_{0,2} - 1_{0,1}$ A &0.623(0.007) &8.162(0.004) &0.947(0.012) &0.617 &0.0085\\
& CH$_3$OH 2$_{-1,2} - 1_{-1,1}$ E  &0.423(0.006) &8.139(0.006) &0.853(0.015) &0.466 &0.0085\\ 
&CH$_3$CHO $5_{(0,5)} - 4_{(0,4)}$ A &0.015(0.004) &8.120(0.06) &0.507(0.163) &0.0277 &0.0066\\ 
&CH$_3$CHO $5_{(0,5)} - 4_{(0,4)}$ E &0.0144(0.003) &7.979(0.053) &0.462(0.144) &0.0292 &0.0066\\
&CH$_2$CHCN $10_{0,10} - 9_{0,9}$ & -- &-- &-- &-- & 0.0065 \\
&CH$_2$CHCN $10_{1,9} - 9_{1,8}$& -- &-- &-- &-- & 0.0068 \\

317&CH$_3$OH 2$_{0,2} - 1_{0,1}$ E  &0.166(0.007) &8.383(0.002) &0.771(0.039) &0.202 &0.0098\\
& CH$_3$OH 2$_{0,2} - 1_{0,1}$ A &1.07(0.008) &8.343(0.003) &0.885(0.008) &1.14 &0.0098\\ 
& CH$_3$OH 2$_{-1,2} - 1_{-1,1}$ E &0.797(0.008) &8.340(0.002) &0.887(0.01) &0.845 &0.0098\\
&CH$_3$CHO $5_{0,5} - 4_{0,4}$ A &0.0599(0.005) &8.388(0.034) &0.835(0.086) &0.0674 &0.0073\\
&CH$_3$CHO $5_{0,5} - 4_{0,4}$ E &0.0617(0.006) &8.301(0.048) &1.04(0.124) &0.0557 &0.0073\\
&CH$_2$CHCN $10_{0,10} - 9_{0,9}$ & -- &-- &-- &-- & 0.0073 \\
&CH$_2$CHCN $10_{1,9} - 9_{1,8}$& -- &-- &-- &-- & 0.0083 \\

321& CH$_3$OH 2$_{0,2} - 1_{0,1}$ E  &0.198(0.055) &8.523(0.242) &0.819(0.242) &0.227&0.0080\\ 
& CH$_3$OH 2$_{0,2} - 1_{0,1}$ A&1.55(0.055) &8.530(0.242) &1.01(0.242) &1.44&0.0080 \\
& CH$_3$OH 2$_{-1,2} - 1_{-1,1}$ E  &1.11(0.055) &8.519(0.242) &0.966(0.242) &1.08 &0.0080\\
&CH$_3$CHO $5_{0,5} - 4_{0,4}$ A &0.0599(0.005) &8.505(0.034) &0.835(0.086) &0.0674 &0.0073 \\
&CH$_3$CHO $5_{0,5} - 4_{0,4}$ E &0.0617(0.006) &8.528(0.048) &1.04(0.124) &0.0557 &0.0073\\
&CH$_2$CHCN $10_{0,10} - 9_{0,9}$ & -- &-- &-- &-- & 0.0080\\
&CH$_2$CHCN $10_{1,9} - 9_{1,8}$& -- &-- &-- &-- & 0.0077\\

326& CH$_3$OH 2$_{0,2} - 1_{0,1}$ E  &0.399(0.096) &7.885(0.242) &1.4(0.242) &0.267 &0.0104\\  
& CH$_3$OH 2$_{0,2} - 1_{0,1}$ A &3.05(0.096) &7.941(0.242) &1.37(0.242) &2.1 &0.0104\\
& CH$_3$OH 2$_{-1,2} - 1_{-1,1}$ E &2.32(0.096) &7.932(0.242) &1.36(0.242) &1.6 &0.0104\\
&CH$_3$CHO $5_{0,5} - 4_{0,4}$ A &0.0885(0.005) &7.917(0.037) &1.45(0.09) &0.0574 &0.0052\\
&CH$_3$CHO $5_{0,5} - 4_{0,4}$ E & 0.0805(0.003) &7.826(0.033) &1.24(0.06) &0.061 &0.0052\\
&CH$_2$CHCN $10_{0,10} - 9_{0,9}$ & -- &-- &-- &-- & 0.0055\\
&CH$_2$CHCN $10_{1,9} - 9_{1,8}$& -- &-- &-- &-- & 0.0076\\

339& CH$_3$OH 2$_{0,2} - 1_{0,1}$ E &0.109(0.008) &8.568(0.037) &1.08(0.091) &0.095  &0.0094\\
& CH$_3$OH 2$_{0,2} - 1_{0,1}$ A &0.759(0.007) &8.638(0.005) &1.04(0.011) &0.683  &0.0094\\ 
& CH$_3$OH 2$_{-1,2} - 1_{-1,1}$ E &0.549(0.007) &8.621(0.007) &1.05(0.017) &0.491 &0.0094 \\
&CH$_3$CHO $5_{0,5} - 4_{0,4}$ A & -- &-- &-- &-- &0.0065\\
&CH$_3$CHO $5_{0,5} - 4_{0,4}$ E & -- &-- &-- &-- &0.0065\\
&CH$_2$CHCN $10_{0,10} - 9_{0,9}$ & -- &-- &-- &-- & 0.0061\\
&CH$_2$CHCN $10_{1,9} - 9_{1,8}$& -- &-- &-- &-- & 0.0079\\

344& CH$_3$OH 2$_{0,2} - 1_{0,1}$ E  &0.0146(0.003) &7.543(0.059) &0.526(0.113) &0.0262&0.0062\\
& CH$_3$OH 2$_{0,2} - 1_{0,1}$ A &0.157(0.004) &7.510(0.012) &0.822(0.025) &0.179 &0.0062\\
& CH$_3$OH 2$_{-1,2} - 1_{-1,1}$ E &0.109(0.004) &7.459(0.015) &0.793(0.033) &0.129 &0.0062\\
&CH$_3$CHO $5_{0,5} - 4_{0,4}$ A & -- &-- &-- &-- &0.0046\\
&CH$_3$CHO $5_{0,5} - 4_{0,4}$ E & -- &-- &-- &-- &0.0046\\
&CH$_2$CHCN $10_{0,10} - 9_{0,9}$ & -- &-- &-- &-- & 0.0048 \\
&CH$_2$CHCN $10_{1,9} - 9_{1,8}$& -- &-- &-- &-- & 0.0054 \\

355 &CH$_3$OH 2$_{0,2} - 1_{0,1}$ E&0.0285(0.003) &7.314(0.006) &0.402(0.062) &0.0666 &0.0064\\ 
& CH$_3$OH 2$_{0,2} - 1_{0,1}$ A &0.286(0.001) &7.314(0.003) &0.485(0.005) &0.553&0.0064 \\ 
& CH$_3$OH 2$_{-1,2} - 1_{-1,1}$ E&0.207(0.003) &7.305(0.003) &0.478(0.009) &0.407 &0.0064\\
&CH$_3$CHO $5_{0,5} - 4_{0,4}$ A & -- &-- &-- &-- &0.0056\\
&CH$_3$CHO $5_{0,5} - 4_{0,4}$ E & -- &-- &-- &-- &0.0056\\
&CH$_2$CHCN $10_{0,10} - 9_{0,9}$ & -- &-- &-- &-- & 0.0055\\
&CH$_2$CHCN $10_{1,9} - 9_{1,8}$& -- &-- &-- &-- & 0.0060\\

398 &CH$_3$OH 2$_{0,2} - 1_{0,1}$ E&-- &-- &-- &-- & 0.0061 \\ 
&CH$_3$OH 2$_{0,2} - 1_{0,1}$ A &0.0557(0.003) &6.150(0.014) &0.496(0.037) &0.106 & 0.0061\\ 
& CH$_3$OH 2$_{-1,2} - 1_{-1,1}$ E & 0.0351(0.003) &6.143(0.02) &0.461(0.051) &0.0715 &0.0061\\
&CH$_3$CHO $5_{0,5} - 4_{0,4}$ A & -- &-- &-- &-- &0.0056\\
&CH$_3$CHO $5_{0,5} - 4_{0,4}$ E & -- &-- &-- &-- &0.0056\\
&CH$_2$CHCN $10_{0,10} - 9_{0,9}$ & -- &-- &-- &-- & 0.0053 \\
&CH$_2$CHCN $10_{1,9} - 9_{1,8}$& -- &-- &-- &-- & 0.0055 \\

413 &CH$_3$OH 2$_{0,2} - 1_{0,1}$ E&0.0179(0.003) &7.767(0.037) &0.439(0.081) &0.0383 & 0.0069\\
& CH$_3$OH 2$_{0,2} - 1_{0,1}$ A  &0.300(0.004) &7.738(0.004) &0.661(0.01) &0.426& 0.0069\\
& CH$_3$OH 2$_{-1,2} - 1_{-1,1}$ E  &0.225(0.004) &7.708(0.001) &0.669(0.015) &0.316 &0.0069 \\
&CH$_3$CHO $5_{0,5} - 4_{0,4}$ A &0.012(0.003) &7.722(0.062) &0.511(0.128) &0.022 & 0.0055\\
&CH$_3$CHO $5_{0,5} - 4_{0,4}$ E & -- &-- &-- &-- &0.0055\\
&CH$_2$CHCN $10_{0,10} - 9_{0,9}$ & -- &-- &-- &-- & 0.0054\\
&CH$_2$CHCN $10_{1,9} - 9_{1,8}$& -- &-- &-- &-- & 0.0057\\

414 &CH$_3$OH 2$_{0,2} - 1_{0,1}$ E&-- &-- &-- &-- & 0.0073\\ 
& CH$_3$OH 2$_{0,2} - 1_{0,1}$ A &0.0755(0.004) &7.731(0.01) &0.433(0.023) &0.164 & 0.0073\\ 
& CH$_3$OH 2$_{-1,2} - 1_{-1,1}$ E &0.0463(0.003) &7.723(0.01) &0.321(0.061) &0.136 &0.0073\\
&CH$_3$CHO $5_{0,5} - 4_{0,4}$ A & -- &-- &-- &-- &0.0066\\
&CH$_3$CHO $5_{0,5} - 4_{0,4}$ E & -- &-- &-- &-- &0.0066\\
&CH$_2$CHCN $10_{0,10} - 9_{0,9}$ & -- &-- &-- &-- & 0.0061\\
&CH$_2$CHCN $10_{1,9} - 9_{1,8}$& -- &-- &-- &-- & 0.0078\\

479 &CH$_3$OH 2$_{0,2} - 1_{0,1}$ E&-- &-- &-- &--&0.0105\\ 
& CH$_3$OH 2$_{0,2} - 1_{0,1}$ A &0.522(0.007) &6.792(0.005) &0.753(0.012) &0.651&0.0105\\
& CH$_3$OH 2$_{-1,2} - 1_{-1,1}$ E  &0.391(0.007) &6.785(0.007) &0.743(0.016) &0.494 &0.0105\\
&CH$_3$CHO $5_{0,5} - 4_{0,4}$ A & -- &-- &-- &-- &0.0075\\
&CH$_3$CHO $5_{0,5} - 4_{0,4}$ E & -- &-- &-- &-- &0.0075\\
&CH$_2$CHCN $10_{0,10} - 9_{0,9}$ & -- &-- &-- &-- & 0.0080\\
&CH$_2$CHCN $10_{1,9} - 9_{1,8}$& -- &-- &-- &-- & 0.0084\\

491&CH$_3$OH 2$_{0,2} - 1_{0,1}$ E &-- &-- &-- &-- &0.0075\\
& CH$_3$OH 2$_{0,2} - 1_{0,1}$ A &0.131(0.005) &6.926(0.011) &0.646(0.028) &0.191&0.0075\\ 
& CH$_3$OH 2$_{-1,2} - 1_{-1,1}$ E &0.0953(0.005) &6.931(0.016) &0.667(0.043) &0.134 &0.0075\\
&CH$_3$CHO $5_{0,5} - 4_{0,4}$ A & -- &-- &-- &-- &0.0055\\
&CH$_3$CHO $5_{0,5} - 4_{0,4}$ E & -- &-- &-- &-- &0.0055\\
&CH$_2$CHCN $10_{0,10} - 9_{0,9}$ & -- &-- &-- &-- & 0.0054\\
&CH$_2$CHCN $10_{1,9} - 9_{1,8}$& -- &-- &-- &-- & 0.0067\\

504 &CH$_3$OH 2$_{0,2} - 1_{0,1}$ E&0.0332(0.005) &6.778(0.03) &0.459(0.075) &0.068 & 0.0061 \\& CH$_3$OH 2$_{0,2} - 1_{0,1}$ A &0.446(0.006) &6.756(0.003) &0.591(0.009) &0.709 & 0.0061\\& CH$_3$OH 2$_{-1,2} - 1_{-1,1}$ E &0.333(0.006) &6.748(0.005) &0.598(0.013) &0.524 &0.0061\\
&CH$_3$CHO $5_{0,5} - 4_{0,4}$ A & -- &-- &-- &-- &0.0052\\
&CH$_3$CHO $5_{0,5} - 4_{0,4}$ E & 0.00985(0.002) &6.763(0.051) &0.329(0.094) &0.0281 &0.0052\\
&CH$_2$CHCN $10_{0,10} - 9_{0,9}$ & -- &-- &-- &-- & 0.0054 \\
&CH$_2$CHCN $10_{1,9} - 9_{1,8}$& -- &-- &-- &-- & 0.0055\\

543&CH$_3$OH 2$_{0,2} - 1_{0,1}$ E &-- &-- &-- &-- &0.0076\\
& CH$_3$OH 2$_{0,2} - 1_{0,1}$ A&0.156(0.004) &7.149(0.007) &0.513(0.017) &0.287&0.0076\\
& CH$_3$OH 2$_{-1,2} - 1_{-1,1}$ E &0.113(0.004) &7.152(0.009) &0.506(0.022) &0.21 &0.0076 \\
&CH$_3$CHO $5_{0,5} - 4_{0,4}$ A & -- &-- &-- &-- &0.0061\\
&CH$_3$CHO $5_{0,5} - 4_{0,4}$ E & -- &-- &-- &-- &0.0061\\
&CH$_2$CHCN $10_{0,10} - 9_{0,9}$ & -- &-- &-- &-- & 0.0062\\
&CH$_2$CHCN $10_{1,9} - 9_{1,8}$& -- &-- &-- &-- & 0.0063 \\

615 &CH$_3$OH 2$_{0,2} - 1_{0,1}$ E&0.0244(0.018) &8.352(0.242) &0.554(0.242) &0.0414 &0.0052\\
& CH$_3$OH 2$_{0,2} - 1_{0,1}$ A &0.352(0.018) &8.394(0.242) &0.539(0.242) &0.613 &0.0052\\
& CH$_3$OH 2$_{-1,2} - 1_{-1,1}$ E  &0.267(0.018) &8.379(0.242) &0.569(0.242) &0.44 &0.0052 \\
&CH$_3$CHO $5_{0,5} - 4_{0,4}$ A &0.0124(0.003) &8.308(0.063) &0.602(0.176) &0.0194 & 0.0045\\
&CH$_3$CHO $5_{0,5} - 4_{0,4}$ E &0.0106(0.002) &8.110(0.041) &0.379(0.066) &0.0263 &0.0045\\
&CH$_2$CHCN $10_{0,10} - 9_{0,9}$ & -- &-- &-- &-- & 0.0052 \\
&CH$_2$CHCN $10_{1,9} - 9_{1,8}$& -- &-- &-- &-- & 0.0049\\

627 &CH$_3$OH 2$_{0,2} - 1_{0,1}$ E&0.0269(0.007) &8.654(0.052) &0.49(0.171) &0.0516&0.0056\\
& CH$_3$OH 2$_{0,2} - 1_{0,1}$ A &0.23(0.005) &8.653(0.005) &0.421(0.011) &0.513&0.0056\\
& CH$_3$OH 2$_{-1,2} - 1_{-1,1}$ E  &0.162(0.005) &8.645(0.006) &0.393(0.012) &0.387 &0.0056\\
&CH$_3$CHO $5_{0,5} - 4_{0,4}$ A &0.0124(0.003) &8.551(0.057) &0.494(0.126) &0.0236 & 0.0054 \\
&CH$_3$CHO $5_{0,5} - 4_{0,4}$ E &0.0172(0.003) &8.372(0.077) &0.686(0.162) &0.0235 &0.0054 \\
&CH$_2$CHCN $10_{0,10} - 9_{0,9}$ & -- &-- &-- &-- & 0.0075\\
&CH$_2$CHCN $10_{1,9} - 9_{1,8}$& -- &-- &-- &-- & 0.0044\\

642&CH$_3$OH 2$_{0,2} - 1_{0,1}$ E &-- &-- &-- &-- &0.0069\\ 
& CH$_3$OH 
2$_{0,2} - 1_{0,1}$ A&0.0521(0.003) &8.518(0.012) &0.41(0.036) &0.119&0.0069\\
& CH$_3$OH 2$_{-1,2} - 1_{-1,1}$ E  &0.0406(0.003) &8.488(0.018) &0.432(0.041) &0.0883 &0.0069\\
&CH$_3$CHO $5_{0,5} - 4_{0,4}$ A & -- &-- &-- &-- &0.0072\\
&CH$_3$CHO $5_{0,5} - 4_{0,4}$ E & -- &-- &-- &-- &0.0072\\
&CH$_2$CHCN $10_{0,10} - 9_{0,9}$ & -- &-- &-- &-- & 0.0110\\
&CH$_2$CHCN $10_{1,9} - 9_{1,8}$& -- &-- &-- &-- &  0.0078\\

656 &CH$_3$OH 2$_{0,2} - 1_{0,1}$ E&0.012(0.011) &9.661(0.242) &0.3(0.242) &0.0377&0.0054\\& CH$_3$OH 2$_{0,2} - 1_{0,1}$ A &0.168(0.011) &9.669(0.242) &0.386(0.242) &0.409 &0.0054\\
& CH$_3$OH 2$_{-1,2} - 1_{-1,1}$ E &0.124(0.011) &9.664(0.242) &0.364(0.242) &0.32 &0.0054\\
&CH$_3$CHO $5_{0,5} - 4_{0,4}$ A & -- &-- &-- &-- &0.0049\\
&CH$_3$CHO $5_{0,5} - 4_{0,4}$ E & -- &-- &-- &-- &0.0049\\
&CH$_2$CHCN $10_{0,10} - 9_{0,9}$ & -- &-- &-- &-- & 0.0068\\
&CH$_2$CHCN $10_{1,9} - 9_{1,8}$& -- &-- &-- &-- & 0.0048\\

657 &CH$_3$OH 2$_{0,2} - 1_{0,1}$ E&-- &-- &-- &-- &0.0073\\
& CH$_3$OH 2$_{0,2} - 1_{0,1}$ A&0.127(0.004) &8.451(0.007) &0.497(0.019) &0.239&0.0073\\
& CH$_3$OH 2$_{-1,2} - 1_{-1,1}$ E  &0.0903(0.004) &8.462(0.011) &0.5(0.029) &0.17 &0.0073\\
&CH$_3$CHO $5_{0,5} - 4_{0,4}$ A & -- &-- &-- &-- &0.0064\\
&CH$_3$CHO $5_{0,5} - 4_{0,4}$ E & -- &-- &-- &-- &0.0064\\
&CH$_2$CHCN $10_{0,10} - 9_{0,9}$ & -- &-- &-- &-- & 0.0097 \\
&CH$_2$CHCN $10_{1,9} - 9_{1,8}$& -- &-- &-- &-- & 0.0065\\

658* 
&CH$_3$OH 2$_{0,2} - 1_{0,1}$ E& 0.0257(0.016) & 9.656(0.242) & 0.441(0.242) & 0.0547 & 0.0038 \\
& CH$_3$OH 2$_{0,2} - 1_{0,1}$ A& 0.2792(0.016) & 9.712(0.242) &  0.407(0.242) &  0.6451& 0.0038\\
& CH$_3$OH 2$_{-1,2} - 1_{-1,1}$ E  & 0.2049(0.016) & 9.704(0.242) & 0.399(0.242) &  0.4821 & 0.0038 \\
&CH$_3$CHO $5_{0,5} - 4_{0,4}$ A & 0.0056(0.001) & 9.533(0.036) & 0.389(0.151) & 0.0134 & 0.0025 \\
&CH$_3$CHO $5_{0,5} - 4_{0,4}$ E & 0.0074(0.001) & 9.494(0.041) &  0.463(0.102) & 0.0149 & 0.0025  \\
&CH$_2$CHCN $10_{0,10} - 9_{0,9}$ & -- &-- &-- &-- & 0.0026\\
&CH$_2$CHCN $10_{1,9} - 9_{1,8}$& -- &-- &-- &-- & 0.0028\\

709 &CH$_3$OH 2$_{0,2} - 1_{0,1}$ E&0.0913(0.004) &8.903(0.013) &0.694(0.031) &0.124&0.0059\\
& CH$_3$OH 2$_{0,2} - 1_{0,1}$ A &0.661(0.004) &8.954(0.002) &0.841(0.006) &0.739 &0.0059 \\
& CH$_3$OH 2$_{-1,2} - 1_{-1,1}$ E &0.464(0.004) &8.938(0.003) &0.827(0.008) &0.527 &0.0059\\
&CH$_3$CHO $5_{0,5} - 4_{0,4}$ A &0.0353(0.004) &8.919(0.044) &0.826(0.098) &0.0401 & 0.0080\\
&CH$_3$CHO $5_{0,5} - 4_{0,4}$ E &0.0433(0.004) &8.771(0.041) &0.89(0.086) &0.0457 &0.0080\\
&CH$_2$CHCN $10_{0,10} - 9_{0,9}$ & -- &-- &-- &-- & 0.0069\\
&CH$_2$CHCN $10_{1,9} - 9_{1,8}$& -- &-- &-- &-- & 0.0048\\

715&CH$_3$OH 2$_{0,2} - 1_{0,1}$ E &0.0265(0.015) &9.068(0.242) &0.713(0.242) &0.0349 &0.0059\\
& CH$_3$OH 2$_{0,2} - 1_{0,1}$ A &0.328(0.015) &9.158(0.242) &0.655(0.242) &0.47 &0.0059\\
& CH$_3$OH 2$_{-1,2} - 1_{-1,1}$ E &0.24(0.015) &9.146(0.242) &0.672(0.242) &0.335 &0.0064\\
&CH$_3$CHO $5_{0,5} - 4_{0,4}$ A &0.0106(0.003) &9.152(0.075) &0.521(0.172) &0.0191 & 0.0054\\
&CH$_3$CHO $5_{0,5} - 4_{0,4}$ E &0.00776(0.003) &9.159(0.056) &0.366(0.253) &0.0199 &0.0054\\
&CH$_2$CHCN $10_{0,10} - 9_{0,9}$ & -- &-- &-- &-- & 0.0069\\
&CH$_2$CHCN $10_{1,9} - 9_{1,8}$& -- &-- &-- &-- & 0.0048 \\

739&CH$_3$OH 2$_{0,2} - 1_{0,1}$ E &-- &-- &-- &--&0.0066\\
& CH$_3$OH 2$_{0,2} - 1_{0,1}$ A &0.25(0.004) &8.443(0.005) &0.621(0.013) &0.379&0.0066\\ 
& CH$_3$OH 2$_{-1,2} - 1_{-1,1}$ E &0.18(0.004) &8.439(0.008) &0.622(0.019) &0.271 &0.0066\\
&CH$_3$CHO $5_{0,5} - 4_{0,4}$ A & -- &-- &-- &-- &0.0058\\
&CH$_3$CHO $5_{0,5} - 4_{0,4}$ E & -- &-- &--  &-- &0.0058\\
&CH$_2$CHCN $10_{0,10} - 9_{0,9}$ & -- &-- &-- &-- & 0.0070\\
&CH$_2$CHCN $10_{1,9} - 9_{1,8}$& -- &-- &--   &-- & 0.0051\\

746 &CH$_3$OH 2$_{0,2} - 1_{0,1}$ E&0.0249(0.003) &9.117(0.035) &0.552(0.092) &0.0424 &0.0058\\
& CH$_3$OH 2$_{0,2} - 1_{0,1}$ A&0.447(0.004) &9.100(0.003) &0.65(0.006) &0.646 &0.0058\\
& CH$_3$OH 2$_{-1,2} - 1_{-1,1}$ E &0.332(0.004) &9.091(0.003) &0.647(0.008) &0.483 &0.0058\\
&CH$_3$CHO $5_{0,5} - 4_{0,4}$ A &0.0252(0.003) &8.812(0.072) &0.966(0.128) &0.0245  & 0.0051\\
&CH$_3$CHO $5_{0,5} - 4_{0,4}$ E &0.0207(0.004) &8.894(0.053) &0.699(0.162) &0.0278 &0.0051\\
&CH$_2$CHCN $10_{0,10} - 9_{0,9}$ & -- &-- &-- &-- & 0.0074\\
&CH$_2$CHCN $10_{1,9} - 9_{1,8}$& -- &-- &-- &-- & 0.0054 \\

747&CH$_3$OH 2$_{0,2} - 1_{0,1}$ E &0.0423(0.004) &8.177(0.032) &0.528(0.078) &0.0752 &0.0083\\
& CH$_3$OH 2$_{0,2} - 1_{0,1}$ A&0.346(0.005) &8.137(0.004) &0.527(0.008) &0.618&0.0083\\
& CH$_3$OH 2$_{-1,2} - 1_{-1,1}$ E  &0.241(0.005) &8.125(0.005) &0.523(0.012) &0.432 &0.0083\\
&CH$_3$CHO $5_{0,5} - 4_{0,4}$ A & -- &-- &-- &-- &0.0074\\
&CH$_3$CHO $5_{0,5} - 4_{0,4}$ E & -- &-- &-- &-- &0.0074\\
&CH$_2$CHCN $10_{0,10} - 9_{0,9}$ & -- &-- &-- &-- & 0.0131\\
&CH$_2$CHCN $10_{1,9} - 9_{1,8}$& -- &-- &-- &-- & 0.0078\\

752 &CH$_3$OH 2$_{0,2} - 1_{0,1}$ E&0.0589(0.003) &7.793(0.016) &0.536(0.036) &0.103&0.0064\\
& CH$_3$OH 2$_{0,2} - 1_{0,1}$ A &0.558(0.004) &7.793(0.002) &0.677(0.006) &0.774&0.0064 \\
& CH$_3$OH 2$_{-1,2} - 1_{-1,1}$ E &0.408(0.004) &7.786(0.003) &0.669(0.008) &0.573 &0.0064\\
&CH$_3$CHO $5_{0,5} - 4_{0,4}$ A &0.0229(0.004) &7.778(0.086) &0.92(0.23) &0.0234 & 0.0054\\
&CH$_3$CHO $5_{0,5} - 4_{0,4}$ E &0.023(0.003) &7.825(0.060) &0.771(0.129) &0.0281 &0.0054\\
&CH$_2$CHCN $10_{0,10} - 9_{0,9}$ & -- &-- &-- &-- & 0.0091\\
&CH$_2$CHCN $10_{1,9} - 9_{1,8}$& -- &-- &-- &-- & 0.0049 \\

768 &CH$_3$OH 2$_{0,2} - 1_{0,1}$ E&0.0179(0.004) &9.327(0.037) &0.424(0.11) &0.0397 &0.0064\\& CH$_3$OH 2$_{0,2} - 1_{0,1}$ A&0.308(0.004) &9.296(0.004) &0.684(0.012) &0.424&0.0064\\
& CH$_3$OH 2$_{-1,2} - 1_{-1,1}$ E &0.226(0.004) &9.279(0.007) &0.706(0.017) &0.3 &0.0064\\
&CH$_3$CHO $5_{0,5} - 4_{0,4}$ A &0.025(0.005) &9.132(0.100) &1(0.277) &0.0235 & 0.0073\\
&CH$_3$CHO $5_{0,5} - 4_{0,4}$ E &0.0117(0.005) &8.947(0.091) &0.559(0.373) &0.0196 &0.0073\\
&CH$_2$CHCN $10_{0,10} - 9_{0,9}$ & -- &-- &-- &-- & 0.0100 \\
&CH$_2$CHCN $10_{1,9} - 9_{1,8}$& -- &-- &-- &-- & 0.0069 \\

780&CH$_3$OH 2$_{0,2} - 1_{0,1}$ E &0.0207(0.003) &10.31(0.042) &0.513(0.098) &0.038&0.0051\\
& CH$_3$OH 2$_{0,2} - 1_{0,1}$ A &0.167(0.003) &10.26(0.003) &0.49(0.009) &0.319&0.0051\\
& CH$_3$OH 2$_{-1,2} - 1_{-1,1}$ E  &0.118(0.003) &10.26(0.005) &0.478(0.013) &0.233 &0.0051\\
&CH$_3$CHO $5_{0,5} - 4_{0,4}$ A & -- &-- &-- &-- &0.0054\\
&CH$_3$CHO $5_{0,5} - 4_{0,4}$ E & -- &-- &-- &-- &0.0054\\
&CH$_2$CHCN $10_{0,10} - 9_{0,9}$ & -- &-- &-- &-- & 0.0076\\
&CH$_2$CHCN $10_{1,9} - 9_{1,8}$& -- &-- &-- &-- & 0.0052 \\

799&CH$_3$OH 2$_{0,2} - 1_{0,1}$ E &0.0236(0.003) &10.38(0.022) &0.394(0.065) &0.0562 &0.0061\\
& CH$_3$OH 2$_{0,2} - 1_{0,1}$ A&0.351(0.003) &10.40(0.002) &0.416(0.005) &0.792&0.0061\\
& CH$_3$OH 2$_{-1,2} - 1_{-1,1}$ E  &0.257(0.003) &10.40(0.002) &0.398(0.005) &0.605 &0.0061\\
&CH$_3$CHO $5_{0,5} - 4_{0,4}$ A &0.0136(0.003) &10.14(0.056) &0.535(0.103) &0.0239 & 0.0053\\
&CH$_3$CHO $5_{0,5} - 4_{0,4}$ E &0.00919(0.002) &10.23(0.055) &0.382(0.112) &0.0226 &0.0053\\
&CH$_2$CHCN $10_{0,10} - 9_{0,9}$ & -- &-- &-- &-- & 0.0077\\
&CH$_2$CHCN $10_{1,9} - 9_{1,8}$& -- &-- &-- &-- & 0.0053\\

800&CH$_3$OH 2$_{0,2} - 1_{0,1}$ E &0.0382(0.003) &10.51(0.026) &0.531(0.052) &0.0676&0.0056\\
& CH$_3$OH 2$_{0,2} - 1_{0,1}$ A &0.493(0.003) &10.45(0.002) &0.612(0.005) &0.758 &0.0056\\
& CH$_3$OH 2$_{-1,2} - 1_{-1,1}$ E &0.37(0.003) &10.44(0.003) &0.62(0.007) &0.56 &0.0056\\
&CH$_3$CHO $5_{0,5} - 4_{0,4}$ A &0.028(0.005) &10.37(0.104) &1.3(0.29) &0.0203 & 0.0050 \\
&CH$_3$CHO $5_{0,5} - 4_{0,4}$ E &0.0164(0.003) &10.14(0.083) &0.7(0.174) &0.022 &0.0050\\
&CH$_2$CHCN $10_{0,10} - 9_{0,9}$ & -- &-- &-- &-- & 0.0083\\
&CH$_2$CHCN $10_{1,9} - 9_{1,8}$& -- &-- &-- &-- & 0.0060\\
\enddata
\tablecomments{Errors reported in parentheses next to the number. 
}.
\end{deluxetable}

\clearpage

\section{Supplemental Material from Yebes 40m Observations} \label{yebesappendix}

As mentioned in section\,\ref{subsec:yebesred}, the Q-band receiver on the Yebes 40m spans 18.5 GHz, from 31.5 -- 50 GHz, and the beam size therefore ranges from 36 -- 56 arcsec. The resolution of 38.0kHz (0.38 km/s -- 0.23 km/s across the band) was sufficient to sample the COM lines that have linewidths $> 0.4$\,km/s. The pointing corrections were obtained using these specific sources: TXCAM; 05:00:50.39, +56:10:52.5, J2000.0, IKTAU; 03:53:28.87, +11:24:21.7, J2000.0, or V11110PH; 18:37:19.26, +10:25:42.2, J2000.0. 

Publicly available Python-based scripts (see\,Data Availability) developed by \cite{2023MNRAS.519.1601M} were used to reduce the data. With this combination of python scripts we loaded in for each source all the data files, searched for lines above a set noise (RMS) level, used those to perform initial baselines, doppler corrected for the line-of-sight velocity across the band, and then combined all the spectral windows for each polarization into single CLASS data file. In total there were 16 spectral windows to combine (8 windows and 2 polarizations) and in some cases multiple data files to combine if a source was observed over multiple days. During this process the main beam efficiency, $\eta_\mathrm{mb}$, measured by Yebes to be  0.66(0.65) at 32.4\,GHz, 0.62(0.62) at 34.5\,GHz, 0.62(0.60) at 36.9\,GHz, 0.59(0.58) at 39.2\,GHz, 0.58(0.56) at 41.4\,GHz, 0.56(0.54) at 43.7\,GHz, 0.54(0.51) at 46.0\,GHz and 0.51(0.49) at 48.4\,GHz for the horizontal(vertical) polarizations, were used to scale our observations to the main beam temperature, $T_\mathrm{mb}$. In this initial reduction we set a high intensity threshold (or RMS cutoff) of 200\,mK because the baseline ripples are high due to the frequency switching technique and if a lower threshold was set, this could lead to `ripple-peaks' mistaken as lines.

Additional CLASS scripts for each source were run to select specific COM transitions of interest by frequency (Table\,\ref{LineList} and Table\,\ref{linelistnondec}), which were then re-baselined and fit with a Gaussian (if a line was detected at $>3 \sigma_{T_\mathrm{mb}}$). A polynomial baseline of degree 10 needed to be used, due to the large baseline ripples produced from the frequency switching technique not completely erased from the initial pipeline procedure and as done in other studies (e.g., \citealt{2023A&A...677L..13A}).

We note that we selected only those transitions that would be `energetically favorable' for cold starless cores, i.e., these transitions have upper energies or $E_{u} <$ 25\,K and Einstein $A_{ul}$ values $> 1.0 \times 10^{-7}$\,s$^{-1}$ (and for CH$_3$CHO, CH$_2$CHCN and HCOOCH$_3$ only $a$-type transitions). For CH$_2$CHCN, HCOOCH$_3$, and CH$_3$OCH$_3$ there are several additional lines within the bandpass that fit this criteria. Yet, these COMs were not detected in any of our sources and therefore we do not include them in Table\,\ref{LineList} for clarity. Instead, in Table\,\ref{linelistnondec} we list additional non-detected transitions of CH$_2$CHCN, HCOOCH$_3$, and CH$_3$OCH$_3$ that we consider energetically favorable, i.e., transitions have upper energies, E$_{u} <$ 25\,K and Einstein $A_{ul}$ values $> 1.0\mathrm{E}-07$\,s$^{-1}$. For CH$_2$CHCN and HCOOCH$_3$ we also only consider $a$-type states for these asymmetric top molecules. 

\begin{figure*}
\includegraphics[width=155mm]{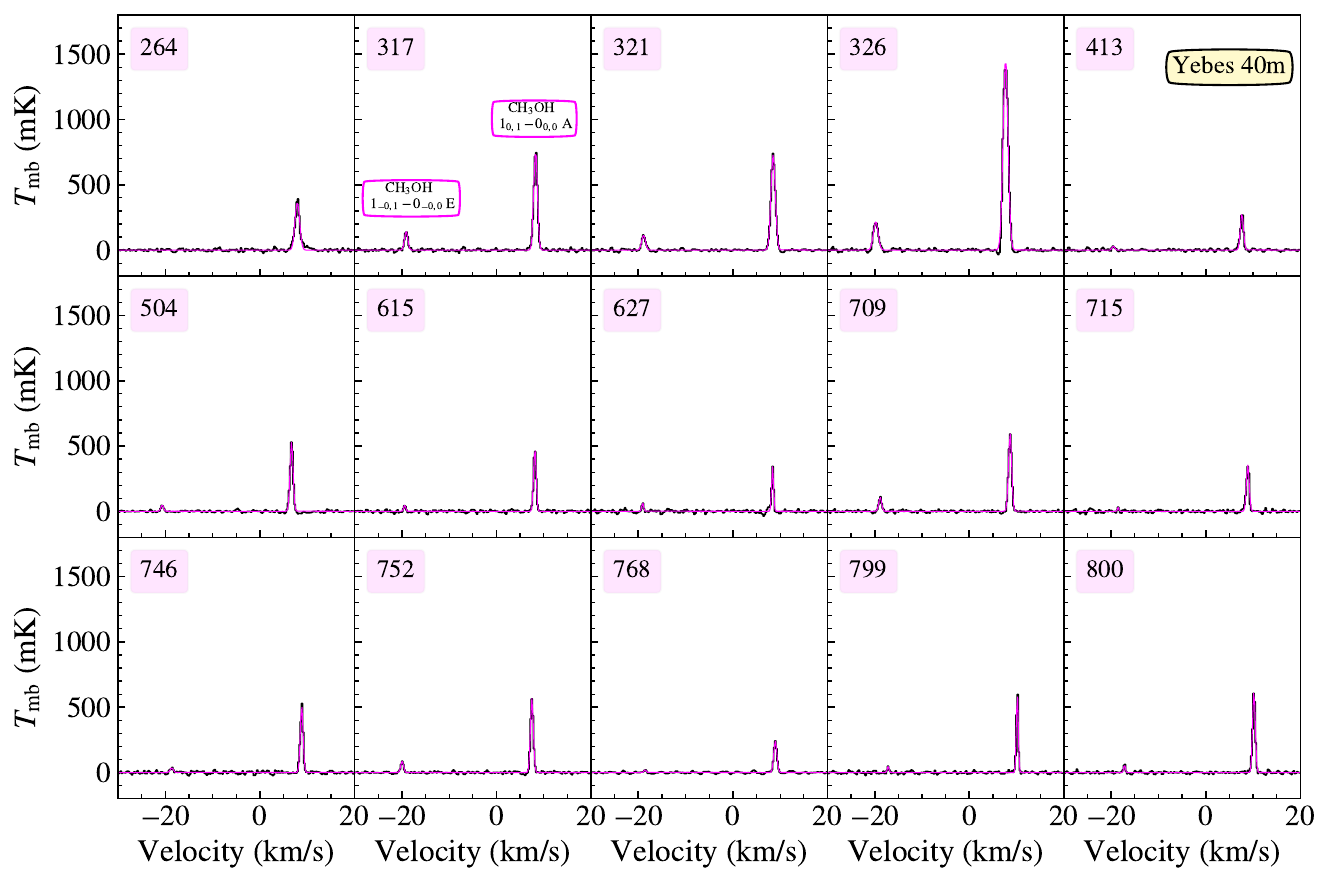} 
\caption{ \label{meth_spec_yebes40m} Methanol, CH$_3$OH, spectrum (in black) in units of $T_\mathrm{mb}$ (K) versus velocity (km/s) from the Yebes 40m for the 15 core sub-sample. Gaussian fits are plotted in magenta. There are two $1-0$ transitions observable, and the A state transitions is centered on the $v_\mathrm{lsr}$ of the core. }
\end{figure*}

\begin{figure*}
\includegraphics[width=155mm]{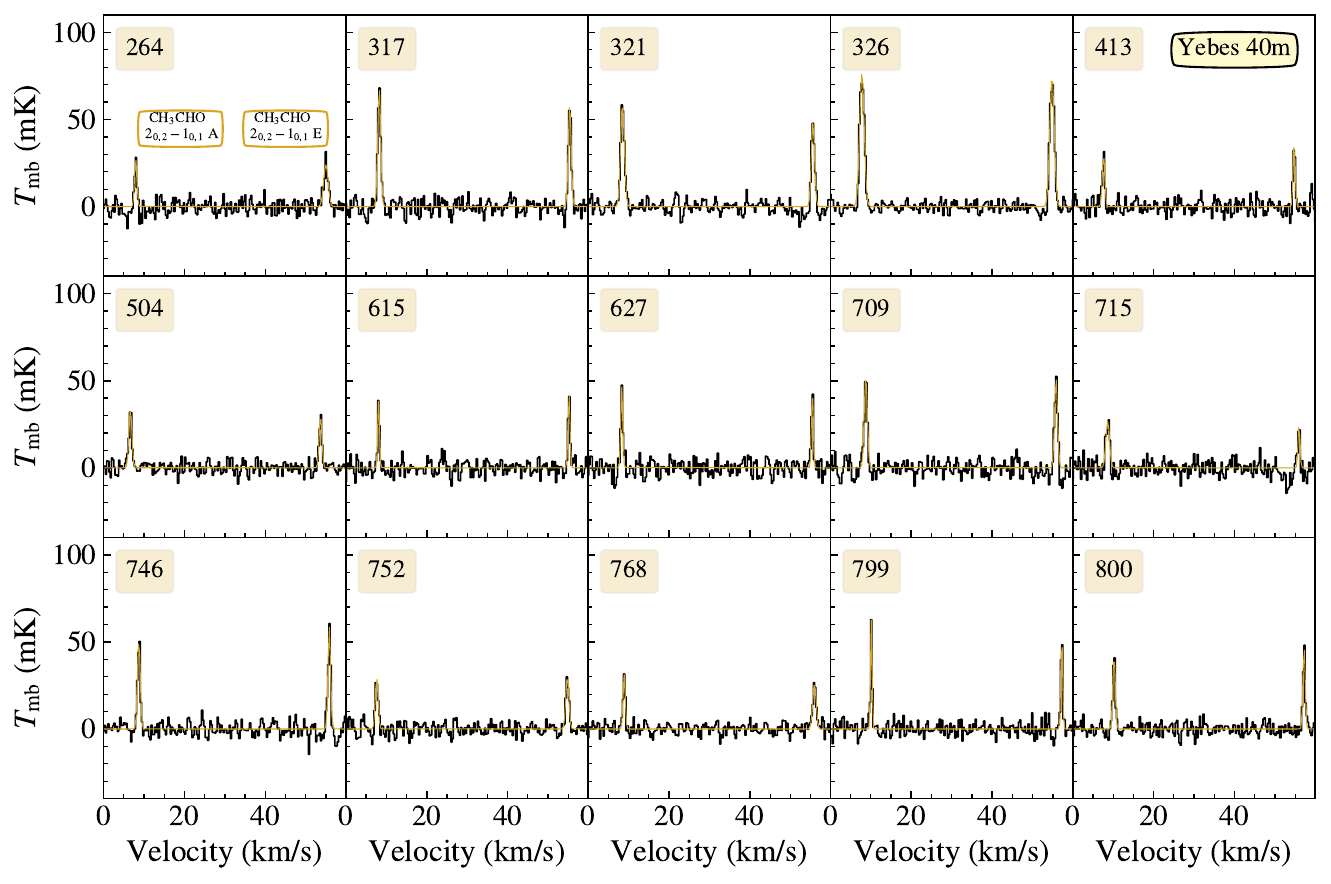} 
\caption{ \label{acet_spec_yebes40m} Acetaldehyde, CH$_3$CHO, spectrum (in black) in units of $T_\mathrm{mb}$ (K) versus velocity (km/s) from the Yebes 40m for the 15 core sub-sample. Gaussian fits are plotted in gold. There are two $2-1$ transitions observable, and the A state transitions is centered on the $v_\mathrm{lsr}$ of the core.  }
\end{figure*}

\begin{figure*}
\includegraphics[width=155mm]{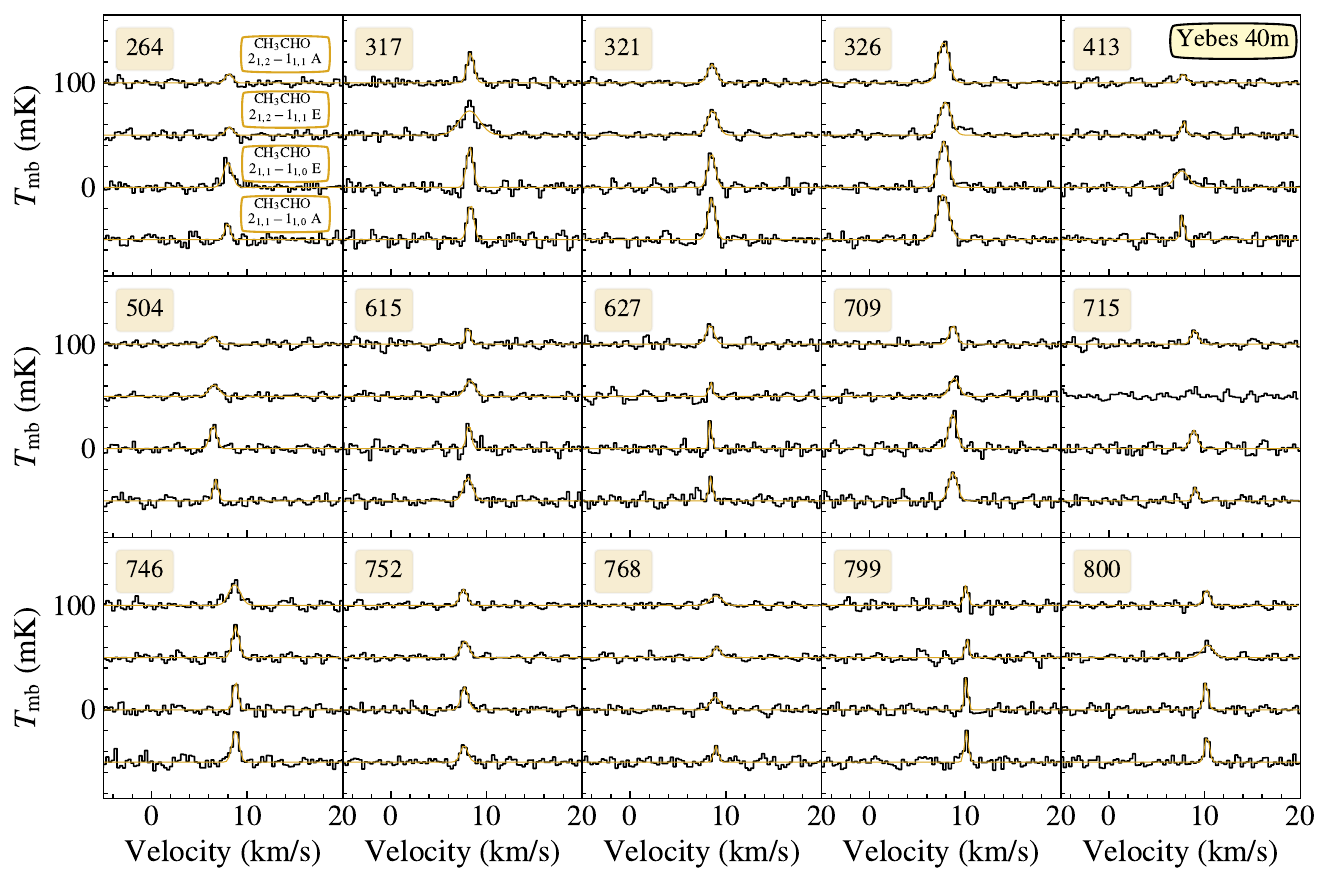} 
\caption{ \label{acetaddl_spec_yebes40m} Additional acetaldehyde, CH$_3$CHO, spectrum (in black) in units of $T_\mathrm{mb}$ (K) versus velocity (km/s) from the Yebes 40m for the 15 core sub-sample. Gaussian fits are plotted in gold. There are four separate $2-1$ transitions observable, and all transitions are centered on the $v_\mathrm{lsr}$ of the core. Spectra are offset by intervals of 50\,mK for easier viewing.  }
\end{figure*}

In Table\,\ref{GaussFits_Yebes} we list the Gaussian fit line parameters, as well as RMS values, calculated using the CLASS software for the Yebes 40m spectral data towards the 15-core sub-sample. For the higher complexity COMs CH$_2$CHCN, HCOOCH$_3$, and CH$_3$OCH$_3$ only the detected transitions are listed and in bold. Line velocities have been re-shifted to the corresponding rest frequencies. The spectra of the additional CH$_3$OH and CH$_3$CHO lines are also presented here in Figures\,\ref{meth_spec_yebes40m},\ref{acet_spec_yebes40m} and \ref{acetaddl_spec_yebes40m}.

\begin{table*}
	\caption{Non-detected Energetically Favorable COM Transitions in Yebes 40m Band}
 \setlength{\tabcolsep}{13pt}
	\label{linelistnondec}
	\begin{tabular}{lllllllc} 
    \tablecolumns{8}
     \tablewidth{0pt}
     \tabcolsep=0.4cm
        Molecule & Transition & Rest Frequency, $\nu$ & E$_u$/k & g$_u$  & $A_{ul}$ & $\theta_{b}^{1}$ & Average $\sigma_{T_\mathrm{mb}}$${^2}$ \\ 
      &   &  (GHz)  & (K)  &  & (s$^{-1}$) & arcsec & (mK) \\
 \hline  
CH$_2$CHCN & $4_{3, 2}- 3_{3, 1}$ &  37.952627	& 24.0 & 27 & 1.8E-06 & 47.7 &  2.52(0.44)\\
           & $4_{3, 1}- 3_{3, 0}$	 & 37.952726 & 	24.0 &	27 & 1.8E-06 & 47.7 &   2.52(0.40)\\
           & $4_{2, 2}- 3_{2, 1}$	 & 37.974365 & 13.2 & 27 & 3.1E-06 & 47.6 & 2.77(0.40)\\
           & $4_{1, 3}- 3_{1, 2}$	 & 38.847735 & 6.8 & 27 & 4.1E-06 & 46.6 & 2.76(0.42)\\
           & $5_{1, 5}- 4_{1, 4}	$ & 46.266933 & 8.8 & 33 & 7.3e-06 & 39.1 & 5.38(1.07) \\
           & $5_{0, 5}- 4_{0, 4}	$ & 47.354648 & 	6.8 & 33 & 8.2E-06 & 38.2 & 6.01(1.04)\\
           & $5_{2, 3}- 4_{2, 2}$	 & 47.489229 & 15.4 & 	33 & 6.9E-06 & 38.1 & 6.07(1.27)\\
           & $5_{1, 4}- 4_{1, 3}$	 & 48.552562 & 	9.1 &	33 & 8.5E-06 & 37.3 & 7.79(2.10) \\
HCOOCH$_3$ & $3_{1, 3}- 2_{1, 2}$ E &34.156884& 3.9& 	14 & 4.6E-07 & 53.0 & 2.57(0.34)\\
           & $3_{1, 3}- 2_{1, 2}$ A &34.158119 & 3.9 &	14 & 4.6E-07& 53.0 & 2.47(0.39) \\ 
           & $3_{2, 2}- 2_{2, 1}$ E & 36.678607& 6.2 & 14 & 3.3E-07 & 49.3 & 2.41(0.35)\\
           & $3_{2, 2}- 2_{2, 1}$ A & 36.657467&	6.1 & 14 & 3.6E-07 & 49.3 & 2.57(0.36)\\
           & $3_{2, 1}- 2_{2, 0}$ E & 37.182123 & 6.2 & 14 & 3.4E-07 & 48.6 & 2.72(0.40)\\
           & $3_{2, 1}- 2_{2, 0}$ A & 37.209617	&6.2 &	14 & 3.7E-07 & 48.6 & 2.69(0.35)\\
           & $3_{1, 2}- 2_{1, 1}$ A& 38.980809 &	4.4 &	14 & 6.9E-07 & 46.4 & 3.23(0.56)\\
           & $4_{2, 3}- 3_{2, 2}$ E & 48.768304 &  8.5 &	18 & 1.2E-06 & 37.1 & 7.57(0.91)\\
           & $4_{2, 3}- 3_{2, 2}$ A & 48.767016 & 8.5 & 	18 & 1.2E-06 & 37.1 & 7.85(1.12)\\
           & $4_{3, 2}- 3_{3, 1}$ E & 49.151617 &	11.9 & 18 & 7.1E-07 & 36.8 & 8.11(1.49)\\
           & $4_{3, 1}- 3_{3, 0}$ E& 49.155295	& 11.9	& 18 & 7.1E-07 & 36.8 & 7.63(0.98)\\
           & $4_{3, 2}- 3_{3, 1}$ A& 49.134631 & 	11.9 & 18 & 7.1E-07 & 36.8 & 8.13(0.93)\\
           & $4_{3, 1}- 3_{3, 0} $A & 49.180102 &  11.9 &	18 & 7.2E-07 & 36.8 & 7.79(0.81)\\
CH$_3$OCH$_3$ & $5_{1, 4}- 5_{ 0, 5}$ EE &  39.047303 & 15.4 	& 176 &	5.0E-07	 & 46.3 & 2.96(0.29)\\
 &	$6_{1, 5}- 6_{0, 6}$ EE & 43.447572 & 21.0 & 208 & 6.5E-07	& 41.6 & 4.15(0.72)\\
  &		$1_{1, 1} - 0_{0, 0}$ EE & 47.674967 &	2.2 &48 & 1.7E-07& 37.9  & 7.16(1.28)\\
  & $1_{1, 0}- 0_{0, 0}$ EE	& 48.844676 & 2.3 &	48 & 1.9E-07 & 37.0 & 7.55(1.08) \\
  & $4_{0, 4}- 3_{1, 3}$ EE& 49.461855 & 9.0	& 144 & 4.5E-07&36.6 & 8.85(2.26) \\
		\hline
    \end{tabular}
      \begin{description} 
     \item  Values for HCOOCH$_3$ from JPL catalog\footnote{\url{https://spec.jpl.nasa.gov/}} (\citealt{1998JQSRT..60..883P}) and for the remaining transitions from CDMS database\footnote{\url{https://cdms.astro.uni-koeln.de}} (\cite{2001A&A...370L..49M}, \cite{2005JMoSt.742..215M}, \cite{2016JMoSp.327...95E}). ${^1}$The beam size corresponding to the selected molecular transition. ${^1}$ The average noise level for the Yebes 40m 15 core sub-sample, where the standard deviation is listed in parentheses. Note that only the EE states for CH$_3$OCH$_3$ are listed. 
      \end{description}
\end{table*}

\clearpage

\begin{deluxetable}{lllllll} 
\tablecaption{Yebes 40m Line Parameters \label{GaussFits_Yebes}}
\tablewidth{-1pt}
\tabcolsep=0.4cm
\tablehead{ 
\colhead{Core} & \colhead{Line} & \colhead{Area} & \colhead{Vel } & \colhead{FWHM} & \colhead{$T_\mathrm{mb}$} & \colhead{RMS}  \\ 
\colhead{} & \colhead{} &\colhead{ (K-km s$^{-1}$)} & \colhead{(km s$^{-1}$)} & \colhead{(km s$^{-1}$)} & \colhead{(K)} & \colhead{(K)} }
\startdata

264 &  t-HCOOH  2$_{1,2} - 1_{1,1}$ & 0.0462(0.005) &8.292(0.212) &3.55(0.423) &0.01223 &0.0038 \\
&  t-HCOOH  2$_{0,2} - 1_{0,1}$ & -- &  --  &  -- &   -- & 0.0047 \\
&  t-HCOOH  2$_{1,1} - 1_{1,0}$ & -- &  --  &  -- &   -- & 0.0067\\

 &H$_2$CCO  2$_{1,2} - 1_{1,1}$ & -- &  --  &  -- &   -- & 0.0031\\
 &H$_2$CCO  2$_{0,2} - 1_{0,1}$ & -- &  --  &  -- &   -- & 0.0033\\
 &H$_2$CCO   2$_{1,1} - 1_{1,0}$ & -- &  --  &  -- &   -- & 0.0040\\
 
 & CH$_3$OH   1$_{0,1} - 0_{0,0}$  A &0.448(0.008) &7.901(0.009) &1.15(0.026) &0.3650 & 0.0082 \\
&   CH$_3$OH  1$_{-0,1} - 0_{-0,0}$  E & -- &  --  &  -- &   -- & 0.0082 \\

&CH$_3$CN $2_1 - 1_1$  & -- &  --  &  -- &   -- &0.0021  \\
& CH$_3$CN $2_0 - 1_0$    & 0.0295(0.003) &7.656(0.085) &1.83(0.234) &0.01513 &0.0021 \\

&  CH$_3$CHO 2$_{1,2} - 1_{1,1}$ A &0.0082(0.002) &8.106(0.104) &0.865(0.2) &0.008904 &0.0025\\
&  CH$_3$CHO 2$_{1,2} - 1_{1,1}$ E  &0.00709(0.002) &8.157(0.12) &0.823(0.241) &0.008089 &0.0027 \\

&  CH$_3$CHO 2$_{0,2} - 1_{0,1}$ E &0.0332(0.004) &7.927(0.065) &1.3(0.166) &0.02398&0.0041\\
&  CH$_3$CHO 2$_{0,2} - 1_{0,1}$ A &0.0198(0.002) &7.952(0.042) &0.686(0.099) &0.02718 &0.0041 \\

&  CH$_3$CHO 2$_{1,1} - 1_{1,0}$ E &0.0271(0.003) &7.979(0.05) &1.06(0.172) &0.02406 &0.0028\\
&  CH$_3$CHO 2$_{1,1} - 1_{1,0}$ A &0.0111(0.002) &7.948(0.061) &0.629(0.145) &0.01654 &0.0038 \\

& \textbf{CH$_2$CHCN}  $4_{1,4} - 3_{1,3}$ & 0.00812(0.002) &7.221(0.091) & 0.846(0.210)  & 0.00902 & 0.0024 \\

& \textbf{HCOOCH$_3$ $4_{0, 4}- 3_{0, 3}$ A} & 0.0192(0.005)    &  7.086(0.122)   &   0.999(0.301) &  0.01810 & 0.0064 \\
& \textbf{HCOOCH$_3$ $4_{1,4} - 3_{1,3}$ E} & 0.02391(0.005)  & 8.099(0.249) &   2.083(0.520) & 0.01787 & 0.0049 \\
& \textbf{HCOOCH$_3$  $4_{1,4} - 3_{1,3}$ A}&  0.01797(0.003)  & 6.868(0.070)   & 0.758(0.147)  &  0.02276 & 0.0049  \\

& \textbf{CH$_3$OCH$_3$  $3_{1, 2} - 3_{0, 3}$ AE+EA} & 0.00760(0.002) &     8.337(0.160) & 1.223(0.250) &  0.00584  & 0.0016 \\ 
& \textbf{CH$_3$OCH$_3$ $3_{1, 2} - 3_{0, 3}$ EE} & 0.01879(0.002)&  7.837(0.135) & 2.149(0.302) & 0.00821 & 0.0016 \\
& \textbf{CH$_3$OCH$_3$ $3_{1, 2} - 3_{0, 3}$ AA} & 0.01340(0.002) & 8.042(0.178) &  2.072(0.427) &   0.00607 & 0.0016 \\

317 &   t-HCOOH  2$_{1,2} - 1_{1,1}$ & -- &  --  &  -- &   -- & 0.0036 \\
& t-HCOOH  2$_{0,2} - 1_{0,1}$ & 0.0103(0.003) &9.399(0.109) &0.687(0.239) &0.01403 &0.0044\\
&  t-HCOOH  2$_{1,1} - 1_{1,0}$ & -- &  --  &  -- &   -- & 0.0062\\

&H$_2$CCO  2$_{1,2} - 1_{1,1}$ &0.0129(0.002) &8.283(0.064) &0.717(0.148) &0.01689 &0.0032 \\
 &H$_2$CCO  2$_{0,2} - 1_{0,1}$ & 0.0139(0.002) &8.513(0.065) &0.765(0.155) &0.01706 &0.0031\\
 &H$_2$CCO   2$_{1,1} - 1_{1,0}$ & 0.0096(0.002) &8.189(0.065) &0.565(0.13) &0.0161 &0.0038\\
 
 & CH$_3$OH   1$_{0,1} - 0_{0,0}$  A & 0.732(0.006) &8.286(0.001) &0.899(0.008) & 0.7645  & 0.0077 \\
&   CH$_3$OH  1$_{-0,1} - 0_{-0,0}$  E &  0.126(0.006) &8.614(0.017) &0.815(0.041) &0.1446 &0.0077\\

&CH$_3$CN $2_1 - 1_1$  & -- &-- &-- &-- &0.0028  \\
& CH$_3$CN $2_0 - 1_0$    &0.00696(0.002) &7.762(0.062) &0.441(0.236) &0.01482 &0.0028\\

&  CH$_3$CHO 2$_{1,2} - 1_{1,1}$ A  &0.0259(0.002) &8.328(0.035) &0.836(0.09) &0.02905 &0.0028 \\
&  CH$_3$CHO 2$_{1,2} - 1_{1,1}$ E  &0.0611(0.004) &8.222(0.067) &2.46(0.22) &0.02332 &0.0025 \\

&  CH$_3$CHO 2$_{0,2} - 1_{0,1}$ E &0.0529(0.003) &8.312(0.022) &0.846(0.048) &0.05873&0.0035 \\
&  CH$_3$CHO 2$_{0,2} - 1_{0,1}$ A &0.065(0.003) &8.267(0.021) &0.895(0.047) &0.06822 &0.0035  \\

&  CH$_3$CHO 2$_{1,1} - 1_{1,0}$ E &0.0333(0.003) &8.251(0.031) &0.798(0.062) &0.03925 &0.0034\\
&  CH$_3$CHO 2$_{1,1} - 1_{1,0}$ A &0.0284(0.003) &8.334(0.041) &0.78(0.08) &0.03427 &0.0042 \\

321 &  t-HCOOH  2$_{1,2} - 1_{1,1}$ & -- &  --  &  -- &   -- & 0.0033 \\
&  t-HCOOH  2$_{0,2} - 1_{0,1}$ & -- &-- &-- &-- & 0.0038 \\
&  t-HCOOH  2$_{1,1} - 1_{1,0}$ & -- &  --  &  -- &   -- & 0.0037\\

 &H$_2$CCO  2$_{1,2} - 1_{1,1}$ &0.0182(0.002) &8.54(0.063) &1.07(0.141) &0.01602 &0.0026 \\
 &H$_2$CCO  2$_{0,2} - 1_{0,1}$ &0.0073(0.002) &8.268(0.068) &0.61(0.148) &0.01138 &0.0026\\
 &H$_2$CCO   2$_{1,1} - 1_{1,0}$ &0.0334(0.002) &8.407(0.048) &1.42(0.102) &0.02215 &0.0024 \\

& CH$_3$OH   1$_{0,1} - 0_{0,0}$  A &0.839(0.005) &8.520(0.003) &1.08(0.007) &0.7299 &0.0066\\
&   CH$_3$OH  1$_{-0,1} - 0_{-0,0}$  E &0.114(0.005) &8.574(0.018) &0.945(0.046) &0.1134 &0.0066  \\

&CH$_3$CN $2_1 - 1_1$  & -- &-- &-- &-- & 0.0026\\
& CH$_3$CN $2_0 - 1_0$    & -- &-- &-- &-- &0.0026  \\

&  CH$_3$CHO 2$_{1,2} - 1_{1,1}$ A&0.0205(0.002) &8.556(0.045) &1.07(0.107) &0.01795 &0.0020\\
&  CH$_3$CHO 2$_{1,2} - 1_{1,1}$ E  &0.0318(0.002) &8.61(0.039) &1.26(0.1) &0.02362 &0.0021 \\

&  CH$_3$CHO 2$_{0,2} - 1_{0,1}$ E &0.0487(0.003) &8.528(0.025) &0.943(0.059) &0.04846 &0.0035  \\
&  CH$_3$CHO 2$_{0,2} - 1_{0,1}$ A &0.0699(0.003) &8.494(0.023) &1.11(0.056) &0.05911 &0.0035 \\

&  CH$_3$CHO 2$_{1,1} - 1_{1,0}$ E &0.0314(0.003) &8.502(0.04) &0.901(0.088) &0.03276 &0.0033\\
&  CH$_3$CHO 2$_{1,1} - 1_{1,0}$ A &0.0385(0.003) &8.507(0.03) &0.909(0.068) &0.0398 &0.0032 \\

& \textbf{HCOOCH$_3$ $4_{1,4} - 3_{1,3}$ E} & 0.0209(0.004)  &   8.457(0.073)  &  0.883(0.175)  & 0.02231 & 0.0049  \\
& \textbf{HCOOCH$_3$  $4_{1,4} - 3_{1,3}$ A}&   0.0485(0.006)   &   8.992(0.154) &   2.528(0.355) &  0.01804 & 0.0049  \\

& \textbf{CH$_3$OCH$_3$  $3_{1, 2} - 3_{0, 3}$ AE+EA} & 0.0049(0.001) &     8.463(0.111) &    0.689(0.213) & 0.00680 & 0.0014  \\ 
& \textbf{CH$_3$OCH$_3$ $3_{1, 2} - 3_{0, 3}$ EE} & 0.0244(0.003) &  8.107(0.155) &    2.530(0.450) & 0.00907 & 0.0014  \\
& \textbf{CH$_3$OCH$_3$ $3_{1, 2} - 3_{0, 3}$ AA} &  0.0073(0.002) & 8.106(0.098)  &   0.814(0.177) &  0.00853 & 0.0014 \\

326 &  t-HCOOH  2$_{1,2} - 1_{1,1}$ & 0.0519(0.005) &8.682(0.188) &3.69(0.379) &0.0132 &0.0035 \\
&  t-HCOOH  2$_{0,2} - 1_{0,1}$ & 0.0319(0.004) &7.925(0.093) &1.59(0.228) &0.0188 &0.0035 \\
&  t-HCOOH  2$_{1,1} - 1_{1,0}$ & -- &  --  &  -- &   -- & 0.0041\\

 &H$_2$CCO  2$_{1,2} - 1_{1,1}$ &0.044(0.002) &7.77(0.035) &1.26(0.071) &0.03268 &0.0025 \\
 &H$_2$CCO  2$_{0,2} - 1_{0,1}$ &0.0311(0.002) &7.657(0.047) &1.2(0.095) &0.02435 &0.0027 \\
 &H$_2$CCO   2$_{1,1} - 1_{1,0}$ &0.0399(0.002) &7.641(0.027) &1.09(0.058) &0.03427 &0.0022 \\

& CH$_3$OH   1$_{0,1} - 0_{0,0}$  A &1.92(0.006) &7.673(0.002) &1.23(0.005) &1.463 &0.0077 \\
&   CH$_3$OH  1$_{-0,1} - 0_{-0,0}$  E & 0.277(0.006) &7.683(0.014) &1.19(0.03) &0.2181 &0.0077 \\
&CH$_3$CN $2_1 - 1_1$  &0.0103(0.002) &5.577(0.101) &0.904(0.177) &0.01067 &0.0032 \\
& CH$_3$CN $2_0 - 1_0$   &0.0487(0.003) &7.136(0.05) &1.65(0.114) &0.02767 &0.0032 \\

&  CH$_3$CHO 2$_{1,2} - 1_{1,1}$ A &0.0547(0.002) &7.745(0.023) &1.31(0.055) &0.03922 &0.0020 \\
&  CH$_3$CHO 2$_{1,2} - 1_{1,1}$ E  &0.0474(0.002) &7.918(0.034) &1.41(0.086) &0.0317 &0.0019 \\

&  CH$_3$CHO 2$_{0,2} - 1_{0,1}$ E &0.104(0.002) &7.744(0.015) &1.31(0.032) &0.07435 &0.0025\\
&  CH$_3$CHO 2$_{0,2} - 1_{0,1}$ A &0.109(0.002) &7.8(0.015) &1.34(0.032) &0.07626 &0.0025  \\

&  CH$_3$CHO 2$_{1,1} - 1_{1,0}$ E &0.057(0.002) &7.756(0.025) &1.21(0.056) &0.04426 &0.0026\\
&  CH$_3$CHO 2$_{1,1} - 1_{1,0}$ A &0.0655(0.003) &7.696(0.028) &1.41(0.062) &0.04376 &0.0028 \\

& \textbf{HCOOCH$_3$ $3_{2, 2}- 2_{2, 1}$ A} & 0.0091(0.003) &    8.181(0.188) &      1.483(0.939) & 0.0057 &  0.0020\\
& \textbf{HCOOCH$_3$ $3_{0, 3}- 2_{0, 2}$ E} & 0.0063(0.001) &    7.749(0.076) &  0.668(0.137) &  0.0088 &  0.0020 \\
& \textbf{HCOOCH$_3$ $3_{0, 3}- 2_{0, 2}$ A }& 0.0120(0.002)  & 7.584(0.119)  &   1.391(0.219) &  0.0081 &  0.0020 \\

& \textbf{HCOOCH$_3$ $4_{0,4} - 3_{0,3}$ A }& 0.0398(0.005) &  7.753(0.078) & 1.241(0.176) & 0.0302 & 0.0061 \\
& \textbf{CH$_3$OCH$_3$  $3_{1, 2} - 3_{0, 3}$ AE+EA} &  0.01189(0.002) &   7.472(0.179) &    1.717(0.348) & 0.00650  & 0.0023\\ 
& \textbf{CH$_3$OCH$_3$ $3_{1, 2} - 3_{0, 3}$ EE} &  0.01785(0.002) &  7.769(0.074) & 1.254(0.195) &   0.01338 & 0.0023  \\
& \textbf{CH$_3$OCH$_3$ $3_{1, 2} - 3_{0, 3}$ AA} & 0.01878(0.002)  &7.338(0.110) &1.631(0.210) &  0.01081 & 0.0023 \\
& \textbf{CH$_3$OCH$_3$  $4_{1, 3} - 4_{0, 4}$ AE+EA} & 0.00995(0.002) &   7.321(0.113)    &  1.323(0.201) &  0.00706 & 0.0016\\
& \textbf{CH$_3$OCH$_3$  $4_{1, 3} - 4_{0, 4}$ EE} & 0.01909(0.003)  & 7.973(0.132) &      2.107(0.426) &  0.00851 & 0.0016 \\
& \textbf{CH$_3$OCH$_3$  $4_{1, 3} - 4_{0, 4}$ AA} & 0.01617(0.002)   & 8.357(0.085)  &    1.431(0.217) &  0.01061 & 0.0016 \\

413 &  t-HCOOH  2$_{1,2} - 1_{1,1}$ & -- &  --  &  -- &   -- & 0.0035 \\
&  t-HCOOH  2$_{0,2} - 1_{0,1}$ & -- &  --  &  -- &   -- & 0.0037 \\
&  t-HCOOH  2$_{1,1} - 1_{1,0}$ & -- &  --  &  -- &   -- & 0.0062\\

 &H$_2$CCO  2$_{1,2} - 1_{1,1}$ &0.0124(0.002) &7.711(0.04) &0.497(0.107) &0.02341 &0.0028 \\
 &H$_2$CCO  2$_{0,2} - 1_{0,1}$ &0.0086(0.002) &7.767(0.054) &0.489(0.086) &0.01671 &0.0031 \\
 &H$_2$CCO   2$_{1,1} - 1_{1,0}$ &0.0188(0.002) &7.701(0.036) &0.764(0.087) &0.02314 &0.0025 \\
 
 & CH$_3$OH   1$_{0,1} - 0_{0,0}$  A  &0.233(0.005) &7.639(0.007) &0.784(0.02) &0.2793 &0.0072\\
&   CH$_3$OH  1$_{-0,1} - 0_{-0,0}$  E & 0.0204(0.004) &7.917(0.077) &0.646(0.16) &0.02974 &0.0072 \\

&CH$_3$CN $2_1 - 1_1$  &0.0165(0.002) &5.679(0.059) &0.983(0.159) &0.01582 &0.0025 \\
& CH$_3$CN $2_0 - 1_0$  &0.0349(0.002) &7.212(0.034) &1.11(0.085) &0.02961 &0.0025 \\

&  CH$_3$CHO 2$_{1,2} - 1_{1,1}$ A&0.00714(0.002) &7.76(0.101) &0.793(0.257) &0.00846 &0.0025\\
&  CH$_3$CHO 2$_{1,2} - 1_{1,1}$ E  &0.0078(0.002) &7.812(0.055) &0.535(0.15) &0.01371 &0.0026 \\

&  CH$_3$CHO 2$_{0,2} - 1_{0,1}$ E &0.0238(0.002) &7.703(0.03) &0.617(0.061) &0.03631  &0.0035 \\
&  CH$_3$CHO 2$_{0,2} - 1_{0,1}$ A &0.023(0.003) &7.621(0.044) &0.73(0.108) &0.02965 &0.0035 \\

&  CH$_3$CHO 2$_{1,1} - 1_{1,0}$ E &0.0284(0.004) &7.605(0.11) &1.67(0.331) &0.01593 &0.0031\\
&  CH$_3$CHO 2$_{1,1} - 1_{1,0}$ A &0.0115(0.002) &7.671(0.042) &0.429(0.107) &0.02519 &0.0035 \\

& \textbf{CH$_2$CHCN  $4_{1,4} - 3_{1,3}$} & 0.0103(0.002)   &  7.486(0.083) & 0.882 (0.180) &  0.01106 & 0.0025 \\
&  \textbf{CH$_2$CHCN  $4_{0,4} - 3_{0,3}$}& 0.0041(0.001)   & 7.713 (0.041)  & 0.302 (0.556)  & 0.01304 & 0.0025 \\

504 &  t-HCOOH  2$_{1,2} - 1_{1,1}$ & -- &  --  &  -- &   -- & 0.0033 \\
&  t-HCOOH  2$_{0,2} - 1_{0,1}$ &0.0097(0.002) &6.81(0.054) &0.645(0.108) &0.01424 &0.0027 \\
&  t-HCOOH  2$_{1,1} - 1_{1,0}$ & -- &  --  &  -- &   -- & 0.0042\\

 &H$_2$CCO  2$_{1,2} - 1_{1,1}$ &0.00855(0.001) &6.731(0.04) &0.449(0.068) &0.0179 &0.0025 \\
 &H$_2$CCO  2$_{0,2} - 1_{0,1}$ &0.0144(0.002) &6.441(0.052) &0.818(0.103) &0.01651 &0.0024 \\
 &H$_2$CCO   2$_{1,1} - 1_{1,0}$ &0.0189(0.003) &6.471(0.111) &1.22(0.206) &0.01456 &0.0029 \\

& CH$_3$OH   1$_{0,1} - 0_{0,0}$  A &0.43(0.004) &6.678(0.004) &0.759(0.009) &0.5324 &0.0077 \\
&   CH$_3$OH  1$_{-0,1} - 0_{-0,0}$  E & 0.0332(0.004) &6.783(0.038) &0.652(0.087) &0.04791 &0.0077 \\

&CH$_3$CN $2_1 - 1_1$  & -- &  --  &  -- &   -- & 0.0030 \\
& CH$_3$CN $2_0 - 1_0$    &0.0369(0.003) &6.078(0.06) &1.45(0.152) &0.02392 &0.0030\\

&  CH$_3$CHO 2$_{1,2} - 1_{1,1}$ A&0.00746(0.002) &6.407(0.115) &0.925(0.189) &0.00757 &0.0022\\
&  CH$_3$CHO 2$_{1,2} - 1_{1,1}$ E  &0.0144(0.002) &6.552(0.081) &1.22(0.163) &0.01106 &0.0020 \\

&  CH$_3$CHO 2$_{0,2} - 1_{0,1}$ E &0.0251(0.002) &6.603(0.033) &0.799(0.081) &0.02947 &0.0029  \\
&  CH$_3$CHO 2$_{0,2} - 1_{0,1}$ A &0.0312(0.002) &6.643(0.032) &0.873(0.085) &0.03356 &0.0029 \\

&  CH$_3$CHO 2$_{1,1} - 1_{1,0}$ E &0.0196(0.002) &6.465(0.044) &0.825(0.126) &0.02231 &0.0028\\
&  CH$_3$CHO 2$_{1,1} - 1_{1,0}$ A &0.013(0.002) &6.709(0.045) &0.591(0.092) &0.0207 &0.0031 \\

615 &  t-HCOOH  2$_{1,2} - 1_{1,1}$ & -- &  --  &  -- &   -- &0.0054 \\
&  t-HCOOH  2$_{0,2} - 1_{0,1}$ & -- &  --  &  -- &   -- & 0.0058 \\
&  t-HCOOH  2$_{1,1} - 1_{1,0}$ & -- &  --  &  -- &   -- & 0.0054\\

 &H$_2$CCO  2$_{1,2} - 1_{1,1}$ &0.0112(0.002) &8.232(0.07) &0.552(0.121) &0.01902 &0.0037 \\
 &H$_2$CCO  2$_{0,2} - 1_{0,1}$ &0.0406(0.005) &8.455(0.17) &2.96(0.431) &0.01287 &0.0035 \\
 &H$_2$CCO   2$_{1,1} - 1_{1,0}$ &0.00953(0.002) &8.041(0.05) &0.447(0.101) &0.02005 &0.0038 \\
 
 & CH$_3$OH   1$_{0,1} - 0_{0,0}$  A & 0.325(0.005) &8.115(0.004) &0.634(0.011) &0.4824  &0.0079 \\
&   CH$_3$OH  1$_{-0,1} - 0_{-0,0}$  E & 0.0258(0.004) &8.060(0.041) &0.506(0.079) &0.04794 &0.0079 \\

&CH$_3$CN $2_1 - 1_1$    & -- &  --  &  -- &   -- & 0.0027 \\
& CH$_3$CN $2_0 - 1_0$   & -- &  --  &  -- &   -- & 0.0027\\

&  CH$_3$CHO 2$_{1,2} - 1_{1,1}$ A &0.00983(0.002) &8.067(0.044) &0.525(0.137) &0.0176 &0.0029\\
&  CH$_3$CHO 2$_{1,2} - 1_{1,1}$ E  &0.0199(0.002) &8.289(0.075) &1.21(0.166) &0.01548 &0.0027\\

&  CH$_3$CHO 2$_{0,2} - 1_{0,1}$ E &0.0264(0.002) &8.13(0.025) &0.579(0.057) &0.04284  &0.0036   \\
&  CH$_3$CHO 2$_{0,2} - 1_{0,1}$ A  &0.0222(0.002) &8.107(0.024) &0.512(0.065) &0.04076 &0.0036  \\

&  CH$_3$CHO 2$_{1,1} - 1_{1,0}$ E &0.015(0.003) &8.153(0.066) &0.585(0.161) &0.02416 &0.0039\\
&  CH$_3$CHO 2$_{1,1} - 1_{1,0}$ A &0.0268(0.003) &8.079(0.064) &1.12(0.169) &0.0224 &0.0035 \\

627 &  t-HCOOH  2$_{1,2} - 1_{1,1}$ & -- &  --  &  -- &   -- & 0.0050 \\
&  t-HCOOH  2$_{0,2} - 1_{0,1}$  & -- &  --  &  -- &   -- & 0.0053 \\
&  t-HCOOH  2$_{1,1} - 1_{1,0}$ & -- &  --  &  -- &   -- & 0.0067\\

 &H$_2$CCO  2$_{1,2} - 1_{1,1}$ &0.0174(0.003) &8.48(0.065) &0.772(0.119) &0.02117 &0.0041 \\
 &H$_2$CCO  2$_{0,2} - 1_{0,1}$  &0.00948(0.003) &8.31(0.119) &0.766(0.213) &0.01163 &0.0040 \\
 &H$_2$CCO   2$_{1,1} - 1_{1,0}$ &0.0138(0.002) &8.298(0.05) &0.569(0.101) &0.02275 &0.0041 \\
 
 & CH$_3$OH   1$_{0,1} - 0_{0,0}$  A &0.176(0.004) &8.402(0.007) &0.482(0.013) &0.3425 &0.0106 \\
&   CH$_3$OH  1$_{-0,1} - 0_{-0,0}$  E &0.0307(0.005) &8.378(0.035) &0.44(0.076) &0.06566 &0.0106  \\

&CH$_3$CN $2_1 - 1_1$  & -- &  --  &  -- &   -- &  0.0030 \\
& CH$_3$CN $2_0 - 1_0$  &  0.0279(0.003) &7.796(0.056) &1.16(0.119) &0.02259 &0.0030 \\

&  CH$_3$CHO 2$_{1,2} - 1_{1,1}$ A &0.0186(0.003) &8.357(0.068) &0.957(0.237) &0.01829 &0.0032 \\
&  CH$_3$CHO 2$_{1,2} - 1_{1,1}$ E  &0.00691(0.002) &8.415(0.074) &0.457(0.132) &0.01421 &0.0031 \\

&  CH$_3$CHO 2$_{0,2} - 1_{0,1}$ E &0.0255(0.003) &8.389(0.025) &0.53(0.084) &0.04528  & 0.0039 \\
&  CH$_3$CHO 2$_{0,2} - 1_{0,1}$ A &.0288(0.003) &8.323(0.026) &0.581(0.064) &0.04654 &0.0039 \\

&  CH$_3$CHO 2$_{1,1} - 1_{1,0}$ E &0.0109(0.001) &8.398(0.019) &0.291(0.296) &0.03522 &0.0032\\
&  CH$_3$CHO 2$_{1,1} - 1_{1,0}$ A &0.0115(0.002) &8.416(0.046) &0.463(0.081) &0.02337 &0.0037 \\

& \textbf{CH$_3$OCH$_3$ $3_{1, 2} - 3_{0, 3}$ EE} & 0.0308(0.003) &   7.645(0.158) &      3.024(0.302) & 0.00959 & 0.0022  \\

709 &  t-HCOOH  2$_{1,2} - 1_{1,1}$ & -- &  --  &  -- &   -- & 0.0043 \\
&  t-HCOOH  2$_{0,2} - 1_{0,1}$ &0.0177(0.004) &9.023(0.085) &0.885(0.218) &0.01874 &0.0047 \\
&  t-HCOOH  2$_{1,1} - 1_{1,0}$ & 0.0192(0.004) &8.98(0.076) &0.688(0.185) &0.02628 &0.0063\\

 &H$_2$CCO  2$_{1,2} - 1_{1,1}$ &0.0379(0.003) &8.916(0.044) &1.18(0.088) &0.0301 &0.0032 \\
 &H$_2$CCO  2$_{0,2} - 1_{0,1}$ &0.0287(0.003) &8.873(0.077) &1.29(0.163) &0.02084 &0.0038 \\
 &H$_2$CCO   2$_{1,1} - 1_{1,0}$ &0.0429(0.004) &8.777(0.044) &1.08(0.11) &0.03722 &0.0040 \\
 
 & CH$_3$OH   1$_{0,1} - 0_{0,0}$  A &0.499(0.004) &8.619(0.003) &0.788(0.008) &0.5949 &0.0086 \\
&   CH$_3$OH  1$_{-0,1} - 0_{-0,0}$  E &0.0832(0.004) &8.605(0.018) &0.727(0.045) &0.1075 &0.0086  \\

&CH$_3$CN $2_1 - 1_1$  &0.0421(0.003) &6.88(0.054) &1.44(0.13) &0.02757 &0.0035  \\
& CH$_3$CN $2_0 - 1_0$  &0.0707(0.003) &8.337(0.028) &1.33(0.059) &0.0501 &0.0035  \\

&  CH$_3$CHO 2$_{1,2} - 1_{1,1}$ A  &0.0168(0.002) &8.722(0.049) &0.882(0.117) &0.01795 &0.0022\\
&  CH$_3$CHO 2$_{1,2} - 1_{1,1}$ E &0.021(0.003) &8.855(0.064) &1.12(0.167) &0.01762 &0.0027\\

&  CH$_3$CHO 2$_{0,2} - 1_{0,1}$ E &0.0477(0.003) &8.739(0.034) &0.877(0.065) &0.05106 &0.0050 \\
&  CH$_3$CHO 2$_{0,2} - 1_{0,1}$ A & 0.0506(0.004) &8.75(0.031) &0.903(0.075) &0.05264 &0.0050 \\

&  CH$_3$CHO 2$_{1,1} - 1_{1,0}$ E &0.0343(0.003) &8.702(0.04) &0.958(0.108) &0.03363 &0.0035\\
&  CH$_3$CHO 2$_{1,1} - 1_{1,0}$ A &0.0307(0.003) &8.717(0.055) &1.05(0.116) &0.02738 &0.0039 \\

&  \textbf{CH$_2$CHCN  $4_{0,4} - 3_{0,3}$}&   0.0133(0.002)  & 8.989(0.090)   &  1.008(0.221) & 0.0124 & 0.0027 \\

715 &  t-HCOOH  2$_{1,2} - 1_{1,1}$ & -- &  --  &  -- &   -- & 0.0042 \\
&  t-HCOOH  2$_{0,2} - 1_{0,1}$ &0.0232(0.004) &9.102(0.073) &0.942(0.161) &0.02316 &0.0048 \\
&  t-HCOOH  2$_{1,1} - 1_{1,0}$ & -- &  --  &  -- &   -- & 0.0058\\

 &H$_2$CCO  2$_{1,2} - 1_{1,1}$ &0.0255(0.003) &8.938(0.064) &1.15(0.131) &0.02076 &0.0034 \\
 &H$_2$CCO  2$_{0,2} - 1_{0,1}$ &0.0228(0.002) &8.73(0.042) &0.87(0.089) &0.0246 &0.0029 \\
 &H$_2$CCO   2$_{1,1} - 1_{1,0}$ &0.0173(0.002) &8.901(0.049) &0.685(0.113) &0.02371 &0.0036 \\
 
 & CH$_3$OH   1$_{0,1} - 0_{0,0}$  A &0.265(0.005) &8.89(0.001) &0.704(0.016) &0.3539 &0.0077 \\
&   CH$_3$OH  1$_{-0,1} - 0_{-0,0}$  E &0.011(0.004) & 8.926(0.066) &0.328(0.089) &0.03161 &0.0077  \\

&CH$_3$CN $2_1 - 1_1$  &0.019(0.003) &6.691(0.095) &1.22(0.229) &0.01469 &0.0033 \\
& CH$_3$CN $2_0 - 1_0$&0.0442(0.003) &8.241(0.043) &1.29(0.104) &0.0323 &0.0033 \\

&  CH$_3$CHO 2$_{1,2} - 1_{1,1}$ A  &0.0118(0.002) &8.95(0.071) &0.818(0.173) &0.01351 &0.0028 \\
&  CH$_3$CHO 2$_{1,2} - 1_{1,1}$ E  & -- & -- &-- &--&0.0026 \\

&  CH$_3$CHO 2$_{0,2} - 1_{0,1}$ E &0.0156(0.003) &8.899(0.058) &0.58(0.146) &0.02521 &0.0040\\
&  CH$_3$CHO 2$_{0,2} - 1_{0,1}$ A   &0.0304(0.003) &8.648(0.062) &1.06(0.113) &0.02693 &0.0040 \\

&  CH$_3$CHO 2$_{1,1} - 1_{1,0}$ E &0.0163(0.003) &8.877(0.07) &0.859(0.141) &0.01787 &0.0036\\
&  CH$_3$CHO 2$_{1,1} - 1_{1,0}$ A &0.00805(0.002) &9.029(0.066) &0.573(0.134) &0.0132 &0.0030\\

&  \textbf{CH$_2$CHCN  $4_{2,3} - 3_{2,2}$}& 0.0101(0.002)  &   8.233(0.064)   &  0.761(0.165) & 0.01256 & 0.0024 \\
& \textbf{HCOOCH$_3$  $3_{2, 2}- 2_{2, 1}$ E} & 0.0243(0.004) & 8.363(0.162) &  2.367(0.451) & 0.009643 & 0.0025 \\
& \textbf{HCOOCH$_3$ $3_{0, 3}- 2_{2, 1}$ A} & 0.0068(0.001) & 8.100(0.317) &   0.861(0.317) & 0.007505 &   0.0018\\

746 &  t-HCOOH  2$_{1,2} - 1_{1,1}$ & 0.0644(0.005) &8.839(0.139) &3.3(0.306) &0.01831 &0.0038 \\
&  t-HCOOH  2$_{0,2} - 1_{0,1}$ &0.0199(0.004) &9.08(0.101) &0.956(0.213) &0.01951 &0.0054 \\
&  t-HCOOH  2$_{1,1} - 1_{1,0}$ & 0.0354(0.007) &9.604(0.291) &2.67(0.471) &0.01244 &0.0062 \\

 &H$_2$CCO  2$_{1,2} - 1_{1,1}$ &0.033(0.003) &8.682(0.045) &0.974(0.101) &0.03184 &0.0037 \\
 &H$_2$CCO  2$_{0,2} - 1_{0,1}$ &0.0208(0.003) &8.684(0.042) &0.713(0.101) &0.02742 &0.0037 \\
 &H$_2$CCO   2$_{1,1} - 1_{1,0}$ &0.0336(0.003) &8.617(0.041) &0.939(0.078) &0.03359 &0.0037 \\

& CH$_3$OH   1$_{0,1} - 0_{0,0}$  A &0.3800(0.006) &8.819(0.005) &0.7(0.013) &0.5101 &0.0091 \\
&   CH$_3$OH  1$_{-0,1} - 0_{-0,0}$  E &0.0298(0.006) &8.787(0.084) &0.799(0.166) &0.03504 &0.0091  \\

&CH$_3$CN $2_1 - 1_1$  &0.00669(0.002) &6.633(0.102) &0.692(0.156) &0.009092 &0.0027 \\
& CH$_3$CN $2_0 - 1_0$    &0.0403(0.003) &8.06(0.055) &1.53(0.125) &0.02478 &0.0027 \\

&  CH$_3$CHO 2$_{1,2} - 1_{1,1}$ A &0.0302(0.005) &8.691(0.08) &1.42(0.408) &0.01999 &0.0033 \\
&  CH$_3$CHO 2$_{1,2} - 1_{1,1}$ E  &0.0259(0.002) &8.799(0.035) &0.802(0.092) &0.0303 &0.0031\\

&  CH$_3$CHO 2$_{0,2} - 1_{0,1}$ E &0.0474(0.003) &8.764(0.022) &0.758(0.055) &0.05875  &0.0046 \\
&  CH$_3$CHO 2$_{0,2} - 1_{0,1}$ A &0.0446(0.003) &8.738(0.025) &0.815(0.057) &0.05145 &0.0046  \\

&  CH$_3$CHO 2$_{1,1} - 1_{1,0}$ E &0.018(0.002) &8.784(0.034) &0.606(0.071) &0.0279 &0.0032\\
&  CH$_3$CHO 2$_{1,1} - 1_{1,0}$ A &0.0269(0.003) &8.786(0.046) &0.824(0.12) &0.03069 &0.0040\\

&  \textbf{CH$_2$CHCN  $4_{0,4} - 3_{0,3}$}& 0.0098(0.002) & 9.147(0.117) & 1.078(0.245) &  0.00859 &  0.0024 \\

752 &  t-HCOOH  2$_{1,2} - 1_{1,1}$ & 0.0336(0.005) &7.81(0.213) &2.55(0.402) &0.01237 &0.0042\\
&  t-HCOOH  2$_{0,2} - 1_{0,1}$ &0.00839(0.002) &7.64(0.062) &0.43(0.119) &0.01834 &0.0041 \\
&  t-HCOOH  2$_{1,1} - 1_{1,0}$ & -- &  --  &  -- &   -- & 0.0049\\

 &H$_2$CCO  2$_{1,2} - 1_{1,1}$ &0.0122(0.002) &7.856(0.062) &0.801(0.146) &0.0143 &0.0023 \\
 &H$_2$CCO  2$_{0,2} - 1_{0,1}$ &0.0241(0.004) &7.615(0.144) &1.89(0.357) &0.01197 &0.0031 \\
 &H$_2$CCO   2$_{1,1} - 1_{1,0}$ &0.0131(0.003) &7.429(0.102) &0.951(0.233) &0.01295 &0.0032 \\
 
 & CH$_3$OH   1$_{0,1} - 0_{0,0}$  A &0.434(0.004) &7.503(0.003) &0.719(0.008) &0.5673 &0.0066\\
&   CH$_3$OH  1$_{-0,1} - 0_{-0,0}$  E &0.0631(0.004) & 7.512(0.021) &0.684(0.052) &0.0867 &0.0066  \\

&CH$_3$CN $2_1 - 1_1$ &0.0212(0.002) &5.683(0.058) &1.29(0.108) &0.01543  &0.0023 \\
& CH$_3$CN $2_0 - 1_0$    & 0.0219(0.002) &7.232(0.061) &1.36(0.133) &0.01505 &0.0023 \\

&  CH$_3$CHO 2$_{1,2} - 1_{1,1}$ A &0.0147(0.002) &7.596(0.055) &0.887(0.122) &0.01559 &0.0023 \\
&  CH$_3$CHO 2$_{1,2} - 1_{1,1}$ E  &0.0175(0.002) &7.73(0.052) &1.02(0.106) &0.01606 &0.0021 \\

&  CH$_3$CHO 2$_{0,2} - 1_{0,1}$ E &0.0289(0.002) &7.705(0.038) &0.913(0.086) &0.02969  &0.0030 \\
&  CH$_3$CHO 2$_{0,2} - 1_{0,1}$ A  &0.0255(0.002) &7.63(0.038) &0.819(0.075) &0.02927 &0.0030  \\

&  CH$_3$CHO 2$_{1,1} - 1_{1,0}$ E &0.0204(0.002) &7.713(0.049) &0.866(0.119) &0.02207 &0.0028\\
&  CH$_3$CHO 2$_{1,1} - 1_{1,0}$ A &0.0152(0.002) &7.664(0.07) &0.897(0.17) &0.01595 &0.0027 \\

768 &  t-HCOOH  2$_{1,2} - 1_{1,1}$ & -- &  --  &  -- &   -- & 0.0028 \\
&  t-HCOOH  2$_{0,2} - 1_{0,1}$ & -- &  --  &  -- &   -- &0.0029 \\
&  t-HCOOH  2$_{1,1} - 1_{1,0}$ & -- &  --  &  -- &   -- & 0.0040\\

 &H$_2$CCO  2$_{1,2} - 1_{1,1}$ &0.0129(0.002) &9.106(0.063) &0.787(0.125) &0.01535 &0.0029 \\
 &H$_2$CCO  2$_{0,2} - 1_{0,1}$  &0.0113(0.002) &8.779(0.09) &0.958(0.184) &0.01103 &0.0027 \\
 &H$_2$CCO   2$_{1,1} - 1_{1,0}$ &0.0278(0.003) &8.74(0.079) &1.42(0.248) &0.01831 &0.0030 \\
 
 & CH$_3$OH   1$_{0,1} - 0_{0,0}$  A &0.199(0.004) &8.975(0.008) &0.774(0.019) &0.2418 &0.0063 \\
&   CH$_3$OH  1$_{-0,1} - 0_{-0,0}$  E &0.0125(0.003) & 8.901(0.08) &0.556(0.166) &0.02104 &0.0063  \\

&CH$_3$CN $2_1 - 1_1$ &0.00871(0.002) &7.181(0.156) &1.4(0.28) &0.005846 &0.0019 \\
& CH$_3$CN $2_0 - 1_0$    & 0.0145(0.002) &8.276(0.099) &1.52(0.207) &0.008931 &0.0019 \\

&  CH$_3$CHO 2$_{1,2} - 1_{1,1}$ A &0.0138(0.003) &8.982(0.122) &1.41(0.446) &0.009204 &0.0019\\
&  CH$_3$CHO 2$_{1,2} - 1_{1,1}$ E  &0.0096(0.002) &9.066(0.076) &0.86(0.206) &0.01048 &0.0022 \\

&  CH$_3$CHO 2$_{0,2} - 1_{0,1}$ E &0.0278(0.002) &8.894(0.041) &1(0.11) &0.02602  &0.0027 \\
&  CH$_3$CHO 2$_{0,2} - 1_{0,1}$ A  &0.0219(0.002) &8.901(0.028) &0.653(0.065) &0.03156 &0.0027 \\

&  CH$_3$CHO 2$_{1,1} - 1_{1,0}$ E &0.0153(0.003) &8.88(0.091) &1.14(0.314) &0.01262 &0.0027\\
&  CH$_3$CHO 2$_{1,1} - 1_{1,0}$ A &0.00919(0.002) &8.97(0.062) &0.556(0.144) &0.01552 &0.0034 \\

& \textbf{HCOOCH$_3$ $4_{1,4} - 3_{1,3}$ E} &  0.0101(0.002) &   9.080(0.101)    &  0.783(0.206)  & 0.0121  &  0.0032   \\

799&  t-HCOOH  2$_{1,2} - 1_{1,1}$ & -- &  --  &  -- &   -- & 0.0039 \\
&  t-HCOOH  2$_{0,2} - 1_{0,1}$ & -- &  --  &  -- &   -- & 0.0044\\
&  t-HCOOH  2$_{1,1} - 1_{1,0}$ & -- &  --  &  -- &   -- & 0.0062\\

 &H$_2$CCO  2$_{1,2} - 1_{1,1}$  &0.0206(0.003) &10.19(0.056) &0.799(0.168) &0.02428 &0.0032 \\
 &H$_2$CCO  2$_{0,2} - 1_{0,1}$ &0.0154(0.002) &10.13(0.032) &0.491(0.118) &0.02948 &0.0034 \\
 &H$_2$CCO   2$_{1,1} - 1_{1,0}$ &0.0154(0.002) &10.13(0.032) &0.491(0.118) &0.02948 &0.0034 \\
 
 & CH$_3$OH   1$_{0,1} - 0_{0,0}$  A &0.314(0.004) &10.150(0) &0.47(0.007) &0.6288  &0.0079  \\
&   CH$_3$OH  1$_{-0,1} - 0_{-0,0}$  E &0.0212(0.004) &10.289(0.037) &0.416(0.095) &0.04803 &0.0079  \\

&CH$_3$CN $2_1 - 1_1$  & -- &  --  &  -- &   -- & 0.0034  \\
& CH$_3$CN $2_0 - 1_0$  & 0.0281(0.004) &9.466(0.08) &1.38(0.255) &0.01907 &0.0034 \\

&  CH$_3$CHO 2$_{1,2} - 1_{1,1}$ A &0.0108(0.002) &10.06(0.052) &0.517(0.078) &0.01961 &0.0031 \\
&  CH$_3$CHO 2$_{1,2} - 1_{1,1}$ E  &0.0089(0.002) &10.18(0.051) &0.445(0.125) &0.01879 &0.0031\\

&  CH$_3$CHO 2$_{0,2} - 1_{0,1}$ E &0.0308(0.002) &10.13(0.017) &0.523(0.059) &0.05525 &0.0033 \\
&  CH$_3$CHO 2$_{0,2} - 1_{0,1}$ A &0.0306(0.002) &10.11(0.017) &0.456(0.03) &0.06307 &0.0033 \\

&  CH$_3$CHO 2$_{1,1} - 1_{1,0}$ E &0.0135(0.002) &10.1(0.036) &0.395(0.067) &0.0321 &0.0035\\
&  CH$_3$CHO 2$_{1,1} - 1_{1,0}$ A &0.0151(0.002) &10.12(0.032) &0.475(0.069) &0.02982 &0.0033\\

800 &  t-HCOOH  2$_{1,2} - 1_{1,1}$ & 0.0212(0.005) &9.853(0.169) &1.5(0.404) &0.01333 &0.0041 \\
&  t-HCOOH  2$_{0,2} - 1_{0,1}$ & 0.0199(0.005) &10.16(0.144) &1.26(0.404) &0.01479 &0.0052 \\
&  t-HCOOH  2$_{1,1} - 1_{1,0}$ & -- &  --  &  -- &   -- & 0.0066\\

 &H$_2$CCO  2$_{1,2} - 1_{1,1}$ &0.0134(0.002) &10.35(0.066) &0.743(0.123) &0.01697 &0.0034\\
 &H$_2$CCO  2$_{0,2} - 1_{0,1}$  &0.0179(0.003) &10.29(0.063) &0.843(0.124) &0.01991 &0.0037\\
 &H$_2$CCO   2$_{1,1} - 1_{1,0}$ &0.0181(0.002) &10.1(0.036) &0.596(0.077) &0.02859 &0.0038 \\
 
 & CH$_3$OH   1$_{0,1} - 0_{0,0}$  A &0.448(0.004) &10.19(0.003) &0.671(0.007) &0.6272 &0.0075 \\
&   CH$_3$OH  1$_{-0,1} - 0_{-0,0}$  E &0.0266(0) &10.26(0.022) &0.378(0.049) &0.06617 &0.0075 \\

&CH$_3$CN $2_1 - 1_1$  & -- &  --  &  -- &   -- & 0.0028\\
& CH$_3$CN $2_0 - 1_0$   & -- &  --  &  -- &   -- &0.0028 \\

&  CH$_3$CHO 2$_{1,2} - 1_{1,1}$ A  &0.0121(0.002) &10.2(0.054) &0.752(0.114) &0.01506 &0.0025\\
&  CH$_3$CHO 2$_{1,2} - 1_{1,1}$ E  &0.018(0.003) &10.3(0.115) &1.4(0.395) &0.01211 &0.0027 \\

&  CH$_3$CHO 2$_{0,2} - 1_{0,1}$ E &0.0333(0.002) &10.16(0.021) &0.66(0.059) &0.04747 &0.0032 \\
&  CH$_3$CHO 2$_{0,2} - 1_{0,1}$ A & 0.0334(0.002) &10.22(0.025) &0.729(0.063) &0.04302 &0.0032 \\

&  CH$_3$CHO 2$_{1,1} - 1_{1,0}$ E &0.0162(0.002) &10.12(0.032) &0.558(0.076) &0.02728 &0.0031\\
&  CH$_3$CHO 2$_{1,1} - 1_{1,0}$ A &0.0153(0.002) &10.25(0.036) &0.566(0.1) &0.02543 &0.0034 \\

\enddata
\tablecomments{Errors reported in parentheses next to the number. 
}.
\end{deluxetable}

\clearpage
\section{Details on Column Density Calculations} \label{N_detail_appendix}

Below we outline the specific steps taken for each molecule in the column density calculations reported in section\,\ref{sec:resultsN}. 

\begin{table*}
	\centering
	\caption{Column Densities: RADEX results for CH$_3$OH}
 \setlength{\tabcolsep}{2pt}
	\label{tab:colden_met}
	\begin{tabular}{clllllllllll} 
    \tablecolumns{13}
     \tablewidth{0pt}
     \tabcolsep=0.3cm
Core \#$^{1}$ &  $\theta_\mathrm{src}$ &  $N$ & $T_\mathrm{ex}$ & $\tau$  &  $N$ & $T_\mathrm{ex}$ & $\tau$ &  $N$ & $T_\mathrm{ex}$ & $\tau$ &  $N_\mathrm{sum}$$^{2}$ \\  [1pt]
(\textit{Herschel}) & arcsec & [$10^{13}$ cm$^{-2}$] &  [K] &  & [$ 10^{13}$ cm$^{-2}$] &  [K] & & [$10^{13}$ cm$^{-2}$] &  [K] & & [$10^{13}$ cm$^{-2}$] \\
\hline
 &  & \multicolumn{3}{c|}{2$_{0,2} - 1_{0,1}$E } & \multicolumn{3}{c|}{2$_{0,2} - 1_{0,1}$ A } & \multicolumn{3}{c|}{2$_{-1,2} - 1_{-1,1}$ E } & \\  
 \hline 
54 &38.5 &4.16$\SPSB{+0.10}{-0.10 }$ &4.78$\SPSB{+0.00}{-0.00 }$ &0.132$\SPSB{+0.003}{-0.003 }$ &2.11$\SPSB{+0.10}{-0.15 }$ &7.43$\SPSB{+0.01}{-0.02 }$ &0.293$\SPSB{+0.013}{-0.020 }$ &1.86$\SPSB{+0.05}{-0.15 }$ &6.66$\SPSB{+0.00}{-0.02 }$ &0.246$\SPSB{+0.006}{-0.019 }$ &  3.97 $\SPSB{+ 0.11 }{- 0.21 }$\\ [1pt]
 
\textit{67} & 38.5 &1.41$\SPSB{+0.15}{-0.20 }$ &4.99$\SPSB{+0.00}{-0.01 }$ &0.036$\SPSB{+0.004}{-0.005 }$ &2.16$\SPSB{+0.10}{-0.15 }$ &7.96$\SPSB{+0.01}{-0.02 }$ &0.266$\SPSB{+0.012}{-0.018 }$ &2.01$\SPSB{+0.10}{-0.15 }$ &7.21$\SPSB{+0.01}{-0.01 }$ &0.235$\SPSB{+0.011}{-0.017 }$  & 4.17 $\SPSB{+ 0.14 }{- 0.21 }$\\ [1pt]

130 & 38.5 &--&--&-- &1.71$\SPSB{+0.10}{-0.15 }$ &7.15$\SPSB{+0.02}{-0.03 }$ &0.339$\SPSB{+0.019}{-0.028 }$ &1.66$\SPSB{+0.05}{-0.15 }$ &6.39$\SPSB{+0.01}{-0.02 }$ &0.304$\SPSB{+0.009}{-0.026 }$ & 3.37 $\SPSB{+ 0.11 }{- 0.21 }$\\[1pt]

231 &38.5 &--&--&-- &1.91$\SPSB{+0.10}{-0.20 }$ &7.59$\SPSB{+0.02}{-0.04 }$ &0.408$\SPSB{+0.020}{-0.040 }$ &1.71$\SPSB{+0.05}{-0.15 }$ &6.78$\SPSB{+0.01}{-0.03 }$ &0.374$\SPSB{+0.010}{-0.031 }$ & 3.62 $\SPSB{+ 0.11 }{- 0.25 }$ \\[1pt]

256 &38.5 &--&--&--&1.96$\SPSB{+0.10}{-0.15 }$ &7.54$\SPSB{+0.02}{-0.03 }$ &0.385$\SPSB{+0.019}{-0.028 }$ &1.91$\SPSB{+0.10}{-0.15 }$ &6.75$\SPSB{+0.01}{-0.02 }$ &0.348$\SPSB{+0.017}{-0.026 }$ & 3.87 $\SPSB{+ 0.14 }{- 0.21 }$\\ [1pt]

\textbf{264} &38.5 &9.31$\SPSB{+0.15}{-0.00 }$ &5.68$\SPSB{+0.00}{-0.00 }$ &0.086$\SPSB{+0.001}{-0.000 }$ &5.81$\SPSB{+0.20}{-0.45 }$ &9.12$\SPSB{+0.01}{-0.02 }$ &0.468$\SPSB{+0.015}{-0.034 }$ &4.86$\SPSB{+0.10}{-0.25 }$ &8.36$\SPSB{+0.01}{-0.01 }$ &0.392$\SPSB{+0.008}{-0.019 }$ & 10.67 $\SPSB{+ 0.22 }{- 0.51 }$ \\ [1pt]

\textbf{317} &38.0 &16.06$\SPSB{+0.05}{-0.30 }$ &7.32$\SPSB{+0.00}{-0.01 }$ &0.195$\SPSB{+0.001}{-0.004 }$ &11.11$\SPSB{+0.30}{-0.65 }$ &12.13$\SPSB{+0.01}{-0.02 }$ &0.637$\SPSB{+0.016}{-0.035 }$ &10.01$\SPSB{+0.10}{-0.35 }$ &11.18$\SPSB{+0.00}{-0.01 }$ &0.495$\SPSB{+0.005}{-0.016 }$ & 21.12 $\SPSB{+ 0.32 }{- 0.74 }$  \\ [1pt]

\textbf{321} &56.5 &10.66$\SPSB{+2.40}{-2.60 }$ &7.53$\SPSB{+0.07}{-0.08 }$ &0.122$\SPSB{+0.030}{-0.031 }$ &9.06$\SPSB{+1.95}{-1.85 }$ &11.54$\SPSB{+0.04}{-0.04 }$ &0.481$\SPSB{+0.098}{-0.095 }$ &8.06$\SPSB{+1.95}{-1.85 }$ &10.83$\SPSB{+0.02}{-0.02 }$ &0.375$\SPSB{+0.087}{-0.083 }$ & 17.12 $\SPSB{+ 2.76 }{- 2.62 }$ \\[1pt]

\textbf{326} &51.5 &17.76$\SPSB{+2.25}{-2.15 }$ &9.16$\SPSB{+0.03}{-0.03 }$ &0.116$\SPSB{+0.016}{-0.015 }$ &26.01$\SPSB{+3.30}{-2.55 }$ &11.87$\SPSB{+0.02}{-0.02 }$ &0.910$\SPSB{+0.107}{-0.084 }$ &24.26$\SPSB{+3.95}{-3.45 }$ &11.32$\SPSB{+0.00}{-0.00 }$ &0.663$\SPSB{+0.102}{-0.090 }$  & 50.27 $\SPSB{+ 5.15 }{- 4.29 }$  \\ [1pt]

339  &38.5 &11.76$\SPSB{+0.45}{-0.35 }$ &7.06$\SPSB{+0.01}{-0.01 }$ &0.091$\SPSB{+0.004}{-0.003 }$ &6.41$\SPSB{+0.00}{-0.05 }$ &13.19$\SPSB{+0.00}{-0.00 }$ &0.286$\SPSB{+0.000}{-0.002 }$ &5.76$\SPSB{+0.10}{-0.05 }$ &12.13$\SPSB{+0.00}{-0.00 }$ &0.223$\SPSB{+0.004}{-0.002 }$  & 12.17 $\SPSB{+ 0.10 }{- 0.07 }$\\[1pt]

344 &38.5 &2.01$\SPSB{+0.25}{-0.30 }$ &5.35$\SPSB{+0.01}{-0.01 }$ &0.042$\SPSB{+0.005}{-0.006 }$ &1.21$\SPSB{+0.00}{-0.05 }$ &8.95$\SPSB{+0.00}{-0.01 }$ &0.117$\SPSB{+0.000}{-0.005 }$ &1.06$\SPSB{+0.00}{-0.00 }$ &8.07$\SPSB{+0.00}{-0.00 }$ &0.099$\SPSB{+0.000}{-0.000 }$ & 2.27 $\SPSB{+ 0.00 }{- 0.05 }$ \\[1pt]

355  &38.5 &4.96$\SPSB{+0.25}{-0.25 }$ &4.57$\SPSB{+0.01}{-0.01 }$ &0.164$\SPSB{+0.008}{-0.008 }$ &3.16$\SPSB{+0.30}{-0.45 }$ &7.17$\SPSB{+0.05}{-0.07 }$ &0.684$\SPSB{+0.060}{-0.089 }$ &2.76$\SPSB{+0.20}{-0.35 }$ &6.40$\SPSB{+0.03}{-0.05 }$ &0.585$\SPSB{+0.039}{-0.068 }$ & 5.92 $\SPSB{+ 0.36 }{- 0.57 }$ \\[1pt]

398  &38.5 &--&--&-- &0.51$\SPSB{+0.00}{-0.00 }$ &6.62$\SPSB{+0.00}{-0.00 }$ &0.118$\SPSB{+0.000}{-0.000 }$ &0.41$\SPSB{+0.00}{-0.05 }$ &5.92$\SPSB{+0.00}{-0.01 }$ &0.100$\SPSB{+0.000}{-0.012 }$ & 0.92 $\SPSB{+ 0.0 }{- 0.05 }$\\[1pt]

\textbf{413} & 39.0 &2.81$\SPSB{+0.25}{-0.25 }$ &4.78$\SPSB{+0.01}{-0.01 }$ &0.079$\SPSB{+0.007}{-0.007 }$ &2.76$\SPSB{+0.15}{-0.30 }$ &7.68$\SPSB{+0.02}{-0.04 }$ &0.402$\SPSB{+0.021}{-0.041 }$ &2.56$\SPSB{+0.10}{-0.20 }$ &6.88$\SPSB{+0.01}{-0.02 }$ &0.352$\SPSB{+0.013}{-0.026 }$ & 5.32 $\SPSB{+ 0.18 }{- 0.36 }$  \\ [1pt]

414  &38.5 &-- &--&--&0.91$\SPSB{+0.10}{-0.05 }$ &5.30$\SPSB{+0.02}{-0.01 }$ &0.310$\SPSB{+0.033}{-0.016 }$ &0.71$\SPSB{+0.05}{-0.05 }$ &4.78$\SPSB{+0.01}{-0.01 }$ &0.324$\SPSB{+0.022}{-0.022 }$ & 1.62 $\SPSB{+ 0.11 }{- 0.07 }$ \\[1pt]

479 & 38.5 &--&--&-- &5.11$\SPSB{+0.20}{-0.40 }$ &8.90$\SPSB{+0.01}{-0.02 }$ &0.530$\SPSB{+0.020}{-0.039 }$ &4.61$\SPSB{+0.05}{-0.25 }$ &8.24$\SPSB{+0.00}{-0.01 }$ &0.431$\SPSB{+0.005}{-0.022 }$  & 9.72 $\SPSB{+ 0.21 }{- 0.47 }$\\[1pt]

491 &38.5 &--&--&-- &1.16$\SPSB{+0.05}{-0.05 }$ &7.28$\SPSB{+0.01}{-0.01 }$ &0.185$\SPSB{+0.008}{-0.008 }$ &1.01$\SPSB{+0.00}{-0.05 }$ &6.50$\SPSB{+0.00}{-0.01 }$ &0.150$\SPSB{+0.000}{-0.007 }$  & 2.17 $\SPSB{+ 0.05 }{- 0.07 }$ \\[1pt]

\textbf{504}  &33.0 &5.66$\SPSB{+0.35}{-0.35 }$ &5.23$\SPSB{+0.01}{-0.01 }$ &0.151$\SPSB{+0.010}{-0.010 }$ &6.76$\SPSB{+0.60}{-1.20 }$ &8.28$\SPSB{+0.04}{-0.08 }$ &0.989$\SPSB{+0.081}{-0.159 }$ &5.91$\SPSB{+0.40}{-0.70 }$ &7.52$\SPSB{+0.02}{-0.04 }$ &0.790$\SPSB{+0.049}{-0.085 }$ & 12.67 $\SPSB{+ 0.72 }{- 1.39 }$  \\[1pt]

543 &38.5 &--&--&--&1.51$\SPSB{+0.10}{-0.10 }$ &6.80$\SPSB{+0.02}{-0.02 }$ &0.330$\SPSB{+0.021}{-0.021 }$ &1.36$\SPSB{+0.05}{-0.10 }$ &5.98$\SPSB{+0.01}{-0.02 }$ &0.298$\SPSB{+0.010}{-0.021 }$  & 2.87 $\SPSB{+ 0.11 }{- 0.14 }$\\ [1pt]

\textbf{615} & 36.5 &4.56$\SPSB{+1.45}{-1.70 }$ &4.56$\SPSB{+0.04}{-0.04 }$ &0.107$\SPSB{+0.034}{-0.040 }$ &4.31$\SPSB{+1.30}{-1.55 }$ &7.50$\SPSB{+0.17}{-0.23 }$ &0.797$\SPSB{+0.212}{-0.267 }$ &3.96$\SPSB{+1.10}{-1.35 }$ &6.62$\SPSB{+0.12}{-0.16 }$ &0.674$\SPSB{+0.166}{-0.214 }$ & 8.27 $\SPSB{+ 1.7 }{- 2.06 }$\\[1pt]

\textbf{627} & 40.5 &4.16$\SPSB{+1.05}{-1.20 }$ &4.77$\SPSB{+0.03}{-0.03 }$ &0.102$\SPSB{+0.026}{-0.030 }$ &1.96$\SPSB{+0.05}{-0.15 }$ &8.14$\SPSB{+0.01}{-0.04 }$ &0.420$\SPSB{+0.010}{-0.030 }$ &1.76$\SPSB{+0.10}{-0.10 }$ &7.16$\SPSB{+0.02}{-0.02 }$ &0.392$\SPSB{+0.021}{-0.021 }$  & 3.72 $\SPSB{+ 0.11 }{- 0.18 }$ \\[1pt]

642  &38.5 &--&-- &--&0.46$\SPSB{+0.00}{-0.05 }$ &7.24$\SPSB{+0.00}{-0.02 }$ &0.117$\SPSB{+0.000}{-0.012 }$ &0.41$\SPSB{+0.05}{-0.00 }$ &6.33$\SPSB{+0.01}{-0.00 }$ &0.098$\SPSB{+0.012}{-0.000 }$ & 0.87 $\SPSB{+ 0.05 }{- 0.05 }$ \\[1pt]

656 &38.5 &2.26$\SPSB{+1.45}{-1.75 }$ &4.34$\SPSB{+0.06}{-0.07 }$ &0.104$\SPSB{+0.067}{-0.080 }$ &1.81$\SPSB{+0.85}{-1.05 }$ &6.64$\SPSB{+0.18}{-0.25 }$ &0.540$\SPSB{+0.230}{-0.303 }$ &1.66$\SPSB{+0.85}{-1.00 }$ &5.97$\SPSB{+0.15}{-0.20 }$ &0.507$\SPSB{+0.234}{-0.296 }$  & 3.47 $\SPSB{+ 1.2 }{- 1.45 }$\\[1pt]

657 &38.5 &--&--&-- &1.21$\SPSB{+0.05}{-0.10 }$ &6.63$\SPSB{+0.01}{-0.02 }$ &0.280$\SPSB{+0.011}{-0.022 }$ &1.06$\SPSB{+0.00}{-0.10 }$ &5.90$\SPSB{+0.00}{-0.02 }$ &0.239$\SPSB{+0.000}{-0.022 }$ & 2.27 $\SPSB{+ 0.05 }{- 0.14 }$ \\[1pt]
 
\textit{658} &38.5 &5.16$\SPSB{+2.10}{-2.55 }$ &4.24$\SPSB{+0.06}{-0.07 }$ &0.166$\SPSB{+0.067}{-0.082 }$ &3.71$\SPSB{+1.40}{-1.85 }$ &6.72$\SPSB{+0.24}{-0.37 }$ &1.039$\SPSB{+0.336}{-0.484 }$ &3.31$\SPSB{+1.35}{-1.75 }$ &5.94$\SPSB{+0.20}{-0.30 }$ &0.929$\SPSB{+0.323}{-0.461 }$  & 7.02 $\SPSB{+ 1.94 }{- 2.55 }$ \\[1pt]

\textbf{709} &33.5 &14.26$\SPSB{+0.25}{-0.50 }$ &5.92$\SPSB{+0.01}{-0.02 }$ &0.214$\SPSB{+0.004}{-0.008 }$ &7.81$\SPSB{+0.35}{-0.65 }$ &10.58$\SPSB{+0.03}{-0.05 }$ &0.591$\SPSB{+0.024}{-0.045 }$ &6.61$\SPSB{+0.25}{-0.45 }$ &9.38$\SPSB{+0.02}{-0.03 }$ &0.476$\SPSB{+0.017}{-0.030 }$ & 14.42 $\SPSB{+ 0.43 }{- 0.79 }$  \\[1pt]

 \textbf{715} & 26.5 &6.26$\SPSB{+1.65}{-1.85 }$ &5.87$\SPSB{+0.05}{-0.05 }$ &0.085$\SPSB{+0.023}{-0.025 }$ &5.31$\SPSB{+1.70}{-1.75 }$ &11.38$\SPSB{+0.16}{-0.19 }$ &0.470$\SPSB{+0.136}{-0.146 }$ &4.76$\SPSB{+1.50}{-1.55 }$ &10.09$\SPSB{+0.13}{-0.15 }$ &0.382$\SPSB{+0.109}{-0.118 }$ & 10.07 $\SPSB{+ 2.27 }{- 2.34 }$  \\[1pt]
 
739   &38.5 &--&--&--&2.46$\SPSB{+0.15}{-0.25 }$ &7.18$\SPSB{+0.02}{-0.03 }$ &0.412$\SPSB{+0.024}{-0.040 }$ &2.11$\SPSB{+0.05}{-0.15 }$ &6.54$\SPSB{+0.01}{-0.01 }$ &0.332$\SPSB{+0.008}{-0.023 }$ & 4.57 $\SPSB{+ 0.16 }{- 0.29 }$ \\[1pt]

\textbf{746} & 35.0 &4.26$\SPSB{+0.35}{-0.30 }$ &4.99$\SPSB{+0.01}{-0.01 }$ &0.093$\SPSB{+0.008}{-0.007 }$ &5.46$\SPSB{+0.40}{-0.75 }$ &8.34$\SPSB{+0.04}{-0.07 }$ &0.728$\SPSB{+0.049}{-0.091 }$ &4.91$\SPSB{+0.30}{-0.55 }$ &7.47$\SPSB{+0.03}{-0.04 }$ &0.620$\SPSB{+0.035}{-0.063 }$ & 10.37 $\SPSB{+ 0.5 }{- 0.93 }$ \\[1pt]

747 & 38.5 &7.71$\SPSB{+0.50}{-0.50 }$ &4.59$\SPSB{+0.01}{-0.02 }$ &0.184$\SPSB{+0.012}{-0.012 }$ &3.56$\SPSB{+0.25}{-0.45 }$ &7.83$\SPSB{+0.05}{-0.09 }$ &0.642$\SPSB{+0.040}{-0.072 }$ &3.06$\SPSB{+0.20}{-0.35 }$ &6.75$\SPSB{+0.03}{-0.06 }$ &0.554$\SPSB{+0.033}{-0.056 }$ & 6.62 $\SPSB{+ 0.32 }{- 0.57 }$\\[1pt]
 
\textbf{752} &46.5 &8.36$\SPSB{+0.05}{-0.10 }$ &4.65$\SPSB{+0.00}{-0.00 }$ &0.189$\SPSB{+0.001}{-0.002 }$ &4.21$\SPSB{+0.30}{-0.50 }$ &8.15$\SPSB{+0.05}{-0.09 }$ &0.564$\SPSB{+0.036}{-0.059 }$ &3.91$\SPSB{+0.25}{-0.45 }$ &7.01$\SPSB{+0.04}{-0.07 }$ &0.526$\SPSB{+0.030}{-0.053 }$  & 8.12 $\SPSB{+ 0.39 }{- 0.67 }$ \\[1pt]

\textbf{768} &  48.0 &1.96$\SPSB{+0.35}{-0.35 }$ &5.00$\SPSB{+0.01}{-0.01 }$ &0.055$\SPSB{+0.010}{-0.010 }$ &1.96$\SPSB{+0.05}{-0.15 }$ &8.09$\SPSB{+0.01}{-0.02 }$ &0.258$\SPSB{+0.006}{-0.019 }$ &1.81$\SPSB{+0.05}{-0.05 }$ &7.28$\SPSB{+0.00}{-0.00 }$ &0.218$\SPSB{+0.006}{-0.006 }$  & 3.77 $\SPSB{+ 0.07 }{- 0.16 }$ \\[1pt]

780  &38.5 &3.66$\SPSB{+0.30}{-0.40 }$ &4.54$\SPSB{+0.01}{-0.01 }$ &0.092$\SPSB{+0.008}{-0.010 }$ &1.56$\SPSB{+0.10}{-0.15 }$ &7.16$\SPSB{+0.02}{-0.03 }$ &0.335$\SPSB{+0.020}{-0.030 }$ &1.41$\SPSB{+0.10}{-0.10 }$ &6.37$\SPSB{+0.02}{-0.02 }$ &0.301$\SPSB{+0.020}{-0.020 }$ & 2.97 $\SPSB{+ 0.14 }{- 0.18 }$ \\ [1pt]

\textbf{799} &41.0 &3.81$\SPSB{+0.25}{-0.25 }$ &4.52$\SPSB{+0.01}{-0.01 }$ &0.127$\SPSB{+0.008}{-0.008 }$ &3.81$\SPSB{+0.40}{-0.65 }$ &7.52$\SPSB{+0.07}{-0.11 }$ &0.911$\SPSB{+0.086}{-0.136 }$ &3.41$\SPSB{+0.30}{-0.55 }$ &6.65$\SPSB{+0.05}{-0.08 }$ &0.825$\SPSB{+0.065}{-0.116 }$ & 7.22 $\SPSB{+ 0.5 }{- 0.85 }$ \\[1pt]

\textbf{800}   &29.5 &7.16$\SPSB{+0.15}{-0.10 }$ &5.76$\SPSB{+0.00}{-0.00 }$ &0.147$\SPSB{+0.003}{-0.002 }$ &8.56$\SPSB{+0.65}{-1.20 }$ &9.92$\SPSB{+0.04}{-0.07 }$ &0.953$\SPSB{+0.066}{-0.120 }$ &7.51$\SPSB{+0.40}{-0.80 }$ &8.91$\SPSB{+0.02}{-0.04 }$ &0.758$\SPSB{+0.037}{-0.073 }$  & 16.07 $\SPSB{+ 0.76 }{- 1.44 }$\\
\hline 
  &   & \multicolumn{3}{c|}{ {1$_{0,1} - 0_{0,0}$ A} } & \multicolumn{3}{c|}{  {1$_{-0,1} - 0_{-0,0}$ E} } & \multicolumn{3}{c|}{ } & \\  
\hline 

 \textbf{264} & 38.5 &5.81$\SPSB{+0.25}{-0.25 }$ &6.84$\SPSB{+0.02}{-0.02 }$ &0.195$\SPSB{+0.008}{-0.008 }$   & -- & -- & -- \\[1pt]
 
\textbf{317} & 38.0  &11.11$\SPSB{+0.45}{-0.40 }$ &8.50$\SPSB{+0.03}{-0.03 }$ &0.310$\SPSB{+0.011}{-0.010 }$ &14.46$\SPSB{+0.60}{-0.50 }$ &7.52$\SPSB{+0.05}{-0.04 }$ &0.388$\SPSB{+0.014}{-0.011 }$  \\[1pt]

\textbf{321}  & 56.5 &9.06$\SPSB{+0.40}{-0.35 }$ &7.90$\SPSB{+0.03}{-0.02 }$ &0.232$\SPSB{+0.009}{-0.008 }$ &9.36$\SPSB{+0.35}{-0.20 }$ &7.43$\SPSB{+0.03}{-0.02 }$ &0.239$\SPSB{+0.008}{-0.005 }$    \\[1pt]

\textbf{326}   &  51.5 &26.01$\SPSB{+0.95}{-0.75 }$ &8.88$\SPSB{+0.03}{-0.03 }$ &0.462$\SPSB{+0.014}{-0.011 }$ &18.51$\SPSB{+0.55}{-0.30 }$ &9.94$\SPSB{+0.03}{-0.01 }$ &0.348$\SPSB{+0.009}{-0.005 }$  \\
[1pt]

\textbf{413}   & 39.0 &2.76$\SPSB{+0.10}{-0.10 }$ &6.35$\SPSB{+0.01}{-0.01 }$ &0.165$\SPSB{+0.006}{-0.006 }$ &3.46$\SPSB{+0.65}{-0.70 }$ &4.34$\SPSB{+0.06}{-0.06 }$ &0.203$\SPSB{+0.034}{-0.038 }$    \\[1pt]

\textbf{504}  & 33.0 &6.81$\SPSB{+0.20}{-0.10 }$ &6.66$\SPSB{+0.01}{-0.01 }$ &0.380$\SPSB{+0.010}{-0.005 }$ &5.61$\SPSB{+0.50}{-0.50 }$ &4.92$\SPSB{+0.04}{-0.05 }$ &0.320$\SPSB{+0.025}{-0.026 }$    \\[1pt]

\textbf{615}   & 36.5 &4.31$\SPSB{+0.05}{-0.05 }$ &6.43$\SPSB{+0.01}{-0.01 }$ &0.320$\SPSB{+0.003}{-0.003 }$ &4.61$\SPSB{+0.50}{-0.50 }$ &4.32$\SPSB{+0.05}{-0.05 }$ &0.340$\SPSB{+0.032}{-0.032 }$  \\[1pt]

\textbf{627}  & 40.5 &1.96$\SPSB{+0.05}{-0.05 }$ &6.91$\SPSB{+0.01}{-0.01 }$ &0.169$\SPSB{+0.004}{-0.004 }$ &4.81$\SPSB{+0.65}{-0.65 }$ &4.60$\SPSB{+0.07}{-0.08 }$ &0.367$\SPSB{+0.041}{-0.042 }$  \\[1pt]

\textbf{709} & 33.5 &7.76$\SPSB{+0.25}{-0.20 }$ &8.18$\SPSB{+0.02}{-0.02 }$ &0.288$\SPSB{+0.008}{-0.006 }$ &12.96$\SPSB{+0.60}{-0.50 }$ &5.91$\SPSB{+0.05}{-0.04 }$ &0.444$\SPSB{+0.017}{-0.014 }$   \\[1pt]
 
\textbf{715} & 26.5 &5.31$\SPSB{+0.30}{-0.20 }$ &8.57$\SPSB{+0.04}{-0.03 }$ &0.204$\SPSB{+0.010}{-0.007 }$ &2.56$\SPSB{+0.65}{-0.65 }$ &5.45$\SPSB{+0.14}{-0.15 }$ &0.106$\SPSB{+0.024}{-0.025 }$  \\[1pt]

\textbf{746} & 35.0 &5.41$\SPSB{+0.15}{-0.10 }$ &6.80$\SPSB{+0.02}{-0.01 }$ &0.320$\SPSB{+0.008}{-0.005 }$ &5.21$\SPSB{+0.85}{-0.85 }$ &4.60$\SPSB{+0.06}{-0.07 }$ &0.309$\SPSB{+0.043}{-0.045 }$   \\[1pt]

\textbf{752} &46.5 &4.21$\SPSB{+0.05}{-0.00 }$ &7.58$\SPSB{+0.01}{-0.00 }$ &0.220$\SPSB{+0.002}{-0.000 }$ &9.16$\SPSB{+0.50}{-0.40 }$ &4.51$\SPSB{+0.03}{-0.03 }$ &0.410$\SPSB{+0.017}{-0.014 }$  \\[1pt]

\textbf{768} & 48.0 &1.96$\SPSB{+0.10}{-0.05 }$ &6.46$\SPSB{+0.01}{-0.01 }$ &0.113$\SPSB{+0.005}{-0.003 }$ &1.76$\SPSB{+0.40}{-0.45 }$ &4.43$\SPSB{+0.05}{-0.05 }$ &0.102$\SPSB{+0.022}{-0.025 }$ \\[1pt]

\textbf{799} & 41.0 &3.81$\SPSB{+0.00}{-0.00 }$ &6.52$\SPSB{+0.00}{-0.00 }$ &0.376$\SPSB{+0.000}{-0.000 }$ &3.51$\SPSB{+0.60}{-0.65 }$ &4.24$\SPSB{+0.07}{-0.08 }$ &0.350$\SPSB{+0.051}{-0.057 }$    \\[1pt]

\textbf{800} &29.5 &8.56$\SPSB{+0.25}{-0.10 }$ &7.65$\SPSB{+0.02}{-0.01 }$ &0.414$\SPSB{+0.010}{-0.004 }$ &4.76$\SPSB{+0.50}{-0.45 }$ &5.61$\SPSB{+0.08}{-0.08 }$ &0.249$\SPSB{+0.023}{-0.021 }$   \\
		\hline
    \end{tabular}
      \begin{description} 
     \item $^{1}$ The cores observed with both the ARO 12m and Yebes 40m are bolded. $^{2}$ We chose to calculate the total column density, $N_\mathrm{sum}$, by summing the 2$_{0,2} - 1_{0,1}$ A and 2$_{-1,2} - 1_{-1,1}$ E column densities. Errors quoted as `0.00' or `0.000' are $<0.005$ or $<0.0005$, respectively.
      \end{description}
\end{table*}

\begin{figure*}
\centering
\begin{center}$
\begin{array}{cccc}
\includegraphics[width=85mm]{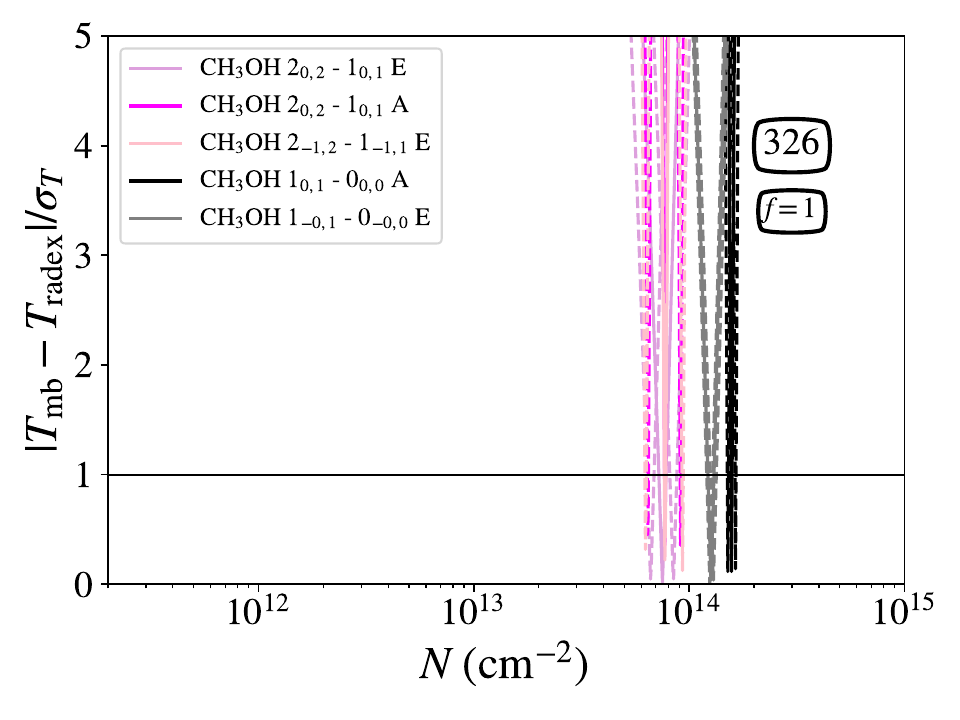} & 
\includegraphics[width=85mm]{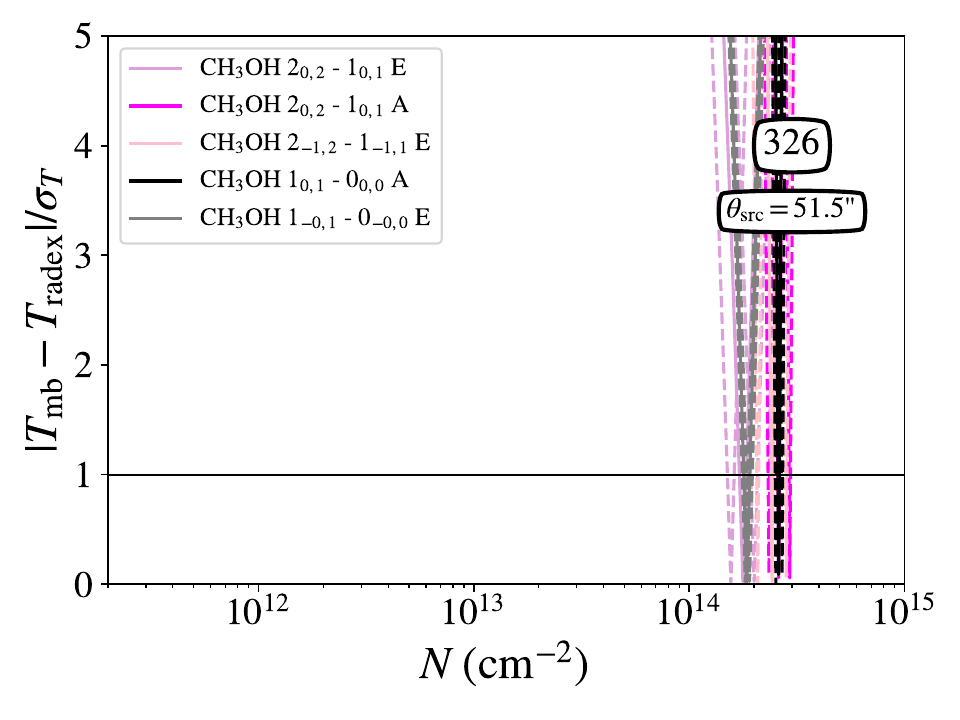}\\ 
\includegraphics[width=85mm]{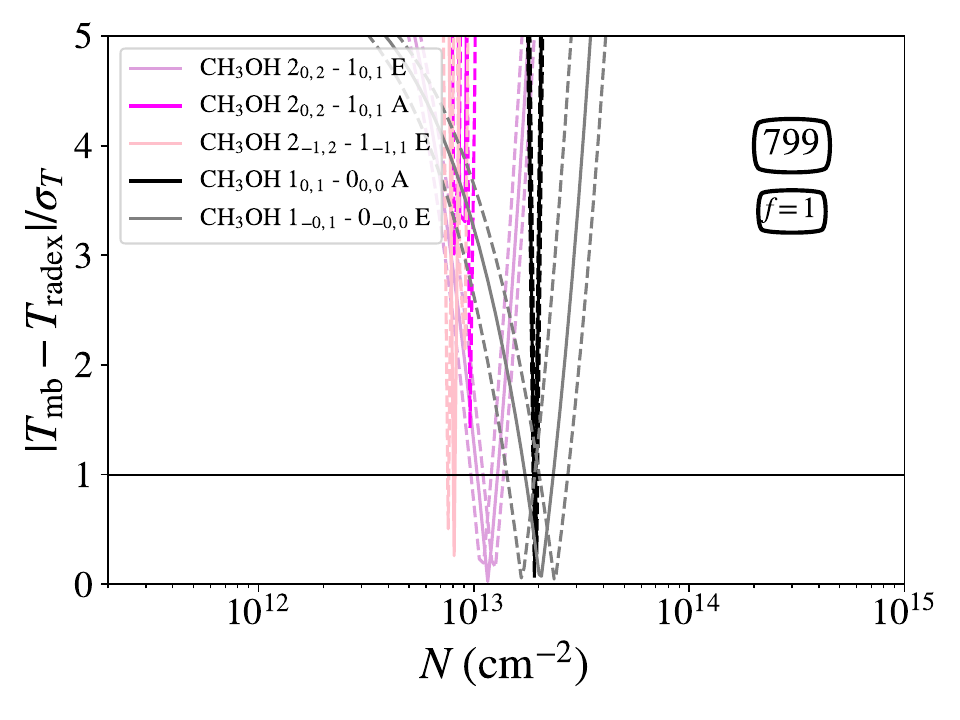} & 
\includegraphics[width=85mm]{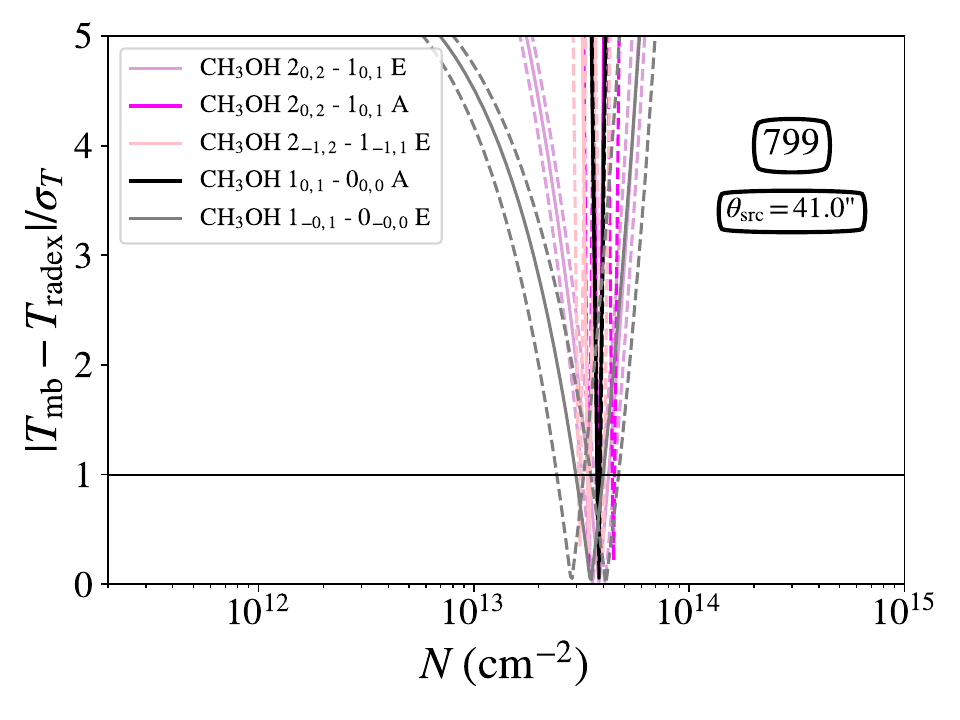}\\
\end{array}$
\end{center}
\caption{\label{meth_radex_fig} Representative RADEX CH$_3$OH minimization plots for cores 326 (top) and 799 (bottom). Plotted in each panel is the difference in the observed brightest temperature, $T_\mathrm{mb}$, and the RADEX-derived brightness temperature, $T_\mathrm{radex}$, divided by the \textit{rms} level, $\sigma_T$, versus the RADEX-derived column density, $N$. When the observations best-match the RADEX model, $|T_\mathrm{mb} - T_\mathrm{radex}|/\sigma_T$ is minimized and the best $N$ is found while $T_\mathrm{ex}$ is being optimized for each transition. Solid lines are color-coded to reflect which CH$_3$OH transition is plotted (ARO 12m transitions in shaded of pink, Yebes 40m transitions in shades of black) and the dashed lines represent the spread in error. In the left panels a standard filling factor, $f = 1$, is assumed and in the right panels the minimization is shown when the best-fit source size, $\theta_\mathrm{src}$, is found (for core 326 $\theta_\mathrm{src}$ = 51.5 arcsec and core 799 $\theta_\mathrm{src}$ = 41.0 arsec).
}
\end{figure*}

\subsection{Methanol, CH$_3$OH} \label{methcoldensection}

Methanol was detected in all 35 cores targeted and therefore we start with this molecule in our column density calculations. The collisional rates for the torsional ground states of the A and E states are available \citep{2010MNRAS.406...95R} and we use the RADEX code to calculate $N$, ${T_\mathrm{ex}}$, and $\tau$ values for the various transitions that were observed (Table\,\ref{tab:colden_met}; spectroscopic work detailed in \cite{2008JMoSp.251..305X}). We note that the $\tau$ value, or optical depth, of the line is not a fitted parameter but a prediction of RADEX for the given parameters and transition. 
The method for calculating $N$ is as described in \cite{2020ApJ...891...73S}, where a grid of RADEX models (2000 total) with $N$ varying from $1 \times 10^{10} - 1 \times 10^{14}$\,cm$^{-2}$ were run for each CH$_3$OH transition. In the calculations we are able utilize the physical conditions already derived for each core, i.e., the average volume density, ${n_\mathrm{H_2}}$, and the kinetic temperature, $T_{k}$ (columns 9 and 10 in Table\,\ref{physparams}). 
The difference in the observed line peak versus the line peak RADEX calculates was then minimized in order to find the best fit column density. In Figure\,\ref{meth_radex_fig} an example is shown where the difference in radiation temperature divided by observed RMS, i.e., $|T_\mathrm{mb}-T_\mathrm{radex}|/\sigma_{T}$, is plotted against the RADEX-derived column density, $N$ for each transition line. For the Yebes 40m sub-sample there are five or four separate transitions minimized (two A-states and two or three E-states), and for the remaining cores two or three transitions minimized (depending on if the weaker $2_{0,2}-1_{0,1}$ E transition was detected). 

Because of the difference in beam sizes of the ARO 12m and Yebes 40m CH$_3$OH transitions (62 arcsec and 37 arcsec, respectively; see Table\,\ref{LineList}), we could constrain the source size, $\theta_\mathrm{src}$, and account for the true filling factor, $f$, when calculating the column density for our 15 core sub-sample. To do this, we ran our minimization procedure for $\theta_\mathrm{src}$ values from 23 to 65 arcsec (intervals of 0.5 arsec) in order to find a `best-fit' $\theta_\mathrm{src}$ value for each core when the two bright CH$_3$OH $2_{0,2} - 1_{0,1}$ A and CH$_3$OH $1_{0,1} - 0_{0,0}$ A are closest in $N$ (see Figure\,\ref{meth_radex_fig}). The $\theta_\mathrm{src}$ ranges from $26.5-56.5$ arcsec where the median $\theta_\mathrm{src}$ is 38.5 arcsec (5 arcsec median deviation).

The error in the observed linewidth, $\Delta v$, as well as a $10\%$ error on ${n_\mathrm{H_2}}$ and $T_{k}$ are propagated through our calculations to produce an error in $N$,  $T_\mathrm{ex}$, and $\tau$ from our RADEX calculations (shown as dashed lines in Figure\,\ref{meth_radex_fig}). The volume density may be more uncertain because it was calculated as an average within the ARO 12m beam and used for all transitions. 
Though, as was found in \cite{2020ApJ...891...73S}, we find that even an order of magnitude difference in volume density leads to only a factor of $\sim$2 in column density. In Table\,\ref{tab:colden_met} we present the minimized column density for each transition separately, as well as a total column density, $N_\mathrm{sum}$, which we calculate from the sum of the brightest A and E states seen for all 35 cores in the sample, i.e., the 96\,GHz $2_{0,2}-1_{0,1}$\,A and $2_{-1,2}-1_{-1,1}$\,E transitions. 

While it is also possible with RADEX to use multiple molecular line transitions to constrain $T_{k}$ and ${n_\mathrm{H_2}}$ for both methanol sub-state, this is only possible for the sub-sample of cores with Yebes 40m data (see Appendix\,\ref{radexappendix}). And, because specifically the A-state transitions are probed by different beams, this would lead to degeneracy if one were to try to also minimize to find a best-fit pair of ${n_\mathrm{H_2}}$ and $T_{k}$ in addition to the source size, $\theta_\mathrm{src}$ . For these reasons we have chosen the standardized approach to minimize for $N$ and (where we can) $f$ while fixing ${n_\mathrm{H_2}}$ and $T_{k}$.

Additionally, since we have two A state transitions of this molecule with different $E_u$ values, i.e., the $2_{0,2} - 1_{0,1}$ at $E_u$ = 7.0\,K and $1_{0,1} - 1_{0,0}$ line at $E_u$ = 2.3\,K, we also employ the CTEX method to calculate $N$ and $T_\mathrm{ex}$ while applying the same corrected filling factors. We find the CTEX-determined column densities are lower by factors of $\sim 2 - 6$ than the RADEX-determined column densities. This is similar to what had been found in the L1544 prestellar core \citep{2014A&A...569A..27B, 2014ApJ...795L...2V}. Looking at Table\,\ref{tab:colden_met}, it is evident that an LTE assumption (which assumes $\tau_\nu \ll 1$) breaks down as the $1-0$ transitions have lower $T_\mathrm{ex}$ values than the $2-1$ transitions, where in some cases $\tau \sim 1$.

\subsection{Acetaldehyde, CH$_3$CHO}

For the total 35 cores from the initial ARO 12m survey, column densities or upper limits were derived for CH$_3$CHO given our observations of the $5_{0,5}-4_{0,4}$ transitions with $E_u/k \sim 14$\,K. We note that original laboratory spectroscopic work is detailed in \cite{1996JPCRD..25.1113K}. Since there are no calculated collisional rate coefficients for this molecule, we are not able to use RADEX. For the 15 cores with Yebes 40m observations, there are additional transitions with varying $E_u/k$ values from $\sim 3 - 5$\,K, which allows us to use the RD method, which we chose to employ for the A-state transitions in order to robustly derive $N$ and ${T_\mathrm{ex}}$ (see Figure\,\ref{acet_rotdiagram} and Table\,\ref{tab:colden_acet}). Overall, we find $N$ for CH$_3$CHO A ranges from  $0.59 - 8.08\,\times\,10^{12}$\,cm$^{-2}$, with a median value (and median standard deviation) of $2.10 \pm 0.61 \times 10^{12}$\,cm$^{-2}$. Additionally, ${T_\mathrm{ex}}$ ranges from $ 4.27 - 7.07$ with a median value of 5.55\,K and median standard deviation of 0.9\,K. This is within the range but slightly higher than the ${T_\mathrm{ex}}$ values derived using the CTEX method for the Taurus cores, which range from 3.09 --5.39\,K, with a median of 3.57\,K \citep{2020ApJ...891...73S}. For the starless core L1521E the ${T_\mathrm{ex}}$ value for CH$_3$CHO A, derived using the same RD method but with different transitions,  is $\sim4.5$\,K \citep{2021MNRAS.504.5754S}, slightly lower than the median for the Perseus core sample.

For cores 67 and 658, for which CH$_3$CHO was detected with the ARO 12m at a $\sigma_{T_\mathrm{mb}}=2.5$\,mK limit and no Yebes 40m data is available, the $N$ values are calculated with the LTE method based on the single A state transition detected with the ARO 12m, assuming the median CH$_3$CHO A $T_\mathrm{ex}$ value of 5.55\,K. For these cores $N$ for CH$_3$CHO A is $\sim0.7\times 10^{12}$\,cm$^{-2}$. For the remaining cores, a 3$\sigma$ upper limit is calculated via the LTE method, with the same mean $T_\mathrm{ex}$ assumed (see Table\,\ref{tab:colden_acet}). 

Because we need to worry again about the differences in beam sizes between the ARO 12m and Yebes 40m CH$_3$CHO transitions, we take this into account by assuming the same source size, $\theta_\mathrm{src}$, that was derived for CH$_3$OH, as these molecules are believed to trace each other spatially and has been done for the Taurus core L1521E\,\citep{2021MNRAS.504.5754S} and predicted by the chemical models of \cite{2017ApJ...842...33V}. The $\theta_\mathrm{src}$ is then plugged into equation\,\ref{fillingfrac}, where $\theta_\mathrm{beam}$ for these A-state transitions of CH$_3$CHO are listed in Table\,\ref{LineList}, in order to account for the filling factor, $f$. We note that if $f$ is set to 1 for all transitions, the $N$ values would decrease by factors from $\sim$1.4 to 3.3. 

\begin{table}
	\centering
	\caption{Column Densities: CH$_3$CHO A}
 \setlength{\tabcolsep}{12pt}
	\label{tab:colden_acet}
	\begin{tabular}{clll} 
    \tablecolumns{10}
     \tablewidth{0pt}
     \tabcolsep=0.1cm
Core \#$^{1}$ &  $\theta_\mathrm{src}$ & \multicolumn{2}{c|}{CH$_3$CHO A}  \\  
(\textit{Herschel}) & arcsec & $N$ [$\times 10^{12}$ cm$^{-2}$] & $T_\mathrm{ex}$  [K]  \\
\hline
54 & 38.5&  $<1.70$&  5.55 \\ [1pt]
\textit{67} &  38.5&  0.68 $\SPSB{+ 0.61 }{- 0.16 }$ &  5.55$\pm$0.9 \\ [1pt]
130 &  38.5&  $< 2.05 $&  5.55 \\[1pt]
231 & 38.5&  $< 1.51$&  5.55 \\[1pt]
256 & 38.5&  $< 1.97$&  5.55 \\[1pt]
\textbf{264} & 38.5& 1.63 $\SPSB{+ 0.20 }{- 0.20 }$&  5.78 $\SPSB{+ 0.34 }{- 0.44 }$ \\ [1pt]
\textbf{317} & 38.0&  5.63 $\SPSB{+ 0.37 }{- 0.38 }$&  6.83 $\SPSB{+ 0.08 }{- 0.08 }$\\ [1pt]
\textbf{321} & 56.5&   3.70 $\SPSB{+ 0.21 }{- 0.21 }$&  6.28 $\SPSB{+ 0.13 }{- 0.14 }$ \\[1pt]
\textbf{326} & 51.5&  8.08 $\SPSB{+ 0.17 }{- 0.17 }$&  6.24 $\SPSB{+ 0.11 }{- 0.11 }$\\ [1pt]
339  & 38.5&  $< 2.07$ &  5.55 \\[1pt]
344  & 38.5&  $< 1.46$ &  5.55 \\[1pt]
355  & 38.5&  $< 1.79$ &  5.55   \\[1pt]
398  & 38.5&  $< 1.79$ &  5.55 \\[1pt]
\textbf{413} & 39.0& 1.49 $\SPSB{+ 0.23 }{- 0.24 }$&  4.97 $\SPSB{+ 0.16 }{- 0.22 }$ \\ [1pt]
414 & 38.5&  $< 2.09$ &  5.55 \\[1pt]
479 &38.5&  $< 2.38 $&  5.55 \\[1pt]
491 &38.5& $< 1.75 $ &  5.55 \\[1pt]
\textbf{504} & 33.0&  2.21 $\SPSB{+ 0.17 }{- 0.14 }$&  5.24 $\SPSB{+ 0.50 }{- 0.62 }$ \\ [1pt]
 543 &38.5&  $< 1.94 $ &  5.55 \\[1pt]
\textbf{615} &36.5&  2.10 $\SPSB{+ 0.18 }{- 0.17 }$&  4.54 $\SPSB{+ 0.22 }{- 0.28 }$\\[1pt]
\textbf{627} &40.5&   1.92 $\SPSB{+ 0.19 }{- 0.18 }$&  4.27 $\SPSB{+ 0.17 }{- 0.23 }$ \\[1pt]
 642  & 38.5&  $< 2.31 $ &  5.55   \\[1pt]
656  & 38.5&  $< 1.56$ &  5.55 \\[1pt]
 657  & 38.5& $< 2.03 $&  5.55 \\[1pt]
\textit{658} &38.5&  0.59 $\SPSB{+ 0.52 }{- 0.14 }$&  5.55 $\SPSB{+ 0.92 }{- 0.92 }$\\[1pt]
\textbf{709} & 33.5&  4.65 $\SPSB{+ 0.41 }{- 0.41 }$&  5.74 $\SPSB{+ 0.06 }{- 0.07 }$ \\[1pt]
 \textbf{715} & 26.5&  2.69 $\SPSB{+ 0.30 }{- 0.29 }$&  4.45 $\SPSB{+ 0.23 }{- 0.31 }$\\[1pt]
739  &38.5&  $<1.84$  &  5.55 \\[1pt]
 \textbf{746} &35.0&  4.27 $\SPSB{+ 0.44 }{- 0.45 }$&  4.68 $\SPSB{+ 0.04 }{- 0.04 }$\\[1pt]
 747  &38.5&  $< 2.37 $ &  5.55 \\[1pt]
 \textbf{752} & 46.5& 1.94 $\SPSB{+ 0.16 }{- 0.16 }$&  5.88 $\SPSB{+ 0.22 }{- 0.27 }$\\[1pt]
\textbf{768} & 48.0&  1.74 $\SPSB{+ 0.25 }{- 0.26 }$&  7.07 $\SPSB{+ 0.23 }{- 0.28 }$\\[1pt]
780   &38.5& $< 1.72 $ &  5.55 \\ [1pt]
\textbf{799} & 41.0&  1.79 $\SPSB{+ 0.14 }{- 0.13 }$&  4.52 $\SPSB{+ 0.21 }{- 0.25 }$\\[1pt]
\textbf{800}  & 29.5&  3.95 $\SPSB{+ 0.32 }{- 0.32 }$&  6.71 $\SPSB{+ 0.33 }{- 0.38 }$ \\
		\hline
    \end{tabular}
      \begin{description} 
     \item $^{1}$ The cores observed with both the ARO 12m and Yebes 40m are bolded, their values calculated from the RD method. For the italicized cores 67 and 658, the values are calculated with the LTE method based on the single A state transition detected with the ARO 12m, assuming the median CH$_3$CHO A-state $T_\mathrm{ex}$ value of 4.94 K. For the remaining cores, a 3$\sigma$ upper limit is calculated via the LTE method.
      \end{description}
\end{table}

\begin{table*}
	\centering
	\caption{Column Densities: The 5-atom COMs, H$_2$CCO and t-HCOOH}
 \setlength{\tabcolsep}{3pt}
	\label{tab:colden_5atom}
	\begin{tabular}{cllllllllll} 
    \tablecolumns{10}
     \tablewidth{0pt}
     \tabcolsep=0.3cm
Core $\#$ &  &   H$_2$CCO  o- (p-) &  t-HCOOH & & H$_2$CCO  o- (p-) &  t-HCOOH  &  & H$_2$CCO  o- (p-) &  t-HCOOH \\  
(\textit{Herschel}) & $T_\mathrm{ex}$  [K] &  $N$ [$\times 10^{12}$ cm$^{-2}$] &  $N$ [$\times 10^{12}$ cm$^{-2}$] & $T_\mathrm{ex}$  [K] &  $N$ [$\times 10^{12}$ cm$^{-2}$] &  $N$ [$\times 10^{12}$ cm$^{-2}$]& $T_\mathrm{ex}$  [K] &  $N$ [$\times 10^{12}$ cm$^{-2}$] &  $N$ [$\times 10^{12}$ cm$^{-2}$] \\

\hline
264 &  5 & $<$5.0 ($<0.8$)  & $14.7$ & 10 & $< 1.8 $ ($<1.1$) & $15.1$ & 20  & $< 2.0$ (2.3)  &  $27.4$  \\ [3pt]
317 &  5 & 5.6 (1.1) & 1.3 & 10 & 2.0 (1.5) & 1.8 & 20 & 2.2 (3.2) & 3.7 \\ [3pt]
321 &  5 & 12.8  (0.6) & $< 1.5$ & 10 & 4.7  (0.8) & $<2.0$   &  20 & 4.3 (1.7)  & $<4.3$ \\[3pt]
326 &  5 & 20.8 (2.5)& 4.1 &  10 & 7.6 (3.4)& 5.7 & 20 & 8.5 (7.3) & 12.1 \\ [3pt]
413 &  5 & 7.7 (0.7)& $< 1.5$ &10 & 2.8 (0.9) & $< 2.1$  & 20 & 3.1 (2.0)  & $< 4.3$  \\ [3pt]
504 &  5 & 6.8 (1.1)& 1.3  &10 & 2.5 (1.6)& 1.8  &  20 & 2.7(3.4) & 3.7 \\ [3pt]
615  &   5 & 5.1(3.3) &$<2.2$   &10 & 1.8 (4.4)& $<3.1$ & 20 & 2.1 (9.5) & $<6.7$ \\[3pt]
627 &    5 & 7.7 (0.7)& $< 2.0$  &10 & 2.8 (1.0)& $< 2.9$ & 20 & 3.1 (2.2) & $< 6.1 $  \\[3pt]
709 &    5 & 20.1 (2.3)& 2.2 &10 & 7.3 (3.1)& 3.1& 20 & 8.1 (6.7) & 6.7   \\[3pt]
715 &  5 & 10.6 (1.8)& 3.0  &10 & 3.9 (2.5)&  4.2& 20 &4.3 (5.3)  & 8.8 \\[3pt]
746  &    5 & 16.5 (1.6)& 2.5 & 10 & 6.0 (2.2)& 3.5& 20 &6.7 (4.8) & 7.5 \\[3pt]
752 &  5 & 6.2 (1.9)& 1.1&  10 & 2.3 (2.6)& 1.5 & 20 & 2.5 (5.6) & 3.2 \\[3pt]
768 &   5 & 10.1 (0.9)& $< 1.1$  &10 & 3.7 (1.2) &  $< 1.6$ & 20 &4.1 (2.6)  & $<3.4$\\[3pt]
799 &    5 & 8.7 (1.2)& $<1.7$ & 10 & 3.2 (1.6) & $<2.4$ & 20 & 3.5 (3.6) & $<5.1$ \\[3pt]
800 &   5 & 7.8 (1.4)& $2.5$ &10 & 2.8 (1.9) & $3.5$ & 20 &3.1 (4.2)  & $7.5$ \\
		\hline
    \end{tabular}
  \begin{description} 
     \item The ortho states and para states (in parentheses) for H$_2$CCO are calculated separately, where $N$ based on each ortho transition ($2_{1,2}-1_{1,1}$ o- and $2_{1,1}-1_{1,0}$ o-) is done using a simultaneous fit to both lines. Estimates for t-HCOOH calculated from the $2_{0,2} - 1_{0,1}$ transition, with the exception of core 264 for which only the $2_{1,2}-1_{1,1}$ transition was detected and thus used in the LTE calculation. 
      \end{description}
\end{table*}

\begin{table*}
	\caption{Column Densities: RADEX results for CH$_3$CN}
 \setlength{\tabcolsep}{20pt}
	\label{tab:colden_cyn}
	\begin{tabular}{clllll} 
    \tablecolumns{5}
     \tablewidth{0pt}
     \tabcolsep=0.3cm
     & \multicolumn{2}{c|}{CH$_3$CN $2_1 - 1_1$  } & \multicolumn{2}{c|}{ CH$_3$CN $2_0 - 1_0$}  \\  
Core $\#$ & $N$ & $T_\mathrm{ex}$ &  $N$ & $T_\mathrm{ex}$ &  $N_\mathrm{avg}$ \\  
(\textit{Herschel}) &  [$\times 10^{12}$ cm$^{-2}$] &  [K]   &  [$\times 10^{12}$ cm$^{-2}$] &  [K] &   [$\times 10^{12}$ cm$^{-2}$] \\
\hline
264&  -- & --  &  0.171 $\SPSB{+ 0.02 }{- 0.02 }$&  8.66$^\mathrm{a}$ &  -- \\[2pt]
317& --& --& 0.106 $\SPSB{+ 0.03 }{- 0.03 }$&  12.3$^\mathrm{a}$&   -- \\[2pt]
321& -- &  -- &   $< 0.05$ &   12.5  & -- \\[2pt]
326&  0.271 $\SPSB{+ 0.04 }{- 0.06 }$&  13.2 $\SPSB{+ 0.6 }{- 0.6 }$&   0.326 $\SPSB{+ 0.05 }{- 0.04 }$&  13.2$^\mathrm{a}$&  0.298(0.028)\\[2pt]
413&  0.336 $\SPSB{+ 0.01 }{- 0.03 }$&  6.54 $\SPSB{+ 1.9 }{- 1.9 }$&   0.196 $\SPSB{+ 0.02 }{- 0.01 }$&  6.53$^\mathrm{a}$&  0.266(0.070)\\[2pt]
504&  --& --&  0.196 $\SPSB{+ 0.02 }{- 0.02 }$&  7.39$^\mathrm{a}$ &   -- \\[2pt]
615& -- &  -- &  $< 0.05 $ &   5.94  & -- \\[2pt]
627& --&  --&    0.166 $\SPSB{+ 0.02 }{- 0.02 }$&  6.30$^\mathrm{a}$ &  --\\[2pt]
709&  0.766 $\SPSB{+ 0.01 }{- 0.04 }$&  8.71 $\SPSB{+ 3.6 }{- 3.6 }$& 0.401 $\SPSB{+ 0.03 }{- 0.02 }$&  8.70$^\mathrm{a}$&  0.584(0.182)\\[2pt]
715&  0.351 $\SPSB{+ 0.03 }{- 0.06 }$&  9.22 $\SPSB{+ 4.4 }{- 4.4 }$&   0.251 $\SPSB{+ 0.03 }{- 0.02 }$&  9.22$^\mathrm{a}$&   0.301(0.050)\\[2pt]
746&  0.146 $\SPSB{+ 0.02 }{- 0.03 }$&  7.02 $\SPSB{+ 2.1 }{- 2.1 }$&   0.221 $\SPSB{+ 0.02 }{- 0.01 }$&  7.02$^\mathrm{a}$&   0.183(0.038)\\[2pt]
752&  0.396 $\SPSB{+ 0.01 }{- 0.01 }$&  5.57 $\SPSB{+ 2.0 }{- 2.0 }$&  0.151 $\SPSB{+ 0.02 }{- 0.01 }$&  5.57$^\mathrm{a}$&   0.274(0.122)\\[2pt]
768&  0.201 $\SPSB{+ 0.02 }{- 0.03 }$&  7.14 $\SPSB{+ 2.2 }{- 2.2 }$&   0.076 $\SPSB{+ 0.01 }{- 0.01 }$&  7.14$^\mathrm{a}$&  0.139(0.063)\\[2pt]
799&  --&  --&   0.081 $\SPSB{+ 0.01 }{- 0.02 }$&  5.81$^\mathrm{a}$&   -- \\[2pt]
800& -- &  -- & $< 0.05 $  & 8.67    & --  \\[2pt]
		\hline
    \end{tabular}
      \begin{description} 
    \item $^\mathrm{a}$ RADEX associated errors for these $T_\mathrm{ex}$ values are $<0.001$.
    \end{description}
\end{table*}

\begin{table}
	\centering
	\caption{Column Densities: CH$_2$CHCN }
 \setlength{\tabcolsep}{22pt}
	\label{tab:colden_vycyn}
	\begin{tabular}{cll} 
    \tablecolumns{10}
     \tablewidth{0pt}
     \tabcolsep=0.1cm
Core \#$^{1}$ & \multicolumn{2}{c|}{CH$_2$CHCN}  \\  
(\textit{Herschel}) &  $N$ [$\times 10^{12}$ cm$^{-2}$] & $T_\mathrm{ex}$  [K]  \\
\hline
54 & $< 0.686 $ &  7.39 \\ [1pt]
\textit{67} & $< 0.286$&  7.39 \\ [1pt]
130 &$< 0.884 $&  7.39  \\[1pt]
231 &  $< 0.720$  &  7.39  \\[1pt]
256 &  $< 0.735$  &  7.39 \\[1pt]
\textbf{264} &  0.227 $\SPSB{+ 0.055 }{- 0.055 }$&  8.66 \\ [1pt]
\textbf{317} & $< 0.212$  &  12.3 \\ [1pt]
\textbf{321} &$ < 0.153$ &  12.5  \\[1pt]
\textbf{326} &  $< 0.192$  &  13.2 \\ [1pt]
339  &  $< 0.833 $ &  7.39  \\[1pt]
344  & $< 0.833$  &  7.39  \\[1pt]
355  & $< 0.752 $ &  7.39    \\[1pt]
398  & $< 0.734$ & 7.39  \\[1pt]
\textbf{413} & 0.283 $\SPSB{+ 0.054 }{- 0.054 }$&  6.54   \\ [1pt]
414 & $<0.840$  &  7.39 \\[1pt]
479 & $< 1.10$  &  7.39 \\[1pt]
491 & $< 0.749 $ &  7.39  \\[1pt]
\textbf{504}  & $< 0.131 $ &  7.39  \\[1pt]
 543 & $< 0.848 $&  7.39 \\[1pt]
\textbf{615} & $< 0.156$  &  5.93 \\[1pt]
\textbf{627} & $< 0.169 $ &   6.29   \\[1pt]
 642  &  $< 1.51 $&  7.39  \\[1pt]
656  & $< 0.929  $&  7.39 \\[1pt]
 657  &  $< 1.33$  &  7.39 \\[1pt]
\textit{658} & $< 0.368 $ &  7.39 \\[1pt]
\textbf{709} & 0.269 $\SPSB{+ 0.040 }{- 0.040 }$&  8.71 \\[1pt]
 \textbf{715} &  0.727 $\SPSB{+ 0.143 }{- 0.143 }$&  9.22 \\[1pt]
739  & $< 0.963  $&  7.39  \\[1pt]
 \textbf{746} & 0.181 $\SPSB{+ 0.036 }{- 0.036 }$&  7.02 \\[1pt]
 747  & $<1.80$  &  7.39\\[1pt]
 \textbf{752} &  $< 0.145 $ &  5.57 \\[1pt]
\textbf{768} &  $<  0.118$  &  7.14  \\[1pt]
780   & $< 1.04  $&  7.39 \\ [1pt]
\textbf{799} &  $< 0.106$&  5.80 \\[1pt]
\textbf{800}  &  $<0.159 $  &  8.67  \\
		\hline
    \end{tabular}
      \begin{description} 
     \item $^{1}$ The cores observed with both the ARO 12m and Yebes 40m are bolded. For the italicized cores 67 and 658, the RMS is lower at $\sim2$\,mK.  
      \end{description}
\end{table}

\begin{figure}
\centering
\begin{center}$
\begin{array}{cc}
\includegraphics[width=80mm]{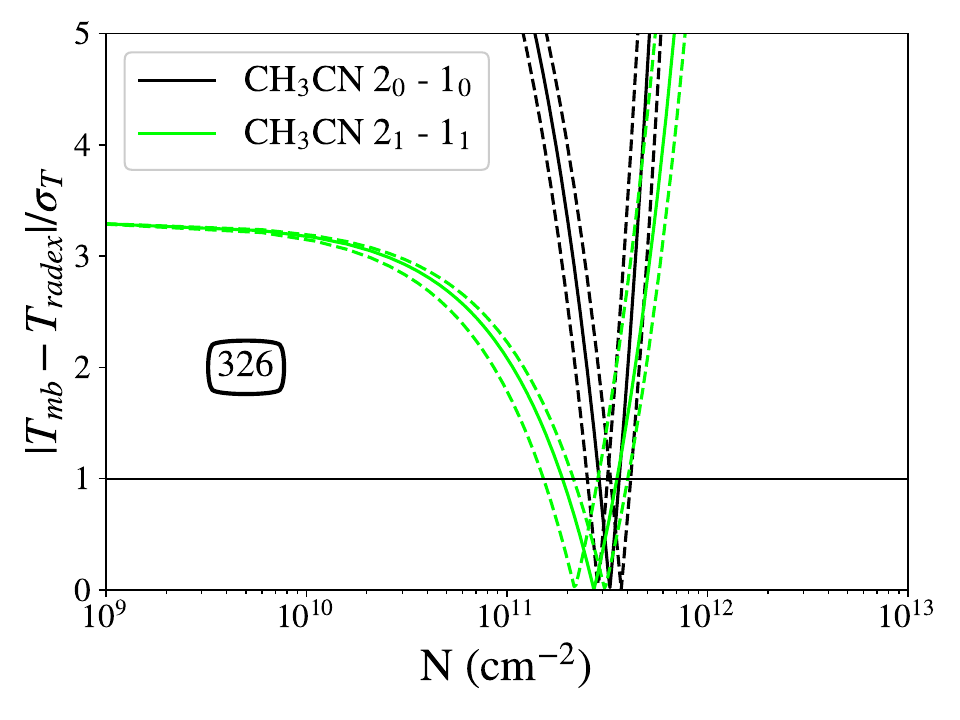} &  
\includegraphics[width=80mm]{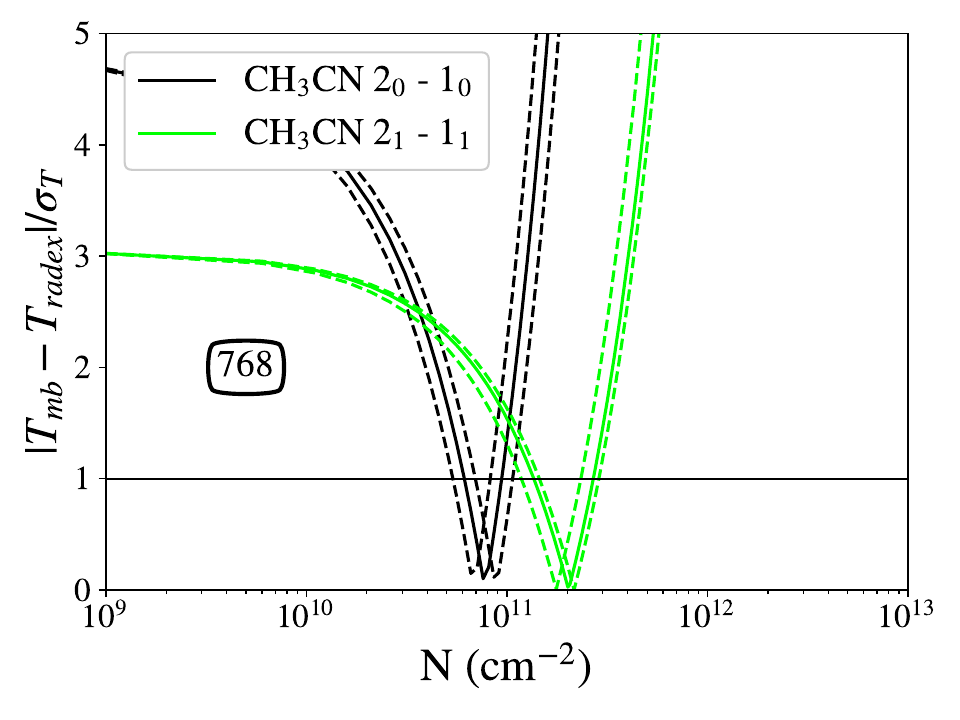} 
\end{array}$
\end{center}
\caption{\label{ch3cn_radex} RADEX minimization plots for CH$_3$CN for the two extreme cases where (top panel) for core 326 both $2-1$ lines agree very well, within a factor of 1.2, and (bottom panel) for core 768 when the $2-1$ lines differ by a factor of 2.6 (bottom panel). Plotted in each panel is the difference in the observed brightest temperature, $T_\mathrm{mb}$, and the RADEX-derived brightness temperature, $T_\mathrm{radex}$, divided by the RMS level, $\sigma_T$, versus the RADEX-derived column density, $N$. When the observations best-match the RADEX model, $|T_\mathrm{mb} - T_\mathrm{radex}|/\sigma_T$ is minimized and the best $N$ is found. The dashed lines represent the spread in error. 
}
\end{figure}
\subsection{Ketene, H$_2$CCO}

A nearly prolate symmetric rotor, H$_2$CCO is a molecule with no calculated collisional rates, whose laboratory spectroscopy at the frequencies cited in this study is detailed in \cite{1990JMoSp.140..340B}. In our Yebes 40m sub-sample, three transitions were observed with different $E_u/k$ values (15.9, 2.9 and 16\,K). However, the two transitions with the similar $E_u/k$ values of 15.9\,K and 16.0\,K are ortho species (meaning $K_a$ = odd with a statistical weight of 3) and the other transition is a para species (meaning $K_a$ = even with a statistical weight of 1). Thus, we use the LTE method to find $N$ for both ortho and para species separately and assume three different excitation temperatures at 5\,K, 10\,K and 20\,K (see Table\,\ref{tab:colden_5atom} for all estimates). 

There is not a straightforward reference to guide an estimate for a true filling factor for H$_2$CCO since each transition was observed with the Yebes 40m and has roughly the same beam size, and we know for the prestellar core L1544 the spatial distribution of H$_2$CCO does not follow that of CH$_3$OH \citep{2017A&A...606A..82S}.
We therefore assume $f = 1$ to find, for $T_\mathrm{ex}$ of 10\,K, $N$(o-H$_2$CCO) range from $1.8 - 7.6 \times 10^{12}$\,cm$^{-2}$ and $N$(p-H$_2$CCO) range from $0.8 - 4.4 \times 10^{12}$\,cm$^{-2}$. Even so, an $f=1$ assumption is still likely underestimating $N$ and if we do assume the average $f$ value from our CH$_3$OH analysis ($f = 0.4$) this would increase our $N$ values by a factor of 2.5. These column densities are still found to be on average lower than updated estimates at the dust peak of prestellar core L1544, where \citealt{2021ApJ...917...44J} find for H$_2$CCO $T_\mathrm{ex}$ = 5-6\,K and $N = 1.5 \pm 0.2 \times 10^{13}$\,cm$^{-2}$. 

Under LTE assumptions the ortho-to-para ratios for the cores in most cases are $>1$, with a median value of 2.4. In some Taurus cores, such as TMC-1 and  L1689B, this ratio is close to the statistical ratio of 3 \citep{1991ASPC...16..387O, 2012A&A...541L..12B} and in others, such as L1517B, they find a value of $\sim$1 \citep{2023MNRAS.519.1601M}. For ratios $ > 3$ this is consistent with the `high temperature limit' whereas for ratios $< 3$ this would be consistent with equilibrium at 10\,K \citep{1991ASPC...16..387O}. These ratios for the Perseus cores should be taken lightly, however, since we are assuming LTE and $f = 1$. It is also likely $f$ varies from source to source (as it does for CH$_3$OH) and for the different ortho and para species of H$_2$CCO, which can alter the true ortho-to-para ratio. 

\subsection{Formic Acid, t-HCOOH}

There are three energetically favorable $2-1$ transitions available to analyze for the \textit{trans} isomer of formic acid, t-HCOOH, from the Yebes 40m observations (Table\,\ref{LineList}; spectroscopic work detailed in \cite{1971JMoSt...9...49B, 2002JMoSp.216..259W}). We first attempted to use the CTEX method to constrain the excitation temperature, $T_\mathrm{ex}$. This method was unsuccessful, however, because unlike H$_2$CCO the $E_u/k$ values for these transitions do not span a wide enough range, i.e., values of 3.2\,K, 6.2\,K and 6.5\,K. 
Instead, we implement the LTE method to calculate $N$ (assuming $f = 1$) by using the more favorable $2_{0,2} - 1_{0,1}$ transition detected in all cores, with the exception of core 264 for which only the $2_{1,2}-1_{1,1}$ transition was detected and thus used in the LTE calculation. We calculate $N$ for three different $T_\mathrm{ex}$ values at 5\,K, 10\,K and 20\,K (Table\,\ref{tab:colden_5atom}). For the $T_\mathrm{ex}$ = 10\,K case $N$ values range from $1.5 - 15 \times 10^{12}$\,cm$^{-2}$. We note that, similar to H$_2$CCO, if we instead assume the average $f$ value from our CH$_3$OH analysis this would increase $N$ by a factor of 2.5. Compared to prestellar core L1544, where \cite{2014ApJ...795L...2V} find $N = 5 \times 10^{11}$\,cm$^{-2}$ for a $T_\mathrm{ex}$=10\,K and \cite{2021ApJ...917...44J} similarly find 4.8$\pm$0.9 $\times10^{11}$\,cm$^{-2}$ for $T_\mathrm{ex} = 12.6 \pm6.9$\,K, the Perseus cores with t-HCOOH detections have an order of magnitude higher $N$ values ($\sim$ a few $\times 10^{12}$\,cm$^{-2}$ for $f =1$).

\subsection{Methyl Cyanide, CH$_3$CN}

The RADEX method is used for CH$_3$CN, since there are collisional rates calculated \citep{1986ApJ...309..331G}. Following the standardized procedure described for CH$_3$OH (section\,\ref{methcoldensection}), we use the two $2-1$ transitions (Table\,\ref{LineList}; spectroscopic work detailed in \cite{1982JChPh..76...97K}) to minimize over a grid of RADEX calculations to find the best fit $N$ values while fixing the volume density, ${n_\mathrm{H_2}}$, and the kinetic temperature, $T_{k}$ (columns 9 and 10 in Table\,\ref{physparams}). Since these transitions were observed close in frequency with the Yebes 40m (Figure\,\ref{methcyn_spec_yebes40m}) and the spatial distribution of this N-bearing molecule is not known in these sources, we assume $f = 1$ while constraining $N$ for this 15 core sub-sample (see Table\,\ref{tab:colden_cyn} and Figure\,\ref{ch3cn_radex}). If one were to assume the average $f = 0.4$ (i.e., from our CH$_3$OH analysis), we find this can increase $N$ by factors of 2.4-2.5.

In cores 264, 317, 504, 627 and 799 only the brighter $2_0-1_0$ transitions is detected above $3\sigma$ and used to constrain $N$. We note that for the remaining cores with constraints for both $2-1$ transitions, the best-fit $N$ value calculated for each separate transition differ at minimum by a factor of 1.2 for core 326 and at maximum a factor of 2.6 for core 768 (see  Figure\,\ref{ch3cn_radex}), and report an average $N_\mathrm{avg}$ value in Table\,\ref{tab:colden_cyn}. For cores 326 and 746 the $2_0-1_0$ line finds a higher $N$ value whereas for the rest of the cores a higher $N$ is found from the $2_1-1_1$ fit. It is shown in Figure\,\ref{methcyn_spec_yebes40m} that the disagreement is likely due to the deviation in peak intensity. For core 326 the $2_0-1_0$ and $2_1-1_1$ ratio is 2.8 whereas in core 768 the ratio is 1.5. 
Note that for the seven cores with more than two transitions, we do explore simultaneous minimization with varying ${n_\mathrm{H_2}}$ and $T_{k}$ and find $N$ values differ by no more than factors of $<3$ from our values in Table\,\ref{tab:colden_cyn} (see Appendix\,\ref{radexappendix} for more detail). For cores 321, 615 and 800, where CH$_3$CN isn't detected at $>3\sigma$ we report an upper limit for $N$ from our RADEX grid where the peak $2_0-1_0$ intensity from the model equals $3\sigma$.  

\subsection{Vinyl Cyanide, CH$_2$CHCN}

For the large N-bearing molecule CH$_2$CHCN, an LTE method needed to be used. We first attempted the CTEX method on core 413 as two transitions of CH$_2$CHCN were detected, the $4_{1,4} - 3_{1,3}$ and $4_{0,4} - 3_{0,3}$ lines (Table\,\ref{LineList}; spectroscopic work detailed in \cite{1985ZNatA..40..998S, 1988JMoSp.130..303C}). However, due to their similar upper energies at $E_{u} = 6.6$\,K and $E_{u} = 4.5$\,K, respectively, this method was unable to constrain an excitation temperature, $T_\mathrm{ex}$. Instead, for this N-bearing species we assume the $T_\mathrm{ex}$ is comparable to CH$_3$CN. In Table\,\ref{tab:colden_vycyn} are the predicted column densities and 3$\sigma$ upper limits based on LTE calculations assuming, for the Yebes 40m 15 core sub-sample, that core's CH$_3$CN $T_\mathrm{ex}$ value, and, for the remaining cores with only ARO 12m constraints, the median CH$_3$CN $T_\mathrm{ex}$ value from that 15 core sub-sample. 

For the cores with at least one transition detected from the Yebes 40m data, the observed peak intensity from the $4_{1,4} - 3_{1,3}$ (for cores 264 and 413), $4_{0,4} - 3_{0,3}$ (for cores 709 and 746) or $4_{2,3}-3_{2,2}$ (core 715) transition is used. For other cores with Yebes spectra the $3\sigma$ value is used to estimate $N$ based on the LTE-predicted brightest $4_{0,4} - 3_{0,3}$ transition. 
And, for the remaining cores, with constraints from just the ARO 12m data, we use $3\sigma$ values based on the noise level from the 94.27\,GHz line, the $10_{0,10}-9_{0,9}$ transition. Our $N$ estimates range from $0.18 - 0.72 \times 10^{12}$\,cm$^{-2}$. The median, however, is 0.27$\times 10^{12}$\,cm$^{-2}$ and in only core 715 is the higher value of 0.72$\times 10^{12}$\,cm$^{-2}$ found from the line intensity of the $4_{2,3}-3_{2,2}$ transition where, according to LTE estimates, should be $\sim 5-8\times$ weaker than the other $4-3$ transitions. As is assumed for CH$_3$CN, we have calculate our $N$ values where $f=1$ and if instead an average value from the CH$_3$OH analysis is applied, i.e., $f=0.4$, this would increase our $N$ values by a factor of 2.5.

\subsubsection{Methyl Formate, HCOOCH$_3$}

HCOOCH$_3$ is an asymmetric rotor with internal rotation due to its methyl group, CH$_3$, which splits rotational levels into A and E substates (as in CH$_3$OH and CH$_3$CHO). Laboratory spectroscopy in the frequency range described here has been detailed in \cite{1979JPCRD...8..583B, 2001JMoSp.210..196K, 2004JMoSp.225...14O, 2009JMoSp.255...32I}. Collisional rates for the A-state transition of HCOOCH$_3$ are also available \citep{2014ApJ...783...72F, 2020Atoms...8...15V}. We therefore use RADEX in our estimates of the column density for the Yebes 40m sub-sample of cores. As done for CH$_3$OH and CH$_3$CN, a grid of RADEX models are run and minimized to find a best-fit $N$ while fixing the volume density, ${n_\mathrm{H_2}}$, and the kinetic temperature, $T_{k}$, (columns 9 and 10 in Table\,\ref{physparams}). 

For core 326 we are again in a position where we have multiple transitions that could potentially span a wide enough frequency range where the beam size varies enough so it is possible to try and account for the true filling factor, $f$. We run the minimization procedure for the detected $3_{0,3}-2_{0,2}$ A transition at 36.102\,GHz, the $3_{2,2}-2_{2,1}$ transition at 36.657\,GHz, and the $4_{0,4} - 3_{0,3}$ transition at 47.536\,GHz, which have beam sizes of 50, 49, and 38 arcsec, respectively. If  we assume the emission is coming from the same source size as found from CH$_3$OH constraints, i.e., for core 326 this is $\theta_\mathrm{src} = 51.5$\,arcsec, the minimized $N$ value for the transitions are in slightly worse agreement (e.g., between the $3_{0,3}-2_{0,2}$ and $4_{0,4} - 3_{0,3}$ transitions by a factor of 1.6) than for the $f = 1$ case (by a factor of 1.3). If the minimization code is then run for a variety of $\theta_\mathrm{src}$ values, we find $\theta_\mathrm{src} \gg \theta_\mathrm{beam}$ is still the best match, i.e., when $f = 1$. This could suggest that the CH$_3$OH emission does not trace the HCOOCH$_3$ emission, at least for the core 326, or that this beam size range is not wide enough to set a strong constraint. Since an $f = 1$ is a better fit to the data for core 326, compared to assuming the methanol source size, and we have no other $f$ constraints from the remaining cores, we adopt the assumption that $f = 1$ in our calculations for all of the cores (note: if $f = 0.4$ is assumed $N$ values increase by factors of 2.4-2.5). 

To help remedy the disagreement between the $3_{0,3}-2_{0,2}$ and $4_{0,4} - 3_{0,3}$ transitions for core 326 (see Table\,\ref{tab:colden_mf}), we also attempt to fit simultaneously all A transitions while letting ${n_\mathrm{H_2}}$ and $T_{k}$ vary, finding elevated $T_{k}$ values more consistent with $\sim 20$\,K (see Appendix\,\ref{radexappendix} for more details). Still, the $N$ values derived in this method are consistent (within less than a factor of 2) with the median $N$ derived calculated from minimizing each transitions separately and thus for consistency we report for each core the $N$  from our fixed ${n_\mathrm{H_2}}$ and $T_{k}$, calculations. 

It is in Table\,\ref{tab:colden_mf} where the RADEX derived column densities, $N$, excitation temperatures, $T_\mathrm{ex}$, and opacity, $\tau$, values are listed for each of the detected transitions in cores 264, 321, 326, 715 and 768. Note for core 768 only the E-state $4_{1,4} - 3_{1,3}$ transition is detected and therefore used to constrain $N$. For the remaining cores, an upper limit for $N$ is derived from the RADEX grid where the peak $3_{0,3} - 2_{0,2}$ intensity from the model equals $3\sigma$ to give the tightest constraint.

\begin{table}
 	\centering
	\caption{Column Densities: RADEX Results for HCOOCH$_3$ A}
 \setlength{\tabcolsep}{8pt}
	\label{tab:colden_mf}
	\begin{tabular}{cllll} 
    \tablecolumns{5}
     \tablewidth{0pt}
     \tabcolsep=0.3cm
     & \multicolumn{3}{c|}{HCOOCH$_3$ A }  \\  
Core $\#$ & Transition & $N$ & $T_\mathrm{ex}$ \\  
(\textit{Herschel}) &  & [$\times 10^{12}$ cm$^{-2}$] &  [K]   \\
\hline
264&  $4_{1,4} - 3_{1,3}$ & 2.51 $\SPSB{+ 0.55 }{- 0.55 }$&  10.6$^\mathrm{a}$  \\[1pt]
   &  $4_{0,4} - 3_{0,3}$ &  2.31 $\SPSB{+ 0.80 }{- 0.75 }$&  10.7$^\mathrm{a}$\\[1pt]
317&  $3_{0,3} - 2_{0,2}$ & $<1.26$  & 14.3 \\[1pt] 
321&  $4_{1,4} - 3_{1,3}$ & 7.26 $\SPSB{+ 1.25 }{- 1.25 }$&  12.6$^\mathrm{a}$\\[1pt]
326&  $3_{2,2} - 2_{2,1}$ & 8.16 $\SPSB{+ 1.15 }{- 1.10 }$&  12.4$^\mathrm{a}$\\[1pt]
   &  $3_{0,3} - 2_{0,2}$ & 1.26 $\SPSB{+ 0.30 }{- 0.30 }$&  12.5$^\mathrm{a}$\\[1pt]
   &  $4_{0,4} - 3_{0,3}$ &  5.36 $\SPSB{+ 0.95 }{- 0.95 }$&  12.3$^\mathrm{a}$\\[1pt]
413& $3_{0,3} - 2_{0,2}$   & $< 1.16$ &  11.1  \\[1pt] 
504&  $3_{0,3} - 2_{0,2}$ & $< 1.16 $  &  10.0  \\[1pt] 
615&  $3_{0,3} - 2_{0,2}$ & $< 1.21 $ &  11.0  \\[1pt] 
627&  $3_{0,3} - 2_{0,2}$ & $< 1.11 $ &  13.9  \\[1pt] 
709&  $3_{0,3} - 2_{0,2}$ & $< 1.21$ &  14.8  \\[1pt] 
715&  $3_{0,3} - 2_{0,2}$ & 1.26 $\SPSB{+ 0.60 }{- 0.50 }$&  16.5$^\mathrm{a}$\\[1pt] 
746&  $3_{0,3} - 2_{0,2}$ & $< 1.16 $ &  11.2  \\[1pt]
752&  $3_{0,3} - 2_{0,2}$ & $< 1.06 $ &  18.2 \\[1pt]
768 & $4_{1,4} - 3_{1,3}$ E$^{*}$  & 1.71 $\SPSB{+ 0.05 }{- 0.05 }$&  10.5$^\mathrm{a}$\\[1pt]
799&  $3_{0,3} - 2_{0,2}$ & $< 0.91 $ &  11.1  \\[1pt][1pt]
800&  $3_{0,3} - 2_{0,2}$ & $< 1.36 $ &  12.3 \\[1pt]
		\hline
    \end{tabular}
      \begin{description} 
     \item $^{*}$ The peak intensity from the detected E-state is used to constrain $N$, as no A-state transition is detected for this core. $^\mathrm{a}$ RADEX associated errors for these $T_\mathrm{ex}$ values are $<0.001$.
      \end{description}
\end{table}

\begin{table}
	\centering
	\caption{Column Densities: CH$_3$OCH$_3$ EE}
 \setlength{\tabcolsep}{14pt}
	\label{tab:colden_dme}
	\begin{tabular}{cll} 
    \tablecolumns{10}
     \tablewidth{0pt}
     \tabcolsep=0.1cm
Core \#$^{1}$ & \multicolumn{2}{c|}{CH$_3$OCH$_3$ EE}  \\  
(\textit{Herschel}) &  $N$ [$\times 10^{12}$ cm$^{-2}$] & $T_\mathrm{ex}$  [K]  \\
\hline
 {264} &    4.54$\pm 0.48$ & 11.4 \\ [1pt]
 {317} &  $<2.04 $ &  11.4 \\ [1pt]
 {321} &   5.91$\pm$0.72 & 11.4  \\ [1pt] 
 {326} &    4.32$\pm$0.57  &   11.4$\pm$5.1 \\ [1pt]
 {413} &   $<1.77$ &  11.4\\ [1pt]
{504}  & $<1.21$  &  11.4\\ [1pt]
{615}  &  $< 1.72$ &   11.4 \\ [1pt]
{627} &   7.46$\pm0.72$  &  11.4  \\ [1pt]
{709} &  $< 1.75$ &   11.4 \\ [1pt]
{715} &    $< 1.90$  &  11.4  \\ [1pt]
{746} & $< 1.60 $   &  11.4  \\ [1pt]
{752} &   $< 1.32$   &  11.4  \\ [1pt]
{768} &  $< 1.24 $ &  11.4  \\ [1pt]
{799} &  $< 1.67$  &  11.4  \\ [1pt]
{800}  &    $< 1.53$ &   11.4 \\ [1pt]
		\hline
    \end{tabular}
      \begin{description} 
     \item $^{1}$ Calculation for Core 326 done via the CTEX method, the remaining with the LTE method assuming the same $T_\mathrm{ex}$ value. For all cores $f = 1$.
      \end{description}
\end{table}

\subsection{Dimethyl Ether, CH$_3$OCH$_3$}

For CH$_3$OCH$_3$, a more complex asymmetric rotor with two methyl groups, the rotational levels are are split into AA, EE, EA and AE substates (spectroscopic details in \cite{1976JMoSp..62..159D, 1979JPCRD...8.1051L, 2009A&A...504..635E}). We use only the brightest EE state to calculate column densities using the CTEX and LTE methods. Because we do not know how extended the emission is, we assume $f=1$. For core 326 there are two bright EE transitions, $3_{1,2} - 3_{0,3}$ and $4_{1,3} - 4_{0,4}$, with a large enough $E_u$ gap, at 7\,K and 11\,K, respectively, that the CTEX method can constrain both $N$ and $T_\mathrm{ex}$ (Table\,\ref{tab:colden_dme}). Note that if instead we do assume the CH$_3$OH source size found for core 326, our $N$ increases by a factor of 2, to $N = 9.47\pm4.05 \times 10^{12}$\,cm$^{-2}$ for a $T_\mathrm{ex} = 6.02\pm1.8$\,K. Using the constraints from the $f=1$ case, at $T_\mathrm{ex} = 11.4$\,K the $N$ for the remaining cores is constrained, which range from $N = 4.54-7.46 \times 10^{12}$\,cm$^{-2}$ (Table\,\ref{tab:colden_dme}).


\clearpage

\section{Physical Parameters from RADEX Calculations} \label{radexappendix}

The sub-sample of 15 cores from the Yebes 40m data have enough transitions ($>$1) for each of the CH$_3$OH A and E substates that we can attempt to independently constrain a volume density, $n{(\mathrm{H{_2}})}$, and kinetic temperature, $T_\mathrm{k}$ while also fitting for $N$. In this exercise we create a model suite with a a range of $T_\mathrm{k}$ from 3 to 23 (intervals of 1\,K), with $n{(\mathrm{H{_2}})}$ sampled evenly in log space from 4 to 6.5 (20 grid points), and $N$ sampled evenly in log space from 12 to 15.5 (20 grid points). Models (8000 total) are minimized over $|T_\mathrm{mb}-T_\mathrm{radex}|/\sigma_T$ (as in section\,\ref{methcoldensection}) for the two $2 - 1$ E-state transitions observed with the same beam size from ARO 12m observations. We also include in our inputs the best-fit source sizes derived from our analysis in the main text (from minimizing over the $2_{0,2} - 1_{0,1}$\,A and $1_{0,1} - 0_{0,0}$\,A transitions). In Table\,\ref{tab:radex_phys} we list the best-fit models. Considering we sample a much coarser grid, we find $N$ values in general consistent, within factors $<$3, when compared to our fixed $n{(\mathrm{H{_2}})}$ and $T_\mathrm{k}$ method (listed in Table\,\ref{tab:colden_met}).

For a handful of cores (seven total) we can simultaneously fit the $2_1 - 1_1$ and $2_0 - 1_0$ lines of  CH$_3$CN by running a grid with a range of $T_\mathrm{k}$ from 3 to 23\,K (intervals of 1\,K), with $n{(\mathrm{H{_2}})}$ sampled evenly in log space from 4 to 6.5 (20 grid points), and $N$ sampled evenly in log space from 10 to 13.5 (20 grid points). We find $N$ values consistent, within factors $<$3, when compared to our fixed $n{(\mathrm{H{_2}})}$ and $T_\mathrm{k}$ method (listed in Table\,\ref{tab:colden_cyn}). Our best-fit $T_\mathrm{k}$ values span a wide range, however, from $6-22$\,K when minimized  (see Table\,\ref{tab:radex_phys}). 

In the case of HCOOCH$_3$, there are two cores (264 and 326) with multiple A state transitions we can use to attempt to constrain $T_\mathrm{k}$ and $n{(\mathrm{H{_2}})}$. We run our grid with a range of $T_\mathrm{k}$ from 3 to 23 (intervals of 1\,K), with $n{(\mathrm{H{_2}})}$ sampled evenly in log space from 4 to 6.5 (20 grid points), and $N$ sampled evenly in log space from 11 to 14.5 (20 grid points). 

We find for cores 264 and 326 they are both minimized for $N$ at 2.9$\times$10$^{12}$ cm$^{-2}$, which is consistent with the median values derived in the main text from Table\,\ref{tab:colden_mf} within a factor of 2 (see Table\,\ref{tab:radex_phys}). We do find 2$\times$ higher $T_\mathrm{k}$ values, $\sim 20$\,K, than what has been derived from NH$_3$ as well as $n{(\mathrm{H{_2}})}$ values varying within a factor of 3 when compared to our beam-averaged assumption (see Table\,\ref{physparams} for reference). 

\begin{table}
 	\centering
	\caption{RADEX Derived Best-fit Physical Parameters}
 \setlength{\tabcolsep}{3pt}
	\label{tab:radex_phys}
	\begin{tabular}{cclllllc} 
    \tablecolumns{8}
     \tablewidth{0pt}
     \tabcolsep=0.3cm
Molecule & Core $\#$ &  $T_\mathrm{k}$ &  $n{(\mathrm{H{_2}})}$  &  $N$ & $T_\mathrm{ex}$ & $\tau$ \\  
 &  (\textit{Herschel}) & [K] & [10$^5$ cm$^{-3}$] & [cm$^{-2}$] &  [K]  &  &  $\frac{|T_\mathrm{mb}-T_\mathrm{radex}|}{\sigma_T}$ \\
\hline
CH$_3$OH E & 264 & 6, 5, 6 & 0.33e5, 1.5e5, 0.13e5 & 16e13, 16e13, 37e13  & 5, 5, 5 & 2.7, 2.7, 6.7  & < 2\\
         & 317 & 10, 20, 21 & 2.8e5, 2.8e5, 2.8e5 & 10e13, 10e13, 10e13 & 9, 19, 20 & 0.637, 0.193, 0.176  & < 1 \\
         & 321 & 7, 6 & 2.8e5, 1.5e5 & 10e13, 16e13 & 6, 6 & 0.992, 1.956  & < 2\\
         & 326 & 9, 10 & 0.45e5, 0.18e5 & 37e13, 57e13 & 7, 7 & 2.3, 3.9  & < 2\\
         & 413 & 13, 11, 7 & 0.33e5, 0.61e5, 0.45e5 & 2.9e13, 2.9e13, 4.5e13 & 8, 8, 5 & 0.36, 0.32, 0.88 &  < 0.8 \\  
         & 504 &   11 & 0.18e5 & 10e13 & 6 & 1.8 & < 1\\
         & 615 & 13, 12, 12, 9 & 0.45e5, 0.61e5, 0.18e5, 1.0e5 & 4.5e13, 4.5e13, 6.9e13, 6.9e13 & 8, 9, 6, 6 & 0.526, 0.507, 1.2, 1.2 & < 1\\
         & 627 & 5, 5, 6, 6 & 1.1e5, 1.5e5, 0.18e5, 0.13e5 & 6.9e13, 6.9e13, 10e13, 16e13 & 5, 5, 5, 5 & 2.6, 2.5, 4.2, 6.3  & < 2\\
         & 709 & 7 & 2.8e5 &  10e13 & 7 & 1.15 & < 1 \\
         & 715 & 14, 20 & 1.1e5, 1.1e5 & 4.5e13, 4.5e13 & 11, 18 & 0.273, 0.146 & < 0.4 \\
         
         & 746 & 21, 17 & 0.33e5, 0.13e5 & 4.5e13, 6.9e13 & 12, 7 & 0.306, 1.0 & < 0.4 \\
         & 752 & 8, 8 & 1.1e5, 0.18e5 & 6.9e13, 16e13 & 7, 5 & 0.910, 3.0 & < 2 \\
         & 768 & 10 & 0.61e5 & 1.9e13 & 7 & 0.232 & < 0.5\\
         & 799 & 12, 11 & 0.33e5, 0.45e5 & 4.5e13, 4.5e13 & 8, 8 & 0.913, 0.8829 & < 1 \\
         & 800 & 16, 11 & 0.61e5, 0.33e5 & 6.9e13, 10e13 & 12, 7 & 0.465, 1.4 & < 1 \\
\hline
CH$_3$CN & 326 & 10, 9 & 2.8e5, 3.7e5 & 0.29e12, 0.29e12 & 10, 9 & 0.003, 0.004 & < 0.15 \\
         & 413 & 16 & 0.13e5 & 0.29e12 & 5 & 0.015 & < 0.2\\
         & 709 &  22, 20 & 0.10e5, 0.13e5 & 0.69e12, 0.69e12 & 4, 5 & 0.028, 0.026 & < 1.2\\
         & 715 & 15, 16 & 0.10e5, 0.10e5 & 0.45e12, 0.45e12 & 4, 4 & 0.025, 0.023 & < 0.15 \\
         & 746 & 6 & 0.13e5 & 0.45e12 & 4 & 0.034 & < 0.15 \\
         & 752 & 21, 22 & 0.10e5, 0.10e5 & 0.29e12, 0.29e12 & 4, 4 & 0.012, 0.011  & < 2\\
         & 768 & 22 & 0.13e5 & 0.13e12 & 5 & 0.003 & < 0.7 \\
\hline 
HCOOCH$_3$ A & 264& 21,19,18 & 1.1e5, 1.5e5, 2.1e5  & 2.9e12, 2.9e12, 2.9e12 &  21, 19, 18 &  0.001, 0.001, 0.001 &  < 0.1 \\[1pt]
  
& 326&  22, 21 & 1.1e5, 1.5e5 & 2.9e12, 2.9e12 & 22, 21 & 0.0007, 0.0008  & < 4.1 \\[1pt]
		\hline
    \end{tabular}
      \begin{description} 
     \item If multiple models fit the minimization criteria (column 8) they are listed sequentially in the table. Note the $T_\mathrm{ex}$ and $\tau$ values for CH$_3$OH from the $2_{-1,2}-1_{-1,1}$ fit, CH$_3$CN from the $2_0 - 1_0$ fit, and for HCOOCH$_3$ A from the $4_{0,4}-3_{0,3}$ fit. 
      \end{description}
\end{table}

\bsp	
\label{lastpage}
\end{document}